\documentclass[11pt,a4paper]{article}
\usepackage{epsfig}
\usepackage[T1]{fontenc}    
\usepackage{graphics}
\usepackage{graphicx}
\usepackage{pstricks,pst-coil,pst-fill,pst-plot}
\usepackage[fleqn]{amsmath}    
\usepackage{amssymb}    
\usepackage{amsfonts}   
\usepackage{verbatim}   
\usepackage{mathrsfs}   
\usepackage{dsfont}
\usepackage{euscript}
\usepackage{yfonts}
\usepackage{enumerate}     
\usepackage{txfonts}
\usepackage{marvosym}
\usepackage{vmargin}        

\setmarginsrb{1.8cm}{2cm}{1.8cm}{2cm}{1cm}{1cm}{1cm}{1.6cm}
 \makeatletter
 \@addtoreset{equation}{section}
 \makeatother

\providecommand{\bysame}{\leavevmode\hbox to3em{\hrulefill}\thinspace}
\providecommand{\MR}{\relax\ifhmode\unskip\space\fi MR }

\providecommand{\href}[2]{#2}

\newcommand{\bra}[1]{\big< \,#1\,|}
\newcommand{\ket}[1]{|\,#1\, \big>}

\let\tend=\rightarrow

\long\def\symbolfootnote[#1]#2{\begingroup%
\def\thefootnote{\fnsymbol{footnote}}\footnote[#1]{#2}\endgroup}

\newtheorem{theorem}{Theorem}[section]
\newtheorem{prop}{Proposition}[section]

\newtheorem{lemme}{Lemma}[section]

\def\Proof{\medskip\noindent {\it Proof --- \ }}

\def\qed{\hfill\rule{2mm}{2mm}}

\newcommand\beq{\begin{equation}}
\newcommand\enq{\end{equation}}
\newcommand\bem{\begin{multline}}
\newcommand\enm{\end{multline}}

\def\beqa{\begin{eqnarray}}
\def\eeqa{\end{eqnarray}}
\def\ba{\begin{array}}
\def\ea{\end{array}}
\def\det{\operatorname{det}}

\newcommand{\f}[2]{{\ensuremath{%
    \mathchoice%
    {\dfrac{#1}{#2}}
    {\dfrac{#1}{#2}}
    {\frac{#1}{#2}}
    {\frac{#1}{#2}}
}}}
\newcommand{\tf}[2]{\ensuremath{#1/#2}}
\newcommand{\pa}[1]{\ensuremath{\left(#1\right)}}

\newcommand{\pab}[2]{\ensuremath{\pa{\ba{c} #1 \\ #2 \ea }}}

\def\a{\alpha}

\def\be{\beta}

\def\Ga{\Gamma}

\def\de{\delta}

\def\De{\Delta}
\def\eps{\epsilon}
\def\veps{\varepsilon}
\def\la{\lambda}

\def\sg{\sigma}
\def\vsg{\varsigma}

\def\th{\theta}

\def\om{\omega}

\def\vsg{\varsigma}

\def\i{\mathrm{i}}

\newcommand{\mc}[1]{\ensuremath{\mathcal{#1}}}
\newcommand{\mf}[1]{\ensuremath{\mathfrak{#1}}}
\newcommand{\msc}[1]{\ensuremath{\mathscr{#1}}}

\newcommand{\bs}[1]{\ensuremath{\boldsymbol{#1}}}

\newcommand{\ov}[1]{\ensuremath{\overline{#1}}}
\newcommand{\wt}[1]{\ensuremath{\widetilde{#1}}}
\newcommand{\wh}[1]{\ensuremath{\widehat{#1}}}

\newcommand{\Int}[2]{\ensuremath{\int\limits_{#1}^{#2}}}
\newcommand{\Oint}[2]{\ensuremath{\oint\limits_{#1}^{#2}}}

\newcommand{\sul}[2]{\ensuremath{\sum\limits_{#1}^{#2}}}
\newcommand{\pl}[2]{\ensuremath{\prod\limits_{#1}^{#2}}}

\newcommand{\R}{\ensuremath{\mathbb{R}}}

\newcommand{\Dp}[1]{\ensuremath{\partial_{#1}}}

\newcommand{\ex}[1]{\ensuremath{\e{e}^{#1}}}

\newcommand{\dd}{\mathrm{d}}
\newcommand{\e}[1]{\ensuremath{\mathrm{#1}}}

\newcommand{\intff}[2]{\ensuremath{ [  #1 \,; #2 ] }}

\newcommand{\intoo}[2]{\ensuremath{ ]  #1 \,; #2 [ }}

\newcommand{\intn}[2]{\ensuremath{[\![ \, #1 \,;\, #2 \,]\!]}}

\begin{document}

\begin{center}
\begin{LARGE}
{\bf Microscopic approach to a class of 1D quantum critical models}
\end{LARGE}

\vspace{30pt}

\begin{large}

{\large Karol K. Kozlowski\footnote{e-mail: karol.kozlowski@u-bourgogne.fr}}%
\\[1ex]
IMB, UMR 5584 du CNRS,
Universit\'e de Bourgogne, France. \\[2.5ex]

{\large Jean Michel Maillet \footnote{e-mail: maillet@ens-lyon.fr}}%
\\[1ex]
Laboratoire de Physique, UMR 5672
du CNRS, ENS Lyon,  France.

\par

\end{large}

\vspace{40pt}

\centerline{\bf Abstract} \vspace{1cm}
\parbox{12cm}{\small
Starting from the finite volume form factors of local operators, we show how and under which hypothesis the $c=1$ free boson conformal field theory
in two-dimensions emerges as an effective theory governing the large-distance regime of multi-point correlation functions for a large class of one dimensional massless
quantum Hamiltonians. 
In our approach, in the large-distance critical regime, the local operators of the initial model are represented by well suited vertex operators associated to  the free
boson model. This provides an effective field theoretic description of the large distance behaviour of correlation functions in 1D quantum critical models. 
We develop this description starting from the first principles and directly at the microscopic level, 
namely in terms of the properties of the finite volume matrix elements of local operators. }

\end{center}

\vspace{40pt}

\section*{Introduction}

It is widely believed  that the spectrum of a quantum Hamiltonian is intimately related to certain overall properties of the large-distance 
asymptotic behaviour of its correlation functions. If, in the infinite volume $L$ limit, a given model's spectrum exhibits a gap 
between a \textit{finitely} degenerated ground state and the tower of excited states above it, then the correlation functions are
expected to decay exponentially fast in the distance of separation between the various operators involved in the correlator. 
This behaviour changes for gapless models, \textit{viz}. those whose ground state, in  the $L\tend +\infty$ limit,  becomes 
directly connected to the continuum of excited states above it. For those models, the correlation functions are expected to decay algebraically in terms of the distance
of separation. The powers of the distance which arise in this algebraic decay are called critical exponents. Models having the same values of 
their critical exponents are said to belong to the same universality class. In fact, it is believed that the features determining a given universality class
are rather sparse in the sense that solely certain overall symmetries of the model should fix it \cite{GriffithsUniversalityAndExponentsParameterDepending}.
Unfortunately, the lack of any exactly solvable and truly interacting many-body quantum critical model in dimensions higher than one did not allow, so far, 
for a direct and explicit check of these properties. However, extensive numerical
data and experimental results do speak in favour of such an interplay \cite{KadanoffGotzeHAmblenHechtLexisPAiCiauskasRaylSwiftComparaisonUniversalitesurResultatsExperiencePlusDiscussion}.

The situation improves drastically in the case of one spatial 
dimension\symbolfootnote[2]{In the language of classical statistical mechanics, this corresponds to the case of a two-dimensional model\cite{SuzukiCorrespondence(D+1)StatPhysDQuantumHamiltonians}.},
where a plethora of exactly solvable models arises:  the Luttinger model (see \cite{HaldaneStudyofLuttingerLiquid} and references therein), $1+1$ dimensional conformal field theories \cite{BeliavinPolyakovZalmolodchikovCFTin2DQFT,DiFrancescoMathieuSenechalCFTKniga} or models that are solvable 
by one of the versions of the Bethe Ansatz \cite{BetheSolutionToXXX} such as the celebrated $XXZ$ spin-$\tf{1}{2}$ chain or the 1D Bose gas at arbitrary repulsive coupling. 
The abundance of exact solutions turns these one-dimensional models into a laboratory allowing one to test the universality principle in many concrete situations. 

It is worth reminding that, in all these models, two scales coexist : 
a microscopic scale $\delta$ which is related to the lattice spacing or, more generally proportional to the inverse of the Fermi momentum, 
and a macroscopic scale $L$ corresponding to the volume or number of sites of the model. Scale invariance can only be realised at distances $\Delta x$ between local operators  
that range  between these two scales but are far from them. This imposes that the ratio of the two scales $\tf{L}{\de}$ is very large in such a way that there exist a whole range of distances verifying $\delta <<\Delta x <<L$. For $\Delta x$ 
close to either the microscopic or the macroscopic scales the scaling properties of the correlation functions get modified in drastic ways.

The very structure of a conformal field theory imposes a simple form for the transformation of its operators under scaling. In its turn, this imposes 
that the correlation function exhibit an algebraic in the distance pre-factor. In particular, two-point functions are purely-algebraic. 
The exponents driving the power-law behaviour of a correlator are then constructed  in terms of the scaling dimensions of the operators that are being averaged. 
The data issuing from conformal field theories can thus be used so as to provide
one with  quite explicit predictions for the large-distance decay of correlators in massless quantum models 
in one spatial dimension. In fact, there, one can be slightly more precise in respect to the two pillars on which these predictions build
\begin{itemize}
 \item in the critical regime $\delta <<\Delta x <<L$, correlation functions should  exhibit conformal invariance as argued by Polyakov \cite{PolyakovArgumentAboutCFTInvarianceCriticalCorrelators}. 
 This suggests that the leading contribution to the long-distance asymptotics should be reproduced 
 by correlation functions of appropriate operators in a two-dimensional conformal field theory. Still, the identification of which conformal 
 field theory is to be used and which are the "appropriate" operators  is left open at this stage.
 \item The $\tf{1}{L}$ corrections to the model's ground and low-lying excited state's energies contain the information on the central charge 
of the conformal field theory describing the asymptotics and the scaling dimensions of the tower of operators 
describing the correlation functions as argued by  Cardy \cite{BloteCardyNightingalePredictionL-1correctionsEnergyAscentralcharge,CardyConformalExponents}.
\item The choice of "appropriate" operators, on the conformal field theory side, is done by advocating that these should inherit the symmetry that is satisfied by the  correlation function one starts with. 
\end{itemize}

To rephrase the above, Polyakov's argument justifies why a conformal field theory should emerge as an effective large-distance theory in the domain $\delta <<\Delta x <<L$ of the model 
while Cardy's observation permits one to extract, from the knowledge of the structure of a model's excitations, the quantities which would characterise
the effective conformal field theory describing the model's large-distance regime. 

A similar line of though is followed by exploiting the Luttinger liquid model as a tool for providing the critical exponents.  Is is argued that a model belongs to the Luttinger liquid universality class if it has the same form
of the low-lying excitations above its ground state. The parameters describing these excitations fix the Luttinger liquid model that 
is pertinent for the model of interest. The critical exponents of the model one starts with are then deduced from the ones computed explicitly 
for the Luttinger liquid model 
\cite{HaldaneCritExponentsAndSpectralPropXXZ,HaldaneLuttingerLiquidCaracterofBASolvableModels,LutherPeschelCriticalExponentsXXZZeroFieldLuttLiquid}. 

Independently of which of the above methods one chooses to employ, one needs to access to the $1/L$ corrections to the energy spectrum of a model's Hamiltonian
so as to be able to carry out effective predictions. The extraction of such corrections from the spectra of various quantum integrable models was initiated 
in the series of works \cite{DestriDeVegaAsymptoticAnalysisCountingFunctionAndFiniteSizeCorrectionsinTBAFiniteMagField,
DeVegaWoynarowichFiniteSizeCorrections6VertexNLIEmethod,KlumperBatchelorNLIEApproachFiniteSizeCorSpin1XXZIntroMethod,
KlumperBatchelorPearceCentralChargesfor6And19VertexModelsNLIE,KlumperWehnerZittartzConformalSpectrumofXXZCritExp6Vertex} 
and led to the identification of the central charge and scaling dimensions, for numerous quantum integrable models. 
It is worth mentioning the work \cite{GiulianiMastropietroProofOfCntralCharge1/2forPertOf2DIsing} where the $1/L$ corrections to the free energy of certain non-integrable perturbations of the two-dimensional
Ising model were obtained, on rigorous grounds through constructive field theory methods. By using Cardy's form of the $1/L$
corrections, this work demonstrated that, for sufficiently small perturbations, the model still has the same central charge $1/2$ as the Ising model.

Although effective in the sense that producing explicit answers,  the above techniques are more of a list of 
prescriptions than a well argued from the first principles line of though that joins, argument after argument, 
the structure of a given microscopic model with the resurgence of an effective field theoretic description in the large-distance regime.
Several attempts have been made in the literature to bring some elements of rigour or, at least, some \textit{ab inicio} justification
of the principle. In a series of works \cite{PolyakovArgumentAbouMicoOriginOfUniversality,PolyakovArgumentAboutpropertiesofLongDistAsympt}, by using the tools
developed in \cite{PatashinskiiPokrovskiiSecondOrderPhaseTransitionBoseFluid}, Polyakov argued the arisal of a universal behaviour on the basis of perturbative field theoretic
calculations. Although constituting an important step forward, these reasonings did not allow for any control on the magnitude of the contributions that he dropped from his calculations 
on the basis of some hand-waving arguments, without mentioning the problems inherent to lacks of convergence in perturbative handling 
in quantum field theory. A rigorous approach allowing one to prove the power-law decay of certain two-point functions in interacting one-dimensional
fermion models that are a sufficiently small perturbation of a free fermion model was first proposed by Pinson and Spencer \cite{PinsonSpencerFirstApproachToUniversalityInPertOf2DIsing}, further
developed by Mastropietro \cite{MastropietroScalingofEnergyEnergyCorrFctsNonIntIsing}
and then generalised in \cite{GiulianiGreenblattMastropietroScalingbehavior2nFctsofEnergyCorrsNonIntIsing} to the case of multi-point energy-energy correlation functions in certain non-integrable perturbations of the two-dimensional Ising model. 
An adaptation of this approach \cite{BenfattoFalcoMastropietroUniversalityOfPerturbedFreefermionsModelsCritExpAndRespondseFcts,BenfattoFalcoMastropietroUniversalityOfPerturbedFreefermionsModelsLuttLiqStructure} 
also allowed to establish the universality of certain properties of Luttinger liquid type  for a class of sufficiently small perturbations of 
a free fermion model. 
The method relies on the possibility to provide a rigorous construction of the path integral for such models in finite volume and then study its
infinite volume and scaling limits through rigorous renormalisation group methods. 
Finally, we should also mention the approach for proving the universality of certain types of bi-dimensional Ising lattcies -which are related to one-dimensional massless quantum Hamiltonians
by means of the correspondence proposed in \cite{SuzukiCorrespondence(D+1)StatPhysDQuantumHamiltonians}- that was developed by Smirnov in \cite{SmirnovProofOfConformalInvarianceInIsing}. 
Finally, we would also like to mention the derivation of the large-distance asymptotic behaviour of spin-spin correlation functions in the massless regime of the XXZ chain on 
the basis of Bethe Ansatz methods by the authors in collaboration with Kitanine, Slavnov and Terras \cite{KozKitMailSlaTerXXZsgZsgZAsymptotics}. The Bethe Ansatz approach was subsequently 
brought to a satisfactory level of rigour in \cite{KozReducedDensityMatrixAsymptNLSE} by one of the authors. The proof of the long-distance asymptotic behaviour solely relies on 
convergence hypothesis for certain auxiliary series of multiple integrals. 

The present paper introduces a convenient microscopic description for the spectrum, space of states \textit{and}
matrix elements of local operators in a class of one-dimensional massless quantum models. Our setting  allows us to construct, in the large-distance regime $\delta <<\Delta x <<L$, 
a one-to-one map (up to higher order corrections in the distance) between the relevant to the large-distance regime sub-space  of the model's Hilbert space and  the 
free boson Hilbert space. The local operators of the original model are then represented in terms of a collection of vertex operators acting in the free boson
Hilbert space.  We should mention that our work takes its roots in the series of papers \cite{KozKitMailSlaTerRestrictedSums,KozKitMailSlaTerRestrictedSumsEdgeAndLongTime,KozKitMailTerMultiRestrictedSums}
where the authors, in collaboration with Kitanine, Slavnov and Terras, developed a form factor series based approach to the analysis of the long-distance and 
large-time asymptotic behaviour of correlation functions in quantum integrable models. 
The setting we introduce in the present paper, as shown in the works
\cite{KozKitMailSlaTerEffectiveFormFactorsForXXZ,KozKitMailSlaTerThermoLimPartHoleFormFactorsForXXZ,LiebExcitationStructureBoseGas}, 
is clearly verified for a large-class of quantum integrable models, 
the XXZ spin-$1/2$ chain being the most prominent example. However, we do trust that the overall hypothesis that we lay down for the 
structure of the model's observable is quite universal and at least encompasses several instances of one dimensional 
quantum liquids \cite{GlazmanImambekovSchmidtReviewOnNLLuttingerTheory}. This is supported, \textit{e}.\textit{g}. by perturbative calculations around a free model
\cite{BenfattoFalcoMastropietroUniversalityOfPerturbedFreefermionsModelsCritExpAndRespondseFcts,BenfattoFalcoMastropietroUniversalityOfPerturbedFreefermionsModelsLuttLiqStructure,CauxGlazmanImambekovShasiNonUniversalPrefactorsFromFormFactors}. 
Furthermore all objects that we handle are standard within the phenomenological approach to interacting Fermi systems \cite{GlazmanImambekovSchmidtReviewOnNLLuttingerTheory,NozieresTheoryIntFermiSyst}.

Going slightly deeper into the details of the physics' jargon, the main assumptions on which our setting builds are that 
\begin{itemize}
\item in the large-volume limit, the relevant part to the critical regime of the model's spectrum is purely constructed in terms of particle-hole excitations\footnote{In the following, 
for simplicity, we shall assume a purely particle-hole excitation spectrum. This allows one to lighten the discussions. We stress that this assumption is not a limitation in that, for instance,
the bound states are expected to solely produce exponentially small corrections in the distance.};
\item the form factors -expectation values of local operators- taken between two states realised in terms of excitations  that are all located in an immediate vicinity of the 
Fermi zone admit a structure descending from a local repulsion principle between the interacting momenta of the particle and hole building up the excited state. 
\end{itemize}
The main result of the paper can be phrased as follows. Let $\mc{O}_1(x_1), \dots, \mc{O}_r(x_r)$ be local operators located at at positions 
$x_1,\dots, x_r$ and associated with some one-dimensional quantum model in finite (but large) volume $L$ and having the Hilbert space $\mf{h}_{\e{phys}}$. 
The operators $\mc{O}_s$ is assumed to induce solely transitions between spin sectors differing by some fixed integer $o_s$. Then, in the large $\de \ll |x_a-x_b| \ll L $
limit, the $r$-point ground-to-ground state expectation value of these operators satisfies
\beq
\Big< \mc{O}_1(x_1) \cdots \mc{O}_r(x_r) \Big>_{\mf{h}_{\e{phys}} } \; \simeq \;  \Big< \msc{O}_1(\om_1) \cdots \msc{O}_r(\om_r) \Big>_{ \mf{h}_{\e{eff}} }
\label{ecriture correspondance asymptotique fct a r pt}
\enq
where the $\simeq$ sign is to be understood as an equality up to the first leading correction arising in each oscillating harmonics in the spacing difference. 
The operator $ \msc{O}_s$, $s=1,\dots,r$, appearing in the \textit{rhs} of \eqref{ecriture correspondance asymptotique fct a r pt} act 
in the Hilbert space $\mf{h}_{\e{eff}}$  associated to the free boson model while the expectation value on the \textit{lhs} of \eqref{ecriture correspondance asymptotique fct a r pt} 
is taken in the Hilbert space $\mf{h}_{\e{phys}}$ of the initial model; they are built up from the free boson vertex operators $\msc{V}(\nu,\kappa; \om)$:
\beq
\msc{O}_s(\om_s) \; = \; \sul{ \kappa \in \mathbb{Z} }{} \mc{F}_{\kappa}\big(\mc{O}_s \big)\cdot \Big( \f{2\pi}{L} \Big)^{\rho(\nu_s(q)-o_s+\kappa) + \rho(\nu_s(-q)+\kappa)} \cdot 
\ex{2\i p_F \kappa x_s} \cdot  \msc{V}_{L}\big(-\nu_s(-q), -\kappa \mid \om_s^{-}  \big)  \msc{V}_{R}\big(\nu_s(q)-o_s,\kappa -o_s\mid \om_s^+  \big) 
\label{ecriture operateur Os cote CFT}
\enq
In this formula, 
\begin{itemize}
\item $\om_s^{\pm}\, =\, \ex{ \pm 2\i \pi  \f{x_s}{L} \a^{\pm} }$ is the variable attached to the left $(-)$ or right $(+)$ end of the Fermi zone which parametrises the position of the operator on the free boson model side. 
The factors $\a^{\pm}$ are certain re-scaling coefficients which take into account the variation of the dressed momentum at the endpoints of the Fermi zone. They are expressed in terms of the physical 
observables of the model, \textit{cf}. \eqref{ecriture formule pour alpha pm}; 
\item the indices $L$ and $R$ refer to the left or right copy of a free boson Hilbert space $\mf{h}$ on which the vertex operator acts, \textit{viz}. $\mf{h}_{\e{eff}}=\mf{h}_L\otimes\mf{h}_R$;
\item  $\msc{V}\big( \nu , \kappa \mid \om  \big) $ is the vertex operator, \textit{cf}. \eqref{definition  r shifted vertex operators}, which connects states belonging to 
sectors differing by the charge $\kappa$;
\item $\nu_{s}(\pm q)$ are the values of the so-called shift function relatively to the ground state, \textit{cf}. \eqref{definition fction shift relative etats gauche et droit}, and 
evaluated at the left (-q) or right (q) end of the Fermi zone. We do stress that it is a function of the charge $\kappa$ and of the operator's spin $o_s$.
\item $\rho(\nu_s(q)-o_s+\kappa)$ and  $\rho(\nu_s(-q)+\kappa)$  is the tower of scaling dimensions associated with the operator $\mc{O}_s$;
\item $\mc{F}_{\kappa}\big(\mc{O}_s \big)$, $\kappa \in \mathbb{Z}$, is the tower of typical form factors of the physical operator $\mc{O}_s$ taken between
the ground state and the lowest-lying excited state belonging to the $\kappa$-Umklapp sector in $\mf{h}_{\e{phys}}$. 
\end{itemize}
 We refer the reader to Sub-Section \ref{SousSection Exemple XXZ}  where explicit examples for the above-listed quantities are given in the case of the fundamental local operators  for the XXZ chain. 
In Sub-Section \ref{SousSoussectionLLStructure} we show how the structure of the Luttinger-liquid critical exponents can be recovered within our formalism. 

What equation \eqref{ecriture correspondance asymptotique fct a r pt} states is that, in the weak sense -\textit{i}.\textit{e}. for expectation values-
and for large distances of separation all the essence of the physical operator $\mc{O}_{a}(x_a)$ acting in the Hilbert space $\mf{h}_{\e{phys}}$ is grasped by the effective operator
$\msc{O}_s(\om_s)$ acting in the free boson model's Hilbert space $\mf{h}_{\e{eff}}$. Formula \eqref{ecriture operateur Os cote CFT} provides an effective map, in the critical regime, between local operators of the physical models 
and vertex operators in the corresponding  conformal field theory. This map is completely determined by the typical form factors $\mc{F}_{\kappa}\big(\mc{O}_s \big)$, $\kappa \in \mathbb{Z}$, 
of the physical operators $\mc{O}_{a}(x_a)$ and the values $\nu_{s}(\pm q)$ taken by the so-called shift function on the left/right endpoint of the Fermi zone. 
A more detailed description of the correspondence and the constituents of the operator $\msc{O}_s$
can be found in the core of the paper. We do stress that the expectation value appearing in the \textit{rhs} of \eqref{ecriture correspondance asymptotique fct a r pt}
is trivial to estimates by means of the free fermion exchange relations, \textit{cf}. \eqref{ecriture relation echange J+ et J -}-\eqref{ecriture calcul valeur moyenne produit op de vertex} and that the correspondence \eqref{ecriture correspondance asymptotique fct a r pt}-\eqref{ecriture operateur Os cote CFT} only provides one with the leading 
order to each oscillatory harmonics. This leading order only involves primary operators. However, should one be interested in sub-leading corrections then
one would need to supplement the expansion \eqref{ecriture operateur Os cote CFT} with additional terms that would involve descendants. 
Obtaining these terms is, in principle, possible within the method developed in the present paper, although would demand more computations.

On top of providing the microscopic origin of the appearance of a $c=1$ conformal field theory as an effective theory in the large-distance regime, 
our setting supports the  interpretation of the universality hypothesis that has been advocated by one of the authors in \cite{KozReducedDensityMatrixAsymptNLSE}. 
Namely, within our approach, the leading power-law contribution associated with a given oscillating harmonic in the large-distance asymptotic expansion of a correlator
stems from a saddle-point like contribution that is extracted from the form factor series expansion of the correlator. In the large-distance regime studied
in the present paper, the saddle-point is located on the two ends of the Fermi zone. In the time-dependent case, one should as well incorporate all the 
possible critical points of the dressed momentum/energy combination $xp-t \veps$, \textit{cf}. \cite{KozKitMailSlaTerRestrictedSumsEdgeAndLongTime}. It is the singular structure, \textit{viz}. local behavior, of an operator's form factor 
in the vicinity of the saddle-point that fixes the value of the critical exponents. Thus, two models belonging to the same universality class -in the sense that 
their appropriate correlators share the same critical exponents- have to share the same singularity structure of their form factors.
Note that the regular part of the form factors can differ from one model to another in that that they solely impact the value of the amplitudes, 
\textit{viz}. the non-universal part of the large-distance asymptotic expansion. 
Recall that the standard formulation of the universality hypothesis states that two models sharing the same overall symmetries share the same critical exponents. 
In this light, it would appear that the set of symmetries of a given model completely fixes the singularity structure of its form factors. However, it is not clear for us at 
the moment how one could derive the singular structure of a model's form factors solely building on its symmetries. 
Nonetheless, it seems more satisfactory to us to think of universality in terms of classes of models sharing the same singular structure of their form factors.
Indeed, this criterion  is precise and can be checked, in certain cases, by explicit calculations, say in a perturbative regime.
The matter is that there is, \textit{a priori}, no criterion allowing one to say that one has identified all the symmetries of a model
that are  pertinent for fixing its universality class. It could well be that two models have apparently the same symmetry structure but that one has missed
some important yet quite oblivious symmetry of one of the models that would imply that, in fact, the models belong to two distinct universality classes.   
It is worth, in this respect, to remind that two apparently very similar models, the 2D Ising and the the eight-vertex model in its lattice model 
formulation belong to fundamentally distinct universality classes.

The paper is organised as follows. The present section is the introduction. In Section \ref{Section modele FF} we recall the free fermion based description of the space of
states of the free boson model. We also present new formulae for the expectation values of exponents of specific current operators. In Section
\ref{Section presentation gle du modele}
we present the general framework -properties of the model's spectrum and form factors- that allow us to derive the correspondence with the free boson model. 
Finally, in Section \ref{Section correspondance modele et boson libre}, we establish the main result of the paper, namely the set of formulae 
\eqref{ecriture correspondance asymptotique fct a r pt}-\eqref{ecriture operateur Os cote CFT} that appeared in the introduction. The paper contains four appendices. 
In Appendix \ref{Appendix fct speciales d interet}, we briefly review the special
functions which arise in the course of our study. In Appendix \ref{Appendix Integrales pertinentes} we compute some two-dimensional integrals which are of 
interest to the probelm. In Appendix \ref{Appendix Preuve proposition valeur expectation ops vertex non shifte}, we prove Proposition 
\ref{Proposition expectation value J-J+ Ops}. Finally, in Appendix \ref{Appendix connection avec sommes restreintes}, we show how one can 
recover, starting from the formalism developed in the present paper, the value of the multi-point restricted sums that were first introduced in \cite{KozKitMailTerMultiRestrictedSums}.




\section{A free fermion description of the free boson model}
\label{Section modele FF}

In the present section, we recall the free fermion based construction of the space of states for the free boson model. Our presentation 
 basically follows the notations and conventions that can be found in the excellent review paper \cite{AlexandrovZaborodinTechniquesOfFreeFermions}. 
The various results found in this review originate from a long series of developments which started with the cornerstone work of Kyoto's school 
(Jimbo, Miwa and Sato) in the late '70's on holonomic quantum fields \cite{JimMiwaSatoQuantumFieldsI}.
See also \cite{OkunkovSomeAlgebraicManipulationWithFreeFermion} for some later developments of the method. 

\subsection{Overall definitions and generalities} 

\subsubsection{The space of states}

We consider a set of fermionic operators $\{\psi_n\}_{n \in \mathbb{Z} }$ and their $*$ associates $\{\psi_n^*\}_{n \in \mathbb{Z}}$
which satisfy the anti-commutation relations
\beq
\{ \psi_n,\psi_m \} \; = \; \{ \psi_n^*,\psi_m^* \} \; = \; 0 \qquad \e{and} \qquad
\{ \psi_n,\psi_m^* \}\;= \; \de_{n,m} \;, 
\enq
where $\de_{n,m}$ is the Kronecker symbol. We also assume the existence of a vacuum vector $\ket{0}$ which satisfies to the properties
\beq
\psi_n \ket{0} \; = \; 0 \; \; \e{for} \; \; n<0  \qquad \e{and} \qquad   \psi_n^{*} \ket{0} \; = \; 0 \; \; \e{for} \; \; n \geq 0  \;. 
\enq
The dual vacuum $\bra{0}$ fulfils the analogous properties  
\beq
\bra{0} \psi_n^{*}  \; = \; 0 \; \; \e{for} \; \; n<0 \qquad \e{and} \qquad  \bra{0} \psi_n  \; = \; 0 \; \; \e{for} \; \; n \geq 0 \;. 
\enq
Starting from the vacuum (resp. the dual vacuum) one constructs vectors (resp. dual vectors) through a repetitive action of the fermion operators.  
Such vectors are parametrised, in a natural way, by the sets 
\beq
\mc{J}_{n_p;n_h} \; = \; \Big\{ \{p_a\}_1^{n_p} \; ; \; \{h_a\}_1^{n_h} \Big\}
\enq
consisting of two collections of ordered integers $1\leq p_1 < \dots < p_{n_p}$ and $1\leq h_1 < \dots < h_{n_h}$. 
It is convenient, for a deeper physical insight, to think of the integers $\{p_a\}_1^{n_p}$ as a labelling of the particle-like excitations and
to think of the integers $\{h_a\}_1^{n_h}$ as a label for the  hole-like excitations. To be more precise, 
to each set $\mc{J}_{n_p;n_h}$ defined as above, one associates the vector 

\beq
\ket{ \mc{J}_{n_p;n_h} } \; = \; \psi_{-h_1}^{*}\cdots \psi_{-h_{n_h}}^{*} \cdot \psi_{p_{n_p}-1} \cdots \psi_{p_1-1} \ket{0}
\enq
and the dual vector 
\beq
\bra{ \mc{J}_{n_p;n_h} } \; = \;  \bra{0}\psi_{p_1-1}^{*} \cdots \psi_{p_{n_p}-1}^{*} \cdot \psi_{-h_{n_h}} \cdots \psi_{-h_1} \;. 
\enq
The Hilbert space $\mf{h}$ of the model is then defined as the span of the vectors introduced above
\beq
\mf{h} \; = \; \e{span} \bigg\{ \ket{ \mc{J}_{n_p;n_h} } \; \e{with} \; n_p, n_h\in \mathbb{N} \; \; \e{and} \; \; 
\ba{c} 1\leq p_1 < \dots < p_{n_p} \\  1\leq h_1 < \dots < h_{n_h} \ea  \; \; 
p_a, h_a \in \mathbb{N}^{*} \bigg\}. 
\label{definition espace de Hilbert boson lilbre}
\enq

Note that, when the number of particle and hole-like integers coincide, \textit{i}.\textit{e}. $n_p=n_h=n$, one can 
identify the set $\mc{J}_{n_p;n_h}$ with a Young diagram. The one-to-one map is obtained by interpreting the integers 
$\Big\{ \{p_a\}_1^{n} \; ; \; \{h_a\}_1^{n} \Big\}$  as the Frobenius coordinates of the Young diagram\symbolfootnote[2]{we adopt
the slightly unusual convention where the origin of the Frobenius coordinates is the diagonal so that  these start from $1$.}. 
Such an identification is also possible when $n_p \not=  n_h$. The bijection is, however, more complicated.
We refer the interested reader to the proof of Lemma \ref{Lemme reecriture de la forme de operateur de translation entre espaces hell} 
where one can find all of the necessary details for constructing such a 
bijection. Note that such a bijection was constructed for the first time, although indirectly, in 
Appendix A.3 of \cite{KozKitMailSlaTerRestrictedSums}. 

The Hilbert space $\mf{h}$ also enjoys of another basis $\ket{ \mathbb{Y} ; \ell}$ which makes the connection with a Young
diagram $\mathbb{Y}$ explicit. This new basis singles out the collection of the so-call vacuum and dual vacuum states having a prescribed charge $\ell$:
\beq
\ket{ \ell } \; = \; \left\{ \ba{cc} \psi_{\ell-1}\cdots \psi_{0}\ket{0} &  \quad \ell >0 \vspace{1mm} \\ 
					\psi_{\ell}^{*}\cdots \psi_{-1}^{*}\ket{0} & \quad  \ell < 0   \ea \right.  \qquad \qquad \e{and} \qquad \qquad 
\bra{ \ell } \; = \; \left\{ \ba{cc} \bra{0} \psi_{0}^{*}\cdots \psi_{\ell-1}^{*} & \quad \ell >0 \vspace{1mm} \\ 
					\bra{0} \psi_{-1}\cdots \psi_{\ell} & \quad \ell < 0   \ea \right.  \;. 
\enq
General states of this basis are build as equal in number particle-hole excitations over the vaccua $\ket{\ell}$, resp.  their dual vaccua $\bra{\ell}$. 
Let 
\beq
\mathbb{Y} \; = \; \Big\{ \{\a_a\}_1^n \; : \; \{\be_a\}_1^n  \Big\} \quad \e{with} \qquad 
\left\{  \ba{c} 		
				  1\leq \a_1 \, < \, \cdots \, <\; \a_n  \vspace{1mm}\\ 
			     1\leq \be_1 \, < \, \cdots \, <\; \be_n   \ea \right. 
\enq
be the Frobenius coordinates of the Young diagram $\mathbb{Y}$. The second basis we mentioned above takes the form 
\beq
\ba{ccc}
\ket{\mathbb{Y};\ell} & = & \psi^{*}_{\ell - \be_1 } \cdots  \psi^{*}_{ \ell - \be_n } \, \psi_{\ell + \a_n - 1 } \cdots  \psi_{ \ell + \a_1 - 1} \, \ket{\ell}  \vspace{2mm} \\
\bra{\ell; \mathbb{Y}} & = & \bra{\ell} \, \psi^{*}_{\ell + \a_1 -1 } \cdots  \psi^{*}_{ \ell + \a_n -1} \, \psi_{\ell - \be_n  } \cdots  \psi_{ \ell - \be_1} 
    \ea \; .   
\label{ecriture bases labelle par les diagrammes de Young}
\enq
In fact, one can even consider mixtures of the basis $\ket{\mathbb{Y};\ell}$ and $\ket{ \mc{J}_{n_p;n_h} }$, namely the basis 
\beq
\ket{ \mc{J}_{n_p;n_h} ;\ell  } \; = \; \psi_{\ell - h_1}^{*}\cdots \psi_{ \ell - h_{n_h}}^{*} \cdot \psi_{p_{n_p}+\ell - 1} \cdots \psi_{p_1+\ell-1} \ket{\ell} \;. 
\enq
It is easy to convince oneself that 
\bem
\e{Span}\Big\{ \ket{\mathbb{Y};\ell+r} \; : \; \mathbb{Y}\; \;  \e{Young} \; \e{diagram} \Big\} \; = \; 
\e{Span}\Big\{  \ket{ \mc{J}_{n_p;n_h}  } \; : \; \e{sets} \; \mc{J}_{n_p;n_h} \; \; n_p-n_h=\ell+r  \Big\}  \\
\; = \; \e{Span}\Big\{  \ket{ \mc{J}_{n_p;n_h} ; \ell } \; : \; \e{sets} \; \mc{J}_{n_p;n_h} \; \; n_p-n_h=r  \Big\}   \;. 
\end{multline}

\subsubsection{The space of operators}

The fermionic operators $\psi_j$ and $\psi_j^*$ can be thought of as building bricks allowing one to construct more general operators on the Hilbert space $\mf{h}$. 
The simplest class of operator is obtained as a linear combination in the fermions:
\beq
\mf{v} \; = \; \sul{m}{} V_{m} \psi_m \qquad \e{resp}. \qquad \mf{w}^* \; = \; \sul{m}{} W_{m} \psi_m^*  \;, 
\label{introduction operateurs fermionique lineaires}
\enq
where $V_{m}$, $W_{m}$ are bounded sequences.  The field and conjugated field operators 
\beq
\Psi(z) \; = \; \sul{j \in \mathbb{Z} }{} \psi_j z^j \qquad  \quad \e{and} \quad  \qquad \Psi^*(z) \; = \; \sul{j \in \mathbb{Z} }{} \psi_j^* z^{-j}
\enq
are an archetype of such operators . 

The second important class of operators is obtained by means of a bilinear pairing of the fermions realised by  an infinite matrix $A$
having bounded entries:
\beq
\sul{i, j \in \mathbb{Z} }{} A_{ij} \psi_i \psi_j^* \;. 
\label{introduction operateurs fermionique bilineaires}
\enq

Expressions of the type \eqref{introduction operateurs fermionique lineaires} or \eqref{introduction operateurs fermionique bilineaires} may look a bit formal 
in particular due to convergence issues of the sums. However, all the sums of interest truncate as soon as one computes matrix elements taken between
the fundamental system of basis vectors labelled by the sets $\mc{J}_{n_p;n_h}$. It is in this sense that all of the above expressions should be understood. 

The most fundamental example of an operator built through a bilinear pairing is the charge operator which takes the explicit form
\beq
J_0 \; = \; \sul{k\geq 0}{} \psi_k \, \psi^{*}_k \; - \; \sul{k<0}{} \psi_k^{*} \, \psi_k \;. 
\enq
It is readily seen that the vector $\ket{ \mc{J}_{n_p;n_h} } $ is associated with the eigenvalue $n_p-n_h$ of the charge operator $J_0$.
We will sometimes say that the vector $\ket{ \mc{J}_{n_p;n_h} } $ has charge  $n_p-n_h$. 

In other words, the charge operator induces a grading of the Hilbert space $\mf{h}$
 into the direct sum of spaces $\mf{h}_{\ell}$ having a fixed charge $\ell$
\beq
\mf{h} \; = \; \bigoplus_{\ell \in \mathbb{Z} } \mf{h}_{\ell}  \; \qquad \e{with}  \qquad
\mf{h}_{\ell} \; = \; \e{span} \Big\{ \ket{ \mc{J}_{n_p;n_h} } \; : \;  n_p-n_h\, = \, \ell \Big\}. 
\enq
The current operators provide one with another example of important operators that are bilinear in the fermions. 
To each half-infinite sequence $\bs{t}\;= \; \{\bs{t}_k \}_{k\in \mathbb{N}^*}$ one associates the current operators  
\beq
J_{\pm}\big( \bs{t} \big) \; = \; \sul{ k \geq 1 }{} \bs{t}_k \cdot J_{\pm k} \quad \qquad \e{with} \qquad  \quad 
J_k \; = \; \sul{j\in \mathbb{Z} }{} \psi_j \psi^*_{j+k} \qquad  \;  \e{for} \; \;  k \not= 0 \; . 
\enq
It is readily checked that the components of the current operator satisfy  the algebra
\beq
\big[ J_k, J_{\ell} \big] \; = \; k \de_{k,-\ell} \;. 
\enq
These commutation relations allow one to interpret the currents $J_k$ as bosonic operator modes (see \textit{e}.\textit{g}. \cite{AlexandrovZaborodinTechniquesOfFreeFermions}). 
Furthermore since, for $k\in \mathbb{N}^{*}$, one has
\beq
J_k \ket{0} \; = \; 0 \qquad \e{and} \qquad \bra{0} J_{-k}  \; = \; 0
\enq
it follows that one can interpret $\{J_k\}_{k\in \mathbb{N}^*}$ as bosonic creation operators while $\{J_{-k}\}_{k\in \mathbb{N}^*}$ as the  bosonic annihilation operators. 
It is on the basis of such an observation that the construction of the free boson field theory is made in the free fermion model
we have introduced so far. 

It is also straightforward to check the commutation relations
\beq
\big[ J_k, \psi_{\ell} \big] \; = \; \psi_{\ell-k} \;     ,  \qquad \qquad \big[ J_k, \psi_{\ell}^{*} \big] \; = \; - \psi_{\ell+k} \;. 
\label{ecriture relation echange pour les courant coordonnees}
\enq
The above commutation relations can be conveniently encoded on the level of the field and conjugated field operators.
As a matter of fact, then, they take the form of the following  exchange relations 
\beq
 \Psi(z) \cdot \ex{J_{\pm}(\bs{t})} \; = \; \ex{-\xi(\bs{t},z^{\pm1})} \cdot  \ex{J_{\pm}(\bs{t})} \cdot \Psi(z)  \qquad \e{and} \qquad 
\Psi^*(z) \cdot \ex{J_{\pm}(\bs{t})} \; = \; \ex{\xi(\bs{t},z^{\pm1})} \cdot \ex{J_{\pm}(\bs{t})}  \cdot  \Psi^*(z) 
\label{ecriture relations echange generales}
\enq
where $\xi(\bs{t},z) \; = \; \sum_{k\geq 1 }^{} \bs{t}_k \cdot z^k$. Given a generic sequence $\bs{t}$,  this is the most compact form of the exchange 
relation since the sum defining $\xi$ cannot be computed explicitly. However, for the specific choice $\bs{t}=\bs{t}_{\pm}$ where
\beq
\big( \bs{t}_+ \big)_k \; = \; - \f{ \nu \om^{-k} }{ k } \qquad \e{and} \qquad 
\big( \bs{t}_- \big)_k \; = \;  \f{ \nu \om^{k} }{ k }  
\enq
the sum reduces to the Taylor expansion of the logarithm. It is the $\bs{t}=\bs{t}_{\pm}$ choice  
that plays a particularly important role in our analysis in that it is associated with the construction of vertex operators.  
We therefore introduce a special notation for the current operators associated with the parameters $\bs{t}_{\pm}$,  namely
\beq
\msc{J}_{\pm}(\nu,\om) \; \equiv \; J_{\pm}(\bs{t}_{\pm}) \; = \; \mp \nu \sul{ k \geq 1}{} \f{  \om^{\mp k} }{ k }  \cdot J_{\pm k } \;. 
\enq
In the case of the operators $\msc{J}_{\pm}(\nu,\om)$, the exchange relations \eqref{ecriture relations echange generales} particularise to 
\beq
\Psi(z) \cdot \ex{\msc{J}_{\pm}(\nu,\om)} \; = \;  \Big( 1- \Big( \f{z}{\om} \Big)^{\pm 1}  \Big)^{\mp \nu}  \ex{\msc{J}_{\pm}(\nu,\om)} \cdot \Psi(z)  \qquad \e{and} \qquad 
\Psi^*(z) \cdot \ex{\msc{J}_{\pm}(\nu,\om)} \; = \; \Big( 1- \Big( \f{z}{\om} \Big)^{\pm 1}  \Big)^{\pm \nu} \ex{\msc{J}_{\pm}(\nu,\om)}  \cdot  \Psi^*(z) \;. 
\label{ecriture relations echange vars Miwa}
\enq

There is also another operator that plays a role in our setting, the so-called shift operator which we shall denote 
as $\ex{P}$. The shift operator maps $\mf{h}_{\ell}$ onto $\mf{h}_{\ell+1}$. In fact, 
the operator $P$ arising in the exponent is the conjugate operator to $J_0$. Although the shift operator has no 
simple expression in terms of the fermions, it takes a particularly simple form in the basis of $\mf{h}$
subordinate to Young diagrams, \textit{cf}. \eqref{ecriture bases labelle par les diagrammes de Young}. 
Indeed, any integer power $\ex{r P}$ of the shift operator satisfies $\ex{r P }\ket{ \mathbb{Y}  ; \ell} \, = \, \ket{ \mathbb{Y} ; \ell + r}$, 
\textit{viz}. 
\beq
\ex{r P } \; = \; \sul{ \mathbb{Y} \, , \, \ell }{} \ket{ \mathbb{Y} ; \ell + r}\bra{ \ell; \mathbb{Y}}
\label{ecriture operateur translation dans base des coord Frob a n parts}
\enq
where the sum runs over all integers $\ell$ and all Young diagramms $\mathbb{Y}$( see \cite{AlexandrovZaborodinTechniquesOfFreeFermions} for more details).

We end this sub-section by introducing the $r$-shifted bosonic vertex operators which play a central role in the correspondence: 
\beq
\msc{V}(\nu, r \mid \om) \; = \; \ex{ \msc{J}_- (\nu + r,\om)  } \cdot \ex{ \msc{J}_+ (\nu + r,\om)  } \cdot \ex{r P} \;. 
\label{definition  r shifted vertex operators}
\enq

\subsubsection{The Wick theorem}

In this sub-section we review the statement of Wick's theorem in the case of an insertion of group-like elements. 
This name refers to a class of operators $G$ satisfying to the so-called basic bilinear condition:
\beq
\sul{ j\in \mathbb{Z} }{}  \Big( G\cdot \psi_j \Big)  \otimes \Big(G\cdot \psi_j^*\Big) 
\; = \; \sul{ j\in \mathbb{Z} }{} \Big( \psi_j\cdot G  \Big) \otimes  \Big( \psi_j^*\cdot G \Big) \;. 
\enq
Group-like elements single out, among other things, because they allow one for a powerful generalisation of the Wick theorem
which we shall recall at the end of this sub-section. 

The operators we have introduced so far provide several examples of group-like elements. 
For instance, it is easily checked that the operators $\ex{J_{\pm}(\bs{t})}$ in general, $\ex{\msc{J}_{\pm}(\nu,\om)}$ in particular,
and $\ex{r P}$ all satisfy to the basic bilinear condition. 
It is likewise trivial to establish that, if $G, G^{\prime}$ are group-like elements, then so is  their product $G\cdot G^{\prime}$. 
These facts ensure that that the $r$-shifted vertex operators $\msc{V}(\nu, r \mid \om)$ are group-like elements.

We stress that group-like elements need not to be invertible. For instance, given any  group element $G$  
and a collection $\mf{v}_1,\dots, \mf{v}_n$, resp. $\mf{w}_1^*, \dots, \mf{w}_n^*$,
of linear combinations of the basic fermion operators $\psi_j$, resp. $\psi_j^*$, \textit{viz}.
\beq
\mf{v}_k \; = \; \sul{m}{} V_{km}\,  \psi_m \qquad \e{resp}. \qquad \mf{w}_k^* \; = \; \sul{m}{} W_{km} \, \psi_m^*  \;, 
\enq
the operators 
\beq
G_{\mf{v};\mf{w}} \; = \; \mf{w}_1^*\cdots \mf{w}^*_n \, \mf{v}_n \cdots \mf{v}_1 \cdot G \qquad \e{and} \qquad 
\wt{G}_{\mf{v};\mf{w}} \; = \; G \cdot  \mf{w}_1^*\cdots \mf{w}^*_n \, \mf{v}_n \cdots \mf{v}_1  
\enq
are also a group-like.  We shall not enter deeper into the detail here and refer the reader to \cite{AlexandrovZaborodinTechniquesOfFreeFermions} for more details. 

 \vspace{3mm}
 
\noindent Group-like elements allow for the following generalisation of Wick's theorem. 

\begin{theorem}

Let $\mf{v}_j$, resp. $\mf{w}_j^*$\, , $j=1,\dots,n$ be any operators linear in the fermions and $G,G^{\prime}$ group-like elements. 
Then the associated expectation value admits a determinant representation:
\beq
\f{ \bra{0}\, G \mf{w}_1^*\cdots \mf{w}_n^* \, \mf{v}_n \cdots \mf{v}_1 G^{\prime} \, \ket{0} }{  \bra{0} \, G  \cdot G^{\prime} \, \ket{0} } \; = \; 
\det_n\Bigg[ \f{ \bra{0}\, G \mf{w}_{p}^* \, \mf{v}_k G^{\prime} \, \ket{0} }{  \bra{0} \, G \cdot G^{\prime} \, \ket{0} }   \Bigg] \;. 
\enq

\end{theorem}

\subsection{Expectation values of exponents of current operators}

In the present sub-section, we compute the expectation value
of the operators $\ex{\msc{J}_-(\nu,\om)}\ex{\msc{J}_+(\nu,\om)}$ between any two states $\ket{ \mc{J}_{n_p,n_h} }$ 
$\ket{ \mc{J}_{n_k,n_t} }$ of the basis. The corresponding result will provide a first brick towards the correspondence
with the free boson model that will be built in Section \ref{Section correspondance modele et boson libre}. 

Prior to stating the result, we first need to introduce a few handy notations. Given two sets of integers 
\beq
\mc{J}_{n_p,n_h} \; = \; \Big\{  \{p_a\}_1^{n_p} \; ; \;  \{h_a\}_1^{n_h}  \Big\}  \qquad \e{and} \qquad 
\mc{J}_{n_k,n_t} \; = \; \Big\{  \{k_a\}_1^{n_k} \; ; \;  \{t_a\}_1^{n_t}  \Big\} 
\label{definition ensemble Jnpnh et Jnknt}
\enq
one defines the set functions 
\beq
 \varpi\Big( \mc{J}_{ n_{p}; n_{h} } ;  \mc{J}_{ n_{k}; n_{t} }    \mid \nu \Big)  
			\; = \;  \pl{a=1}{n_h  }   \Bigg\{  \f{  \pl{b=1}{ n_k } \big( 1-k_b-h_a+\nu  \big)    }
{  \pl{b=1}{n_t } \big( t_b-h_a+\nu \big)  }  \Bigg\}  \cdot 
\pl{a=1}{n_p}  
\Bigg\{  \f{ \pl{b=1}{n_t} \big(p_a+t_b+\nu  -1 \big)    }
{  \pl{b=1}{n_k} \big( p_a - k_b + \nu \big)  }  \Bigg\}		\;. 
\label{definition de la fonction  varpi}
\enq
and 
\bem
 \msc{D}\Big( \mc{J}_{ n_{p}; n_{h} }   \mid \nu , \om \Big)  \; = \; \bigg( \f{\sin [\pi \nu] }{ \pi } \bigg)^{  n_{h}  }    
\cdot \pl{a=1}{n_p} \Bigg\{ \om^{p_a-1} \,  \Ga\Bigg( \ba{c}  p_{a}+\nu    \\
						    p_{a} \ea \Bigg) \Bigg\}  \cdot 
\pl{a=1}{n_h} \Bigg\{ \om^{h_a} \, \Ga\Bigg( \ba{c}  h_{a}-\nu    \\
						    h_{a} \ea \Bigg) \Bigg\}  \\
\times \f{ \pl{a>b}{ n_{p} } (p_{b} - p_{a} ) \cdot \pl{a>b}{ n_{h}  } ( h_{b} - h_{a} ) }
{ \pl{a=1}{ n_{p} } \pl{b=1}{ n_{h} } (p_{a} + h_{b} - 1) } \;. 
\end{multline}
These functions appear as the building blocks of the so-called discrete form factor
\bem
\msc{F}\Big( \mc{J}_{ n_{p}; n_{h} } ;  \mc{J}_{ n_{k}; n_{t} }  \mid \nu ,\om \Big)
\; = \;  (-1)^{ n_{p} + n_{t} }\cdot (-1)^{ \f{(n_p-n_h)(n_p-n_h+1)}{2} } \cdot \bigg( \f{\sin [\pi \nu] }{ \pi } \bigg)^{ n_{p} \, - \, n_{h}  }   
\cdot  \msc{D}\Big( \mc{J}_{ n_{p}; n_{h} }   \mid \nu , \om \Big)  \\
 \times \msc{D}\Big( \mc{J}_{ n_{k}; n_{t} }   \mid -\nu , \om^{-1} \Big) \cdot  \varpi\Big( \mc{J}_{ n_{p}; n_{h} } ;  \mc{J}_{ n_{k}; n_{t} }    \mid \nu \Big)  \;.   
\label{definition microscopic form factor}
\end{multline}

In \eqref{definition microscopic form factor} we made use of the hypergeometric-like notations for ratios of $\Ga$-functions, \textit{cf}. Appendix \ref{Appendix fct speciales d interet}. 

It is the above discrete form factor that enters in the description of the expectation values of the current operators  $\ex{ \msc{J}_-(\nu,\om) } \ex{  \msc{J}_+(\nu,\om)  }$
taken between the states $ \ket{  \mc{J}_{ n_{k}; n_{t} } }$ and $\bra{ \mc{J}_{ n_{p}; n_{h} } } $.

\begin{prop}
\label{Proposition expectation value J-J+ Ops}

Let $\mc{J}_{ n_{p}; n_{h} }$ and $\mc{J}_{ n_{k}; n_{t} }$ be as given in \eqref{definition ensemble Jnpnh et Jnknt}. 
Then form factor of the product of exponents of  current operators admits the following representation
\beq
\bra{ \mc{J}_{ n_{p}; n_{h} } } \ex{ \msc{J}_-(\nu,\om) } \ex{  \msc{J}_+(\nu,\om)  } \ket{  \mc{J}_{ n_{k}; n_{t} } }
\; = \;  \de_{n_p-n_h,n_k-n_t}\cdot  \msc{F}\Big( \mc{J}_{ n_{p}; n_{h} } ;  \mc{J}_{ n_{k}; n_{t} }  \mid \nu ,\om \Big) \;. 
\label{ecriture representation VO pour la valeur moyenne a shift r nul}
\enq

\end{prop}

Note that the Kronecker symbol in \eqref{ecriture representation VO pour la valeur moyenne a shift r nul} is there to emphasise that 
the form factor is zero unless $n_p-n_h=n_k-n_t$. This is due to the fact  that the exponents of current operators preserve the charge of a state, \textit{viz}.
can only connect states belonging to the same charge sector.  

The above statement is well known in the case $n_p=n_h=0$. Indeed, then, the representation \eqref{ecriture representation VO pour la valeur moyenne a shift r nul}
follows, for instance, from the Giambelli determinant representation for Schur functions along 
with the the possibility to compute explicitly the Schur functions $s_{\a,\be}(\bs{t}_{\pm})$ associated
with hook diagrams $(\a,\be)$ in Frobenius notations (\textit{cf}. Appendix A of \cite{AlexandrovZaborodinTechniquesOfFreeFermions} for more
details). We do however stress that the general case given above is, to the best of our knowledge, new. 
In particular, the proof of the general case relies on new ideas not related to the theory of Schur functions. The proof, being slightly technical, 
is postponed to Appendix \ref{Appendix Preuve proposition valeur expectation ops vertex non shifte}. 
We would also like to point out that Proposition \ref{Proposition expectation value J-J+ Ops} yields a new type of explicit 
representation for the skew Schur functions associated with the parameters $\bs{t}_{\pm}$.
 
In order to identify the emergence of an effective field theory description in the large-distance spacing between the operators, one needs
to determine the matrix elements of the general $r$-shifted vertex operators that have been defined in \eqref{definition  r shifted vertex operators}. 
These will be given in Theorem \ref{Theorem FF Vop entre etats generaux} below and can be deduced 
from \eqref{ecriture representation VO pour la valeur moyenne a shift r nul}. 
 
\begin{theorem}
\label{Theorem FF Vop entre etats generaux}

The form factor of the $r$-shifted vertex operator $\msc{V}(\nu, r\mid \om)$  reads
\beq
\bra{ \mc{J}_{ n_{p}; n_{h} } } \msc{V}(\nu, r\mid \om) \ket{  \mc{J}_{ n_{k}; n_{t} } }
\; = \; (-1)^{\frac{r(r+1)}{2}} \cdot  \f{ \de_{n_p-n_h, n_k-n_t+r} }{  \big(\om\big)^{\frac{r(r-1)}{2}+r(n_k-n_t)} }\cdot G\bigg( \ba{c} 1-\nu \\ 1-\nu -r \ea\bigg)
\cdot  \msc{F}\Big( \mc{J}_{ n_{p}; n_{h} } ;  \mc{J}_{ n_{k}; n_{t} }  \mid \nu,\om \Big) \;. 
\label{ecriture representation VO pour la valeur moyenne a shift r general}
\enq
The symbol $G$ appearing above stands for the hypergeometric-like notation for the ratio of two Barnes functions, see Appendix \ref{Appendix fct speciales d interet}. 
\end{theorem}

The proof of the theorem builds, in particular, on the form of the action of the $r$-shifted shift operator $\ex{rP}$
on the states $\ket{ \mc{J}_{n_k;n_t} }$. We shall determine the form of this action in Lemma \ref{Lemme reecriture de la forme de operateur de translation entre espaces hell} given below.
Still, first of all, we need to introduce some notations relative to so-called 
$r$-translates $\mc{J}^{(r)}_{ n_{k}^{\prime}; n_{t}^{\prime} }$ of the set $\mc{J}_{ n_{k}; n_{t} }$.  
 
Given a set $\mc{J}_{n_k;n_t}$, we define $q$, $0 \leq q \leq r$, to be the unique integer such that 
\beq
1 \, \leq \, t_1 \, < \, \cdots  \, < \,  t_q  \, \leq \,  r  \, <  \, t_{q+1} \,  < \, \cdots  \, < \, t_{n_t} \;. 
\label{definition entier q et partition ensemble des ta}
\enq
Then, the set $\{\wt{t}_a \}_1^{r-q}$ stands for the complement of the set $\{r+1-t_a\}_1^q$ in $\intn{1}{r}$, namely   
\beq
1 \, \leq \, \wt{t}_1 \, < \, \cdots  \, < \,  \wt{t}_{r-q}  \, \leq \,  r \qquad \e{and} \qquad 
\{ r+1-t_a \}_1^q \cup \{ \wt{t}_a \}_1^{r-q} \; = \; \intn{1}{r} \;. 
\label{definition ensemble t tilde a}
\enq
The $r$-translate $\mc{J}^{ (r) }_{ n_k^{\prime} ; n_t^{\prime} }$ of the set  $\mc{J}_{ n_{k}; n_{t} }$ corresponds to the set
\beq
\mc{J}^{ (r) }_{ n_k^{\prime} ; n_t^{\prime} }\, = \, \Big\{  \{ k_a^{\prime} \}_1^{ n^{\prime}_k } \; : \; \{ t_a^{\prime} \}_1^{ n^{\prime}_t }  \Big\}
\label{ecriture ensemble r shifte}
\enq
The sequence $k_a^{\prime}$ appearing in \eqref{ecriture ensemble r shifte} reads 
\beq
 k_a^{\prime}\; = \; \wt{t}_{a} \qquad \e{for} \qquad a=1,\dots r-q \qquad \e{and} \qquad k_{a+r-q}^{\prime}\; = \; k_{a} \, + \, r \qquad \e{for} \qquad a=1,\dots n_k 
\enq
and has a total of $n_k^{\prime}\, = \, n_k+r-q$ elements while the sequence $t_a^{\prime}$ is defined as 
\beq
t_a^{\prime}\; = \; t_{a+q} -r \qquad \e{for} \qquad a=1,\dots  n_t-q   
\enq
 and has a total of $n_t^{\prime}\, = \, n_t - q $ elements.

 \Proof \textit{(of Theorem \ref{Theorem FF Vop entre etats generaux})}
 Let $r\geq 0$. Then the representation \eqref{ecriture representation VO pour la valeur moyenne a shift r general} follows from the form taken by the action of the shift operator 
\eqref{ecriture action operateur translation sur etat J nk nt} on the states $\ket{ \mc{J}_{n_k;n_t} }$ and from the identity 
\bem
(-1)^{\sg_r(\{t_a\})+q(n_k+n_t-q)} \cdot \msc{F}\Big( \mc{J}_{ n_{p}; n_{h} } ;  \mc{J}^{(r)}_{ n_{k}^{\prime}; n_{t}^{\prime} }  \mid \nu,\om \Big)  \\
\; = \; G\bigg( \ba{c} 1+r-\nu \\ 1-\nu \ea\bigg)\cdot  (-1)^{r(n_k+n_h-n_p-n_t)} \cdot \big(\om\big)^{-\frac{r(r-1)}{2}-r(n_k-n_t)} \cdot 
\msc{F}\Big( \mc{J}_{ n_{p}; n_{h} } ;  \mc{J}_{ n_{k}; n_{t} }  \mid \nu-r,\om \Big) \;. 
\end{multline}
There $q$ is as defined in \eqref{definition entier q et partition ensemble des ta} 
while $(-1)^{\sg_r(\{t_a\})}$ refers to the signature of the permutation introduced in \eqref{definition permutation sg r des ta}.  

The formula \eqref{ecriture representation VO pour la valeur moyenne a shift r general} holds, in fact, for $r<0$ as well. To establish this fact, observe that the 
whole construction is invariant under the transformation $\big( \psi_k, \psi_k^{*} \big) \hookrightarrow \big( \psi_{-1-k}^*, \psi_{-1-k} \big)$. It is easy to see by using the 
properties of the states and operators involved that, under this transformation, one has
\beq
\bra{ \mc{J}_{ n_{p}; n_{h} } } \msc{V}(\nu, r\mid \om) \ket{  \mc{J}_{ n_{k}; n_{t} } } \hookrightarrow 
(-1)^{ n_h - n_t + \f{r(r+1)}{2} + (n_p-n_h) }   \bra{ \mc{J}_{ n_{h}; n_{p} } } \msc{V}(-\nu, -r\mid \om) \ket{  \mc{J}_{ n_{t}; n_{k} } } \;. 
\enq
After some algebra, one can convince oneself that the \textit{rhs} of \eqref{ecriture representation VO pour la valeur moyenne a shift r general}
satisfies to the same transformations. 

\qed

\begin{lemme} 
\label{Lemme reecriture de la forme de operateur de translation entre espaces hell}

For any $r\geq 0$, it holds
\beq
\ex{r P} \cdot \ket{ \mc{J}_{ n_k ; n_t }  } \; = \; (-1)^{q(n_k+n_t-q)}\cdot (-1)^{\sg_r(\{t_a\})} \cdot (-1)^{\f{r(r+1)}{2}} \cdot \ket{ \mc{J}^{ (r) }_{ n_k^{\prime} ; n_t^{\prime} } } \;. 
\label{ecriture action operateur translation sur etat J nk nt}
\enq
Above, $\sg_r(\{t_a\})$ corresponds to the permutation
\beq
\sg_r(\{t_a\}) \; : \; 1,\cdots,r \; \mapsto \; r-t_q+1,\cdots , \, r-t_1+1\, , \, \wt{t}_{r-q}, \cdots , \, \wt{t}_{1}
\label{definition permutation sg r des ta}
\enq
$ (-1)^{\sg_r(\{t_a\})}$ stands for its signature, while $q$ and $\{\wt{t}_a\}$ are as defined in \eqref{definition entier q et partition ensemble des ta}-\eqref{definition ensemble t tilde a}

\end{lemme}

 \Proof 
 
 We shall prove the lemma in the case when  $n_k-n_t\equiv \ell \geq 0$. The $\ell<0$ case follows from a similar reasoning so that its proof is left as an exercise. Let $ s \in \intn{0}{\ell}$ be the unique integer such that
\beq
1 \, \leq \, k_1 \, < \, \cdots  \, < \,  k_s  \, \leq \,  \ell  \, <  \, k_{s+1} \,  < \, \cdots  \, < \, k_{n_k} \;. 
\enq
We then introduce the complementary sequence  $\{\ell+1-\wt{k}_a \}_1^{\ell-s}$ to $\{k_a\}_1^s$ in $\intn{1}{\ell}$. The sequence 
$\{\wt{k}_a \}_1^{\ell-s}$ is such that 
\beq
1 \, \leq \, \wt{k}_1 \, < \, \cdots  \, < \,  \wt{k}_{\ell-s}  \, \leq \,  \ell \qquad \e{and} \quad 
\{ k_a \}_1^s \cup \{ \ell + 1 - \wt{k}_a \}_1^{\ell-s} \, = \, \intn{ 1 }{ \ell } \;. 
\enq
The sequence $\{\a_a\}_1^{n_k-s}$ is then defined as
\beq
\a_a\, = \; k_{s+a} - \ell \quad \e{for} \quad a=1,\dots, \ell-s 
\enq
and the sequence $\{\be_a\}_1^{n_k-s}$ as
\beq
\be_a\, = \; \wt{k}_a  \quad \e{for} \quad a=1,\dots, \ell-s \qquad \e{and} \qquad
\be_{\ell-s + a} \, = \; t_a+\ell \quad \e{for} \quad a=1,\dots, n_t \; . 
\enq
Note that there is, in total, $n_k-s$ labels for the $\be$'s due to the condition $n_t\,=\,n_k-\ell$. 
Finally, for convenience, we introduce the sequence $\{ k_a^{\e{c}} \}_1^{n_p-s}$ which we define as
$k_a^{\e{c}} \, = \, \ell+1-\wt{k}_{\ell-s+1-a}$, so that one has the ordering 
\beq
1\, \leq \,  k_1^{\e{c}} \, < \,  \cdots \, < \, k_{\ell-s}^{\e{c}} \, \leq \, \ell . 
\enq
Defining $ \ov{\sg}_{\ell}\big(\{k_a\}\big)$ as  the permutation
\beq
\ov{\sg}_{\ell}\big(\{k_a\}\big) \, : \, 1,\dots,  \ell \quad \mapsto  \quad k_s,\dots,k_1 , k_1^{\e{c}}, \dots, k_{\ell-s}^{\e{c}} \;. 
\enq
one has
\beq
\psi_{\ell-1}\cdots \psi_0\ket{0} \; = \; (-1)^{ \frac{\ell(\ell-1)}{2} +\ov{\sg}_{\ell}\big(\{k_a\}\big)} 
\cdot \psi_{k_s-1}\cdots \psi_{k_1-1}\, \psi_{k_1^{\e{c}}-1} \cdots \psi_{k_{\ell-s}^{\e{c}}-1}\ket{0}
\enq
We  denote by $\mathbb{Y}$ the Young diagram with Frobenius coordinates $ \Big\{ \{\a_a\}_1^{n_k-s} \, ; \, \{\be_a\}_1^{n_k-s} \Big\}$. 
The vector $\ket{\mathbb{Y};\ell}$ can then be recast as 
\bem
\ket{\mathbb{Y};\ell} \; = \; \big( -1 \big)^{ \ov{\sg}_{\ell}\big(\{k_a\}\big) +\frac{\ell(\ell-1)}{2} } \cdot \psi_{k_{\ell-s}^{\e{c}}-1}^*\cdots  \psi_{k_{1}^{\e{c}}-1}^*\, 
\psi_{-t_{1}}^*\cdots \psi_{ - t_{n_t} }^*\, \psi_{k_{n_k}-1}\cdots  \psi_{k_{s+1}-1} \cdot  \psi_{k_{s}-1}\cdots  \psi_{k_{1}-1} \cdot
  \psi_{k_{1}^{\e{c}}-1} \cdots \psi_{k_{\ell-s}^{\e{c}}-1} \ket{0}  \\
\; = \; \big( -1 \big)^{ \ov{\sg}_{\ell}\big(\{k_a\}\big)   } \cdot (-1)^{ \f{ \ell(\ell-1) }{ 2 } } \cdot \big(-1\big)^{(n_t+n_k)(\ell-s)} \cdot \ket{ \mc{J}_{n_k;n_t} } \;. 
\end{multline}
We now recast the vector $\ket{ \mathbb{Y} ; \ell+r }$ where the partition $ \mathbb{Y} $ is defined in terms of exactly the same Frobenius coordinates. 
Permuting the fermionic operators labelled by the indicies $\{ r+1 , \dots, r+\ell \}$ and $\{ 1,\dots, r \}$ we are led to
%
%
\bem
\ket{\mathbb{Y};\ell+r} \; = \; \big( -1 \big)^{ \ov{\sg}_{\ell}(\{k_a\}) +\frac{\ell(\ell-1)}{2}} \cdot \big(-1)^{\frac{r(r-1)}{2} + \sg_{r}(\{t_a\})  } \cdot
\psi_{r+k_{\ell-s}^{\e{c}}-1}^*\cdots  \psi_{r+k_{1}^{\e{c}}-1}^*\cdot
\psi_{r-t_{1}}^*\cdots \psi_{ r- t_{n_t} }^*   \\
\times  \psi_{r+k_{n_k}-1}\cdots  \psi_{r+k_{s+1}-1} \cdot   \psi_{r+k_{s}-1} \cdots \psi_{r+k_{1}-1}  \cdot \psi_{r+k_{1}^{\e{c}}-1} \cdots \psi_{r+k_{\ell-s}^{\e{c}}-1}
\cdot  \psi_{ r- t_{q} }\cdots \psi_{ r- t_{1} }\, \psi_{ \wt{t}_{r-q}-1 }\cdots \psi_{  \wt{t}_{1}-1 } \ket{0}  \\
\; = \; \big( -1 \big)^{ \ov{\sg}_{\ell}(\{k_a\}) +\frac{\ell(\ell-1)}{2}} \cdot \big(-1)^{\frac{r(r-1)}{2} + \sg_{r}(\{t_a\})  } 
\cdot \big(-1\big)^{(n_t+n_k)(\ell-s)+q(n_t+n_k-q)} \cdot \ket{ \mc{J}_{n_k^{\prime};n_t^{\prime}}^{(r)} } \;. 
\end{multline}
The action \eqref{ecriture action operateur translation sur etat J nk nt} follows straightforwardly from the above and from the 
representation \eqref{ecriture operateur translation dans base des coord Frob a n parts} of the shift operator.\qed




\section{The model}
\label{Section presentation gle du modele}

In the present section, we describe the setting in which we assume the one-dimensional quantum Hamiltonian to fit in. More precisely, we describe the assumptions we make on 
its spectrum, its states and on certain general features satisfied by the form factors of the model's local operators. 
Such a description is standard for quantum integrable systems in one-dimension, see \textit{e}.\textit{g}. \cite{BogoliubiovIzerginKorepinBookCorrFctAndABA} 
in what concerns the spectrum part and the papers \cite{KozKitMailSlaTerEffectiveFormFactorsForXXZ,KozKitMailSlaTerThermoLimPartHoleFormFactorsForXXZ} in what concerns 
the form factors. The validity of the picture has been also established, on a perturbative level, in \cite{CauxGlazmanImambekovShasiNonUniversalPrefactorsFromFormFactors}. 
We thus believe that this setting is common, in particular, to all models belonging to the universality class of the Luttinger model.

\subsection{General hypothesis on the spectrum}

We consider a one-dimensional quantum Hamiltonian $\mc{H}$ representing a physical system in \textit{finite} volume $L$. 
The volume $L$ represents, for instance, the number of sites in the case of a lattice model or the overall volume of the 
occupied space in the case of a model already formulated in the continuum. 

We shall assume that the eigenstates of the Hamiltonian can be organised into sectors with a fixed bare-particle number $N^{\prime}$. 
In practical situations, the integer $N^{\prime}$ may be related to the total longitudinal spin of a state (in the case of spin chains
with total longitudinal spin conservation) or may simply correspond to the number of bare particles building up the 
many body eigenstate of $\mc{H}$ (in the case of models enjoying a conservation of the total number of bare particles). 

In such a setting, the ground state of the model $\ket{ \Psi_{g;N} }$ is located within the sector with $N$ bare-particles. 
This number does depend on $L$ and is such that, in the thermodynamic limit $L\tend + \infty$, one has 
$\lim_{L\tend +\infty} \big( \tf{N}{L}  \big) \, = \, D>0$. Furthermore, taking the thermodynamic limit restricts the space of states
to the sector corresponding to excitations having a \textit{finite}, when $L\tend+\infty$,  energy relatively to the ground state. 
We assume that the eigenstates having this property are located in sectors with $N^{\prime}$ bare particles where 
$N^{\prime}$ is such that  the difference $s=N^{\prime}-N$ remains finite in the thermodynamic limit. 
Having in mind the CFT-based interpretation of the integer $s$, just as the spin-chain setting, we shall refer to $s$ as the spin of the 
excited state.  

We further assume that the excited states are only built up from particle-hole excitations. 
This means that we assume that it is possible to label the eigenstates, within each sector built up from $N+s$ bare-particles,
by a set of integers 
\beq
\mc{I}_{n}^{(s)} \; = \;\Big\{  \{ p_a^{(s)} \}_1^{n} \quad ;  \quad \{ h_a^{(s)} \}_1^{n} \Big\} 
\label{definition ensemble total pour parametriser etats}
\enq
containing two collections of integers which label the so-called particle $\{ p_a^{(s)} \}_1^{n}$ and hole $\{ h_a^{(s)} \}_1^{n}$ excitations. 
In this parametrisation, the integer $n$ may run through $ 0, 1, \dots N + s$ while :
\beq
p_1^{(s)} < \dots < p_n^{(s)} \qquad \e{and}  \qquad h_1^{(s)} < \dots < h_n^{(s)} \qquad \e{with} \qquad 
\left\{ \ba{c}  p_a^{(s)} \in  \intn{-M_{L}^{(1)} }{ M_{L}^{(2)} } \setminus \intn{ 1 }{ N +s }   \vspace{2mm} \\ 
				h_a^{(s)}  \in \intn{ 1 }{ N + s}   \ea \right. \;. 
\enq
The precise values of the integers $M_{L}^{(a)}$ defining the range of the $p_a$'s vary from one model to another. 
Typically for models having no upper bound on their energy, one has $M_{L}^{(a)}=+\infty$ while
for model having an upper bound, $M_{L}^{(1)}, M_{L}^{(2)}$ are both finite but such that $M_{L}^{(a)}-N$, $a=1,2$,
both go to $+\infty$ sufficiently fast with $L$. According to this setting, we shall denote the 
 eigenstates of the model as $\ket{ \mc{I}_{n}^{(s)} } $. 
 
The vectors $\ket{ \mc{I}_{n}^{(s)} }$  provide one with the so-called microscopic description of the model, namely a complete parametrisation of the space of states
in terms of discrete integers. However, it is the macroscopic description that is 
pertinent for describing the thermodynamic limit of the observables in the model. This macroscopic description arises
by means of the so-called counting function $\wh{\xi}_{ \mc{I}_{n}^{(s)} }$ associated with each given excited state $\ket{ \mc{I}_{n}^{(s)} }$. 
More precisely, the particle, resp. the hole-like excitations in a given eigenstates are described directly by a set of rapidities $\{\wh{\mu}_{p_a}^{(s)} \}_1^n$, resp. 
 $\{\wh{\mu}_{h_a}^{(s)} \}_1^n$. These rapidities are defined  as the unique solutions to 
\beq
\wh{\xi}_{ \mc{I}_{n}^{(s)} }\big( \wh{\mu}_{p_a}^{(s)} \big)  \; = \;  \f{ p_a^{(s)} }{L}  \qquad \e{and} \qquad 
\wh{\xi}_{ \mc{I}_{n}^{(s)} }\big( \wh{\mu}_{h_a}^{(s)} \big)  \; = \;  \f{ h_a^{(s)} }{L}\;. 
\label{ecriture equations definissant rap part et trous}
\enq
Since, as mentioned, the counting function does depend, \textit{a priori} on the set of integers labelling the excited state 
$\ket{ \mc{I}_{n}^{(s)} }$, the system of equations \eqref{ecriture equations definissant rap part et trous} is, in fact, extremely involved.

In the following, we shall build on the assumption that any counting function admits, in the $L\tend + \infty$ limit, the asymptotic expansions
\beq
\wh{\xi}_{ \mc{I}_{n}^{(s)} }\big(\om \big)  \; = \; \xi(\om) \; + \; \f{ 1 }{  L   }  \xi_{-1} (\om)
\; - \;  \f{ 1 }{  L   } F_{\mc{R}_{n}^{(s)}} (\om)   \; + \; \e{O} \Big( \f{ 1 }{ L^2 } \Big) \;. 
\label{ecriture definition fonction comptage}
\enq
The asymptotic expansion of the counting functions involves three "macroscopic" functions.  

\begin{itemize}
\item The function $\xi$ is the asymptotic counting function.  It is the same for \textit{all} excited states and we assume that it
is strictly increasing. This function defines a set of "macroscopic" rapidities $\{ \mu_a \}_{a \in \mathbb{Z} } $ 
\beq
\xi( \mu_a ) \; = \;  \f{ a }{ L } \;.  
\label{ecriture equation definition rapidites macro}
\enq
These macroscopic rapidities provide one with the leading order in $L$ approximation of the rapidities $\{\wh{\mu}_{p_a}^{(s)} \}_1^n$ and 
 $\{\wh{\mu}_{h_a}^{(s)} \}_1^n$:
\beq
\wh{\mu}_{p_a}^{(s)} \; \simeq \;  \mu_{ p_a^{(s)} }  \qquad \e{and} \qquad 
\wh{\mu}_{h_a}^{(s)} \; \simeq \;  \mu_{ h_a^{(s)} }  \;. 
\label{ecriture eqn Asymp pour position reseau rapidites}
\enq
\item  The function $F_{\mc{R}_{n}^{(s)} } $ stands for the shift function (of the given excited state in respect to the model's ground state). 
It is a function of the spin $s$ and of the set macroscopic rapidities  
\beq
\mc{R}_{n}^{(s)} \; = \; \Big\{ \{ \mu_{ p_a^{(s)} } \}_1^n \; ; \; \{ \mu_{h_a^{(s)}} \}_1^n \Big\} \;. 
\label{introduction variables macroscopiques}
\enq
This function measures the small $\e{O}(L^{-1})$ drift in the position of a rapidity in the Fermi sea under the effect of interactions. 
\item Finally, the function $\xi_{-1}$ represents the $1/L$ corrections to the ground state's counting functions. 
 We stress that our way of decomposing the $1/L$ corrections to the counting function 
is such that the ground state's shift function vanishes, \textit{i}.\textit{e}. $F_{\mc{R}_{0}^{(0)} } =0 $.  
\end{itemize}

In the large-$L$ limit and within such a setting, the rapidities for the ground state form a dense distribution 
 on $\intff{-q}{q}$ -- the so-called Fermi zone of the model -- with density 
$\xi^{\prime}$. The endpoints $\pm q$ are called the Fermi boundaries.
The particle and hole excitations in the model\symbolfootnote[2]{Typically, for quantum integrable models, these functions are given as solutions to linear integral equations \cite{GaudinFonctionOndeBethe}. 
Their description, for more complex, non-integrable models, is in principle much more complicated.} carry a dressed momentum $p$ and a dressed energy $\veps$. 
The dressed energy and momentum are smooth and satisfy to the general properties
\beq
p^{\prime}_{\mid \R } >0 \qquad \; \qquad  \veps_{\intoo{-q}{q}}<0  \qquad \e{and} \qquad \veps_{\R\setminus \intff{-q}{q}}>0 \;. 
\label{ecriture proprietes generales de p et veps}
\enq
The relative momentum and energy of an excited state are expressed in terms of the dressed momentum and energy as 
\beqa
\De\mc{E}\big( \mc{I}_{n}^{(s)}  \big) \; \equiv \; \mc{E}\big( \mc{I}_{n}^{(s)} \big) \; - \;  \mc{E}\big( \mc{I}_{0}^{(0)} \big) 
& = &  \sul{a=1}{n} \Big( \veps\big( \mu_{p_a^{(s)}} \big) \; - \;  \veps\big( \mu_{h_a^{(s)}} \big) \Big)
\; + \; \e{O}\Big( \f{1}{L} \Big)  \label{definition relative ex energy} \\
\De\mc{P}\big( \mc{I}_{n }^{(s)}  \big) \; \equiv \; \mc{P}\big( \mc{I}_{n}^{(s)} \big) \; - \;  \mc{P}\big( \mc{I}_{0}^{(0)} \big) & = & 
\sul{a=1}{n} \Big( p\big( \mu_{p_a^{(s)}} \big) \; - \;  p\big( \mu_{h_a^{(s)}} \big) \Big) 
\; + \; \e{O}\Big( \f{1}{L} \Big)\;.
\label{definition relative ex momentum}
\eeqa
Above, $\mc{I}_{0}^{(0)}=\{ \emptyset ; \emptyset \}$ refers to the set of integers which parametrises the ground state
of the model and $\mc{P}\big( \mc{I}_{n}^{(s)} \big)$ and $\mc{E}\big( \mc{I}_{n}^{(s)} \big)$ are respectively the momentum and energy of the state parametrised by the set of integers $\mc{I}_{n}^{(s)}$.  
We remind that the superscript $(s)$ refers to the fact that the excitation labelled by $\mc{I}_{n }^{(s)}$ takes place above the 
lowest lying energy level in the spin $s$ sector. 

The excitation momentum and energy are probably the best examples allowing one to discuss a loss of information due to parametrising the
particles and holes in terms of their "macroscopic" momenta. Indeed, on the level of \eqref{definition relative ex energy}-\eqref{definition relative ex momentum},
there arises a huge degeneracy in what concerns the $\e{O}(1)$ finite parts of the excitation energy and momentum in that numerous choices of the set $\mc{I}_n^{(s)}$
will lead to the \textit{same} value of the finite part. Indeed, the very definition \eqref{ecriture equation definition rapidites macro} of the macroscopic rapidities ensures that
\beq
\mu_{k} - \mu_{k+\ell} \; = \; \e{O}\big( \f{\ell}{L}\big) \;. 
\enq
Thus, a direct computation shows that for any fixed collection $\{k_a, t_a\}_1^n$ of  $L$-independent integers, the two sets of integers
\beq
\mc{I}_n^{(s)} \; = \; \Big\{ \{p_a\}_1^n \; ; \; \{h_a\}_1^n \Big\} \quad \e{and} \quad 
\wt{\mc{I}}_n^{(s)} \; = \; \Big\{ \{p_a + k_a \}_1^n \; ; \;  \{h_a+t_a\}_1^n \Big\}
\label{ecriture ensemble entiers donnant meme vars macro}
\enq
will generate the same finite part of $\De\mc{E}$ or $\De\mc{P}$, namely 
\beq
\De\mc{E}\big( \mc{I}_{n}^{(s)}  \big) \; - \;  \De\mc{E}\big( \wt{\mc{I}}_{n}^{(s)}  \big)  \; = \; \e{O}\Big( \f{1}{L} \Big) \qquad \e{and} \qquad
\De\mc{P}\big( \mc{I}_{n}^{(s)}  \big) \; - \;  \De\mc{P}\big( \wt{\mc{I}}_{n}^{(s)}  \big)  \; = \; \e{O}\Big( \f{1}{L} \Big) \;. 
\enq

Thus, the so-called microscopic variables allow one to distinguish between all of the states of the Hilbert space whereas the macroscopic variables,  
at least in what concerns the leading in $L$ order, naturally give rise to a huge indeterminacy in identifying a given state in the model
- many microscopically distinct states give rise to the same macroscopic variables-. Hence, in the large $L$ limit, there is a loss of information
when passing from the microscopic to the macroscopic description.




\subsection{The operators and their form factors}

We assume that the model is endowed with a collection of local operators $\mc{O}_r$. 
These operators are best characterised in terms of their form factor, namely their 
expectation values taken between two excited states. We shall assume that the operator 
$\mc{O}_r$ only connects those eigenstates which differ by $o_r$ in their spin, namely
%
%
%
%
%
\beq
\mc{F}_{\mc{O}_r} \Big( \mc{I}_{ m }^{(s^{\prime})} \, ; \,   \mc{I}_{ n }^{(s)} \mid x  \Big) \; = \; 
\ex{ \i x [ \mc{P}( \mc{I}_{ m }^{(s^{\prime})} ) -  \mc{P}( \mc{I}_{ n }^{(s)} ) ] } 
\cdot  \bra{ \mc{I}_{ m }^{(s^{\prime})} } \mc{O}_r(0) \ket{ \mc{I}_{ n }^{(s)} }  \; \not= \; 0
\qquad \e{only}\; \e{if} \quad s \, - \, s^{\prime} \;  = \;  o_r\;. 
\enq

Scalar observables introduced in the last subsection were parametrised, in the large-$L$ limit, 
solely by the macroscopic set of rapidities $\mc{R}_{n}^{(s)}$ subordinate to the set of multi-indices $\mc{I}_{n}^{(s)}$. 
This is no longer the case for form factors as we demonstrated in our previous work  \cite{KozKitMailSlaTerThermoLimPartHoleFormFactorsForXXZ}. Within our setting, 
the latter are parametrised, in the large-$L$ limit, \textit{not only} by the sets of macroscopic rapidities $\mc{R}_{ n }^{(s+o_r)}, \mc{R}_{ m }^{(s)} $ 
but also by the sets of \textit{discrete} integers $\mc{I}_{n}^{(s+o_r)}, \mc{I}_{ m }^{ (s) }$. 
Namely, for properly normalised states $\ket{ \Psi\big( \mc{I}_{ n }^{(s+o_r)} \big) } $ and 
their duals $\bra{ \Psi\big( \mc{I}_{ m }^{(s)} \big) } $, the form factors take the form 
\bem
\mc{F}_{\mc{O}_r}\big( \mc{I}_{ m }^{(s)} \, ; \,  \mc{I}_{ n }^{(s+o_r)}  \mid 0 \big) 
\; = \; 
\mc{S}^{( \mc{O}_r)} \Big(   \mc{R}_{ m }^{(s)} \, ; \,   \mc{R}_{n}^{(s+o_r)}     \Big)
     \cdot 	\mc{D}^{(s)}
 \Big(   \mc{R}_{m}^{(s)}\, ; \,  \mc{R}_{n}^{(s+o_r)}     \big| \, 
    \mc{I}_{ m }^{(s)} \, ; \,   \mc{I}_{ n }^{(s+o_r)}   	\Big)	  \cdot 
\bigg(  1+ \e{O}\Big( \f{\ln L }{ L } \Big) \bigg)      		 \;. 
\end{multline}
In this decomposition 
\begin{itemize}
\item $\mc{S}^{( \mc{O}_r )}$ is called the smooth part: it solely depends 
on the macroscopic rapidities $\mc{R}_{n}^{(s+o_r)}, \mc{R}_{ m }^{(s)} $. Furthermore, this dependence is smooth what ensures that 
a change in the value of the integers in the spirit of \eqref{ecriture ensemble entiers donnant meme vars macro} will only affect the value of the 
smooth part in the  $1/L$ corrections.  
\item The smooth part is a set function so that it is invariant under permutation of the particle or hole rapidities. It satisfies to 
particle-hole reduction properties, meaning that 
\beq
\mc{S}^{( \mc{O}_r)} \Big(  \mc{R}_{ m }^{(s)} ; \,  \mc{R}_{ n }^{(s+o_r)}    \Big)
_{\mid \mu_{p_k^{(s)}}= \mu_{h_{\ell}^{(s)}} } \; = \; 
\mc{S}^{( \mc{O}_r)} \Big(  \wh{\mc{R}}_{ m }^{(s)} ; \,  \mc{R}_{ n }^{(s+o_r)}    \Big)
\qquad \e{with} \qquad 
\wh{\mc{R}}_{m}^{(s)}  \; = \; \Big\{ \{ \mu_{ p_a^{(s)} } \}_{1; \not=k}^{ n } 
\; ; \; \{ \mu_{h_a^{(s)}} \}_{1; \not= \ell}^{ n } \Big\} \;. 
\nonumber
\enq
The same type of reduction holds as well for the particle-hole rapidities that belong to the set $\mc{R}_{ n }^{(s+o_r)}$. 

\item $\mc{D}^{(s)}$ is called the discrete part because it depends explicitly on both 
types of parametrisations: the macroscopic rapidities $\mc{R}_{ n }^{(s+o_r)}, \mc{R}_{ m }^{(s)} $ and 
 the sets of integers labelling the excited states $\mc{I}_{ n }^{(s+o_r)}, \mc{I}_{ m }^{(s)} $. 
We do stress that this last dependence is \textit{explicit} in the sense that it goes beyond 
a sole dependence of the integers through the parametrisation by the macroscopic rapidities.  
 The main effect of such a dependence is that a change in the value of the integers in the spirit of \eqref{ecriture ensemble entiers donnant meme vars macro}
 will result in a significant change (of the order of $\e{O}(1)$) in the
value of $\mc{D}^{(s)}$. The discrete part thus keeps track of the microscopic details of the different
excited states. 
\end{itemize}
The smooth part represents, in fact, a non-universal part of the model's form factors. 
Its explicit expressions not only depends on the operator $\mc{O}_r$ but also varies strongly 
from one model to another, see \cite{KozKitMailSlaTerThermoLimPartHoleFormFactorsForXXZ,KozFFConjFieldNLSELatticeSpacingGoes0} 
for concrete examples of the form taken by $\mc{S}_{\mc{O}_r}$ for different quantum integrable models. The part $\mc{D}^{(s)}$
is, however, entirely universal within the present setting of the description of the model's spectrum. 
It solely depends on the values of the spin $o_r$ of the operator $\mc{O}_r$. 
Its general explicit expression plays no role in our analysis, in the sense that we shall only 
need the expression for specific excited states, namely the ones belonging to the so-called $\ell_s$-critical classes that we define bellow. 

We refer the interested reader to \cite{KozKitMailSlaTerRestrictedSums} for a thorougher discussion relative to the 
origin of the discrete part $\mc{D}^{(s)}$ within the framework of Bethe Ansatz  solvable models.

\subsection{The critical $\ell_s$ class}

Observe that, in virtue of \eqref{ecriture proprietes generales de p et veps}, one has $\veps(q)=0$. As a consequence,
 by \eqref{definition relative ex energy}, there exists a possibility to realise zero energy excitations in the thermodynamic limit $L\tend +\infty$ by forming holes and 
particles whose rapidities scale down to $\pm q$ when $L\tend +\infty$. This singles out a class of excited states of the model
which we shall refer to as critical states. An excited state is said to belong to the critical class 
if , in the $L \tend +\infty$ limit,  \textit{all} the macroscopic rapidities of the particles and holes associated with this state 
"collapse" on the Fermi boundary:
\beq
\mu_{p_a^{(s)}} \simeq   \pm q \qquad \e{and} \qquad \mu_{h_a^{(s)}} \simeq   \pm q  \;. 
\enq
There, the $\pm$ sign depends on whether the particle or hole's rapidity collapses on the right or the left Fermi boundary.

A set of integers $\mc{I}^{(s)}_n$ is said to parametrise 
a critical excited state if the associated particle-hole integers $\{ p_a^{(s)} \}_1^{ n }$ and $\{ h_a^{(s)} \}_1^{ n }$  can be represented as
\beq
\big\{ p_a^{(s)}  \big\}_1^{ n } \; = \;  \big\{  N+s + p_{a;+}^{(s)} \big\}_1^{ n_{p;+}^{(s)} } \, \cup  \,
\big\{ 1 -  p_{a;-}^{(s)} \big\}_1^{ n_{p;-}^{(s)} }  \qquad \e{and} \qquad 
 \big\{ h_a^{(s)} \big\}_1^{ n } \; = \;  \big\{1+ N+ s - h_{a;+}^{(s)} \big\}_1^{ n_{h;+}^{(s)} } \, \cup  \, 
\big\{ h_{a;-}^{(s)}  \big\}_1^{ n_{h;-}^{(s)} }   
\label{ecriture decomposition locale part-trou close Fermi zone}
\enq
where the integers $p_{a;\pm}^{(s)}, h_{a;\pm}^{(s)} \in \mathbb{N}^{*}$ are "small" compared to $L$, \textit{i}.\textit{e}. 
\beq
\lim_{L\tend +\infty} \f{ p_{a;\pm}^{(s)} }{ L }  \; = \; \lim_{L\tend +\infty} \f{ h_{a;\pm}^{(s)} }{ L }  \; = \; 0 \;, 
\enq
 and the integers $n_{p/h;\pm}^{(s)}$ satisfy to the constraint
\beq
n_{p;+}^{(s)} \, + \,  n_{p;-}^{(s)} \; = \;  n_{h;+}^{(s)} \, + \,  n_{h;-}^{(s)} \;= \; n \;. 
\enq
Within this setting, one can readily check that the critical excited state described above 
will have $n_{p;+/-}^{(s)}$ particles, resp. $n_{h;+/-}^{(s)}$ holes, on the right/left end of the Fermi zone $\intff{-q}{q}$
associated with the spin $s$ sector. 

In fact, one can distinguish between various critical states by organising them into so-called 
$\ell_s$-critical classes. This classification takes its origin in the fact that all such states have a vanishing 
excitation energy (up to \e{O}(1/$L$) corrections) but can be gathered into classes depending on the value of their
macroscopic momenta $2\ell_s p_F$, where $p_F=p(q)$ is the so-called Fermi momentum and 
\beq
\ell_s \; = \;  n_{p;+}^{(s)}  \; - \;  n_{h;+}^{(s)} \; = \; n_{h;-}^{(s)}  \; - \;  n_{p;-}^{(s)} \;. 
\label{ecriture lien shift ells et differences part trous sur bords zone Fermi}
\enq
Note that the subscript $s$ in $\ell_s$ allows one to localise the given critical class in a specific spin $s$ sector.

The finite part of the relative excitation momentum associated with an excited state belonging to the $\ell_s$ critical class only depends
on $\ell_s$, while the first dependence on the specific representative of the class arises on the level of the $L^{-1}$corrections. 
More precisely, one has:
\bem
\De\mc{P}\big( \mc{I}_{n }^{(s)}  \big) \; =\;  2 \ell_s  p_F 
\; + \; \f{2\pi}{L} \a^{+} \Bigg\{ \sul{a=1}{ n_{p;+}^{(s)} } (p_{a;+}^{(s)}-1) \; + \;  \sul{ a=1 }{ n_{h;+}^{(s)} } h_{a;+}^{(s)}   \Bigg\}
\; - \; \f{2\pi}{L} \a^{-} \Bigg\{ \sul{a=1}{n_{p;-}^{(s)} } (p_{a;-}^{(s)} -1)\; + \;  \sul{a=1}{ n_{h;-}^{(s)} }  h_{a;-}^{(s)}  \Bigg\} \\
\; + \; \f{2\pi}{L} \Big\{ \a^{-} \ell_s \f{(\ell_s+1)}{2} \, - \, \a^{+} \ell_s \f{(\ell_s-1)}{2} \Big\} \; + \; \dots
\label{ecriture ex momentum pour etats ell shiftees}
\end{multline}
where we have set 
\beq
\a^{\pm} \; = \; \f{1}{2\pi} \f{ p^{\prime}(\pm q) }{ \xi^{\prime}(\pm q)  } \;. 
\label{ecriture formule pour alpha pm}
\enq
We \textit{insist} that all the terms included in the dots either
\begin{itemize}
 \item  are of the order of $\e{O}(1/L)$  \textit{but} do \textit{not} depend on the integers $p_{a;\pm}$ and $h_{a;\pm}$ nor on $n_{p/h;\pm}$ (they can, nonetheless depend on $\ell_s$); 
\item depend on these integers \textit{but} are of the order of $\e{O}(1/L^2)$. 
\end{itemize}
Note that, in principle, the terms present in the second line of  \eqref{ecriture ex momentum pour etats ell shiftees} could have been simply included in the $\dots$. We chose to write them down
for normalisation purposes. Namely, the formula  \eqref{ecriture ex momentum pour etats ell shiftees} in its present form allows one for a more straightforward 
correspondence with the free boson model. 

The partitioning of the particle and hole integers of a given state belonging to the $\ell_s$ critical class into two collections of integers of $\pm$
types \eqref{ecriture decomposition locale part-trou close Fermi zone} suggests to parametrise the excited states within the $\ell_s$ critical class directly
in terms of the two subsets $\mc{J}^{(s)}_{ n_{p;+}^{(s)} ; n_{h;+}^{(s)} } \cup \mc{J}^{(s)}_{ n_{p;-}^{(s)} ; n_{h;-}^{(s)} } $ where 
\beq
  \mc{J}^{(s)}_{  n ; m  }  \; = \; \Big\{ \{ p_{a}^{(s)} \}_1^{ n } 
\; ; \;  \{ h_{a}^{(s)} \}_1^{ m } \Big\}  \;. 
\enq
Hence, from now on, for any state belonging to a critical $\ell_s$ class, we shall identify the sets 
\beq
\mc{I}^{(s)}_n \qquad  \e{and} \qquad 
\mc{J}^{(s)}_{ n_{p;+}^{(s)} ; n_{h;+}^{(s)} } \cup \mc{J}^{(s)}_{ n_{p;-}^{(s)} ; n_{h;-}^{(s)} } \;.
\enq
We do stress that, due to \eqref{ecriture decomposition locale part-trou close Fermi zone}, the value of the spin $s$ does play a role in the correspondence between 
$\mc{I}^{(s)}_n$ and $\mc{J}^{(s)}_{ n_{p;+}^{(s)} ; n_{h;+}^{(s)} } \cup \mc{J}^{(s)}_{ n_{p;-}^{(s)} ; n_{h;-}^{(s)} }$. 
We also draw the reader's attention to the fact that the value of $\ell_s$ is encoded in the very notation $\mc{J}^{(s)}_{ n_{p;+}^{(s)} ; n_{h;+}^{(s)} } 
\cup \mc{J}^{(s)}_{ n_{p;-}^{(s)} ; n_{h;-}^{(s)} } $, see \eqref{ecriture lien shift ells et differences part trous sur bords zone Fermi}.

\subsection{Large-$L$ expansion of form factors connecting critical states}
\label{Section dvpt a grand L des FF}

We are finally able to discuss the expression for the form factors of local operators $\mc{O}_r$
taken between two eigenstates belonging to critical classes. It is precisely these form factors that are responsible for the 
emergence of a effective field theory description at large distances of separation between the operators. 
Let 
\beq
\mc{I}^{(s)}_m \; \equiv \; \mc{J}_{ m_{p;+}; m_{h;+} } \cup \mc{J}_{ m_{p;-}; m_{h;-} } 
\quad \e{and} \quad 
\mc{I}^{(s+o_r)}_n \; \equiv \; \mc{J}_{ n_{p;+}; n_{h;+} } \cup \mc{J}_{ n_{p;-}; n_{h;-} } 
\enq
be two sets of integers\symbolfootnote[2]{In this writing we chose not to omit the spin sector label so as to lighten the formulae. Since the 
context of the setting is clear, this ought not lead to any confusion.} parametrising excited states belonging to the 
\beq
 \ell_{\e{out}}\; = \; m_{p;+} -  m_{h;+} \; = \; m_{h;-} -  m_{p;-}   \qquad \e{and} \qquad
 \ell_{\e{in}}\; = \; n_{p;+} -  n_{h;+} \; = \; n_{h;-} -  n_{p;-}
\enq
critical classes. 
 
Within the framework we impose on the model, the form factors of local operators taken between two excited states belonging to the critical classes 
introduced above take the form:  
\bem
\mc{F}_{\mc{O}_r}\bigg(  \mc{I}^{(s)}_m  ;  \mc{I}^{(s+o_r)}_n \mid x_r \bigg)  \; = \; \Big\{ \ex{2 \i p_F x_r }  \Big\}^{ \ell_{ \e{out} }- \ell_{\e{in}}  }
 \cdot C^{(\ell_{ \e{out} } - \ell_{\e{in}})} \big( \nu_r^+, \nu_r^- \big) \cdot \mc{F}_{\ell_{ \e{out} } - \ell_{\e{in}}}\big( \mc{O}_r \big)
\cdot \bigg( \f{ 2\pi }{ L } \bigg)^{ \rho\big(\nu_r^{+}+\ell_{ \e{out} }- \ell_{\e{in}} \big) + \rho\big(\nu_r^-+\ell_{ \e{out} }- \ell_{\e{in}}\big) }   \\
\times   \, 
\msc{F}\Big[  \mc{J}_{ m_{p;+}; m_{h;+}  } ;  \mc{J}_{ n_{p;+} ; n_{h;+} }     \mid   \nu_r^{+}  ,   \om_r^{+} \Big] \cdot 
 \msc{F}\Big[ \mc{J}_{ m_{p;-} ; m_{h;-} } ;  \mc{J}_{ n_{p;-} ; n_{h;-} }  \mid - \nu_r^{-}  , \om_r^{-} \Big] \\
\big(\om_r^+ \big)^{ \ell_{\e{in}}\f{\ell_{\e{in}}-1}{2} - \ell_{\e{out}}\f{\ell_{\e{out}}-1}{2} } \cdot  %
\big(\om_r^- \big)^{ \ell_{\e{in}}\f{\ell_{\e{in}}+1}{2} - \ell_{\e{out}}\f{\ell_{\e{out}}+1}{2} } 
\cdot \bigg( 1 \, + \, \e{O}\Big( \f{ \ln L }{ L } \Big) \bigg)\;. 
\label{ecriture conjecture form general facteur de forme deux etats excites}
\end{multline}

The constituents of the above formula are parametrised by the values 
\beq
\nu_{r}^{+} \; =  \; \nu_r(  q )  \, - \, o_r \qquad \e{and} \qquad
\nu_{r}^{-} \; =  \; \nu_r(  -q )
\enq
that the relative shift function between the $\ell_{\e{in}}, \ell_{\e{out}}$ critical states 
\beq
\nu_{r}(\la) \; = \;  F_{s}(\la) \; - \; F_{s+o_r}(\la)  
\label{definition fction shift relative etats gauche et droit}
\enq
takes on the right/left endpoints of the Fermi zone, up to subtracting the level $o_r$ of the operator $\mc{O}_r$
in the case of the right endpoint. The also depend on 
\beq
\om_r^{+} \; = \; \ex{ 2\i \pi \a^+ \f{x_r}{L}  }  \qquad \e{and} \qquad  \om_r^{-} \; = \; \ex{ - 2\i \pi \a^- \f{x_r}{L}  }   
\enq
representing the exponent of the individual momentum brought by a particle excitation on the left or right Fermi boundary. 
In the above large-$L$ asymptotics, the quantity $\mc{F}_{\ell_{ \e{out} }- \ell_{ \e{in} } }\big( \mc{O}_r \big)$ 
represents the properly normalised finite and non-universal (\textit{i}.\textit{e}. model and operator dependent) part of the large-$L$ behaviour of the 
form factor of the operator $\mc{O}_r$ taken between fundamental representatives of the $\ell_{ \e{in} }$ and $\ell_{ \e{out} }$ critical classes. 
More precisely, it is defined as
\beq
\mc{F}_{\ell_{ \e{out} }- \ell_{ \e{in} } }\big( \mc{O}_r \big) \; = \; \lim_{L\tend +\infty}\Bigg\{
 \bigg( \f{ L }{ 2\pi } \bigg)^{ \rho\big(\nu_r^{+}+\ell_{ \e{out} }- \ell_{\e{in}} \big) + \rho\big(\nu_r^-+\ell_{ \e{out} }- \ell_{\e{in}}\big)  }
\bra{  \mc{L}_{ \ell_{ \e{out} } }^{(s)}  } \mc{O}_r(0)   \ket{  \mc{L}_{ \ell_{ \e{in} } }^{(s+o_r)}  } \Bigg\}
\label{ecriture definition facteur de forme proprement normalise macroscopique}
\enq
in which the sets of integers $\mc{L}_{ \ell_{ \e{out} } }^{(s)}$  and $ \mc{L}_{ \ell_{ \e{in} } }^{(s+o_r)} $ 
parametrising the excited states correspond to the fundamental representatives of the $\ell_{ \e{out} }$ and $\ell_{ \e{in} }$ critical classes. 
Namely,   $ \mc{L}_{ \ell  }^{(s)} $  is the set of particle-hole integers living on the Fermi boundary 
in the spin $s$ sector and such that 
\beq
 \mc{L}_{ \ell }^{(s)}  \; = \; 
\left\{   \ba{cc} \Big\{  \{ p_{a;+}^{(s)}  = a \}_1^{\ell} \; ;  \; \{ \emptyset \}  \Big\} \bigcup
\Big\{  \{ \emptyset \} \; ; \;  \{ h_{a;-}^{(s)} = a \}_1^{\ell}  \Big\}  &
				\e{if} \; \ell \geq 0   \vspace{2mm} \\ 
		\Big\{  \{ \emptyset \} \; ; \;  \{ h_{a;+}^{(s)}  = a \}_1^{-\ell}   \Big\} \bigcup
\Big\{    \{ p_{a;-}^{(s)} = a \}_1^{-\ell}  \; ; \; \{ \emptyset \}  \Big\}   &
				\e{if} \; \ell \leq 0 \ea \right. 		\; . 
\enq
Note that this form factor \textit{solely} depends on the \textit{difference}  $\ell_{ \e{out} }- \ell_{ \e{in} }$. This issues from the fact that, a priori,
in the thermodynamic limit, any fundamental representative can be chosen to be the reference ground state. Yet, relatively to the fundamental representative of the $\ell_{\e{out}}$-class,
the other excited state corresponds to the fundamental representative of the $\ell_{ \e{out} }- \ell_{ \e{in} }$ class.

The power of the volume $L$ arising in \eqref{ecriture conjecture form general facteur de forme deux etats excites} and 
\eqref{ecriture definition facteur de forme proprement normalise macroscopique} involves
the right $\rho(\nu_r^+)$ and left $\rho(\nu_r^-)$ scaling dimensions whose generic expression reads  
\beq
 \rho(\nu) \; = \;  \f{\nu^2}{2}  \;. 
\label{definition scaling dimension}
\enq

The factors $\msc{F}$ have been  defined in \eqref{definition microscopic form factor}
and contain  all the "microscopic" contributions issuing from excitations 
on the right or left  Fermi boundary. These local microscopic form factors depend on the 
value taken on the right or left Fermi boundary by the relative shift function $\nu_r$ associated with the critical excited states of interest
and on the position of the operator. 
These local microscopic form factors also depend on the sets of integers $\mc{J}_{ m_{p;\pm}; m_{h;\pm} } $ and  
$\mc{J}_{ n_{p; \pm }; n_{h; \pm } } $ parametrising the excitations on the boundary of the Fermi zone to which they are associated. 
Those microscopic form factors are as defined in \eqref{definition microscopic form factor}. 

Finally, the normalisation constant  $C^{ (\ell_{ \e{out}} - \ell_{\e{in}} ) } \big( \nu_r^+, \nu_r^- \big)$ is 
chosen in such a way that it cancels out the $L$-independent contributions of the right and left critical form factors
when focusing on the fundamental representative of the $\ell_{\e{out}}$ and $\ell_{ \e{in} }$ critical classes. 
Its explicit form can be computed in terms of the Barnes $G$-function \cite{BarnesDoubleGaFctn1} and reads:
\beq
C^{(\ell_{ \e{out} } - \ell_{\e{in}} )} \big( \nu_r^+, \nu_r^- \big) \; = \; 
  G\bigg( \ba{cc} 1 + \nu_r^- , 1-\nu_r^+   \\
1  + \nu_r^- + \ell_{ \e{out} } - \ell_{\e{in}}  ,   1-\nu_r^+ - \ell_{ \e{out} } + \ell_{\e{in}}    \ea \bigg)   
	\;. 	
\enq
Above we have used the hypergeometric-like  notations for ratios. These are explained in  \eqref{introduction notation produit compact gamma functions}.

\subsection{General remarks on the scope of applicability of the model}

The structure of the spectrum and form factors that we discussed throughout this section already appeared in the paper \cite{KozKitMailTerMultiRestrictedSums}.

 The formula for the form factors of local operators taken between excited states belonging to critical classes
can be proven within the framework of the algebraic Bethe Ansatz for various quantum integrable models on the basis of determinant representations for their
form factors \cite{KMTFormfactorsperiodicXXZ,KorepinSlavnovApplicationDualFieldsFredDets,KozFFConjFieldNLSELatticeSpacingGoes0,OotaInverseProblemForFieldTheoriesIntegrability,SlavnovFormFactorsNLSE}. 
The corresponding calculations are a straightforward generalisation of the method developed in \cite{KozKitMailSlaTerEffectiveFormFactorsForXXZ,KozKitMailSlaTerThermoLimPartHoleFormFactorsForXXZ,SlavnovFormFactorsNLSE}. 
However, we strongly believe that the decomposition \eqref{ecriture conjecture form general facteur de forme deux etats excites} is, in fact, universal. 
More precisely, the properly normalised form factor $\mc{F}_{  \ell_{ \e{out} }- \ell_{ \e{in} }  }\big( \mc{O}_r \big)$ is definitely model dependent
and thus can only be obtained on the basis of exact computations. Its explicit expression is available for many quantum integrable models. 
However, we do trust that the local microscopic pre-factors and the leading power-law behaviour in $L$ is universal: namely that they take the same form for models belonging to the Luttinger liquid universality class. 
Parts of this structure have been confirmed by perturbative calculations around a free model in \cite{CauxGlazmanImambekovShasiNonUniversalPrefactorsFromFormFactors}.

The attentive reader might observe a slight difference between the expression for the form factors taken between critical excited states as given in 
Section \ref{Section dvpt a grand L des FF} and Section 2.4 of paper  \cite{KozKitMailTerMultiRestrictedSums}. 
More precisely, the formulae differ by 
\begin{itemize}
\item[i)] an overall sign 
\beq
(-1)^{n_{p;+}+n_{p;-}+n_{k;+}+n_{k;-} }\cdot (-1)^{ \frac{n_{p;+}}{2}(n_{p;+}-1) } \cdot (-1)^{ \frac{n_{k;+}}{2}(n_{k;+}-1) } \; ;
\enq
\item[ii)] in  \cite{KozKitMailTerMultiRestrictedSums}, the expression for the normalisation constant $C^{(\ell_{ \e{out} } - \ell_{\e{in}} )} \big( \nu_r^+, \nu_r^- \big)$ 
took the form
\beq
C^{(\ell_{ \e{out} }; \ell_{\e{in}} )}_{\e{original}} \big( \nu_r^+, \nu_r^- \big) \; = \; 
  G\bigg( \ba{cc} 1 + \nu_r^- , 1-\nu_r^+ , 1+\ell_{\e{in}}  - \nu_{r}^-, 1-\ell_{\e{in}} + \nu_r^+  \\
1 -  \ell_{\e{in}} + \nu_r^- ,   1+\ell_{\e{in}}-\nu_r^+   , 1 - \ell_{\e{out}}+\ell_{\e{in}} - \nu_r^- , 1+ \ell_{\e{out}}-\ell_{\e{in}} + \nu_r^+  	\ea \bigg)   \;;
\enq

\item[iii)] the presence, in  \cite{KozKitMailTerMultiRestrictedSums}, of the factor 
\beq
C^{(\ell_{ \e{out} }; \ell_{\e{in}} )}_{\e{extra}} \big( \nu_r^+, \nu_r^- \big) \; = \; 
\Big( \f{ \sin[ \pi \nu_{r}^{+} ] }{ \sin[ \pi \nu_{r}^{-} ] } \Big)^{ \ell_{\e{out}} } 
\enq
in the definition of the local microscopic form factors as compared to the definition 
\eqref{ecriture conjecture form general facteur de forme deux etats excites} used in the present paper. 
\end{itemize}
The difference of sign pre-factors is irrelevant in that the sign can be readily absorbed into the normalisation of the eigenstates. 
The present choice of the sign was more natural here since it makes the comparison with the free boson model more direct. 
Finally, in what concerns the normalisation constants, one can check, on the basis of identity \eqref{Formule de la double redcution pour la fonction de Barnes},
that one has
\beq
C^{(\ell_{ \e{out} }; \ell_{\e{in}} )}_{\e{extra}} \big( \nu_r^+, \nu_r^- \big) \cdot C^{(\ell_{ \e{out} }; \ell_{\e{in}} )}_{\e{original}} \big( \nu_r^+, \nu_r^- \big) 
\; = \; C^{(\ell_{ \e{out} } - \ell_{\e{in}} )} \big( \nu_r^+, \nu_r^- \big)  \;. 
\enq
%
%
%



\section{The resurgence of an effective free boson field theory in the large-distance regime.}
\label{Section correspondance modele et boson libre}

\subsection{An effective free boson field theory model}

The effective Hilbert space is defined as the tensor product of two copies $\mf{h}_L$ and $\mf{h}_R$ of the free boson Hilbert space $\mf{h}$
introduced previously:
\beq
\mf{h}_{\e{eff}} \; = \;   \mf{h}_{L}  \otimes \mf{h}_{ R } \;. 
\enq
The first space (resp. second) arising in the tensor product will be called left (resp. right) space and will be associated to modelling 
what happens on the left (resp. right) Fermi boundaries.

To a local operator $\mc{O}_r(x_r)$, we associate the below operator on $\mf{h}_{ \e{eff} }$ 
\beq
\msc{O}_r(\om_r) \; = \; \sul{ \kappa \in \mathbb{Z} }{} \mc{F}_{\kappa}\big(\mc{O}_r \big)\cdot \Big( \f{2\pi}{L} \Big)^{\rho(\nu_r(q) -o_r + \kappa)+\rho(\nu_r(-q)+\kappa)} \cdot 
\ex{2\i p_F \kappa x_r} \cdot  \msc{V}_{L}\big(-\nu_r(-q)  , -\kappa \mid \om_r^{-}  \big) \cdot  \msc{V}_{R}\big(\nu_r(q)-o_r,\kappa -o_r \mid \om_r^{+}  \big)  \;. 
\enq
In this formula, 
\begin{itemize}
 \item $\msc{V}_{L/R}(\nu,\kappa;\om)$ stands for the operator acting non-trivially on the $L/R$ copy of the original 
Hilbert space as the vertex operator $\msc{V}(\nu,\kappa;\om)$ defined in \eqref{definition  r shifted vertex operators}, \textit{viz}. 
\beq
\msc{V}_{L}(\nu,\kappa \mid \om) \; = \; \msc{V}(\nu,\kappa \mid  \om) \otimes \e{id} \qquad \e{and} \qquad 
\msc{V}_{R}(\nu,\kappa \mid  \om) \; = \;  \e{id} \otimes \msc{V}(\nu,\kappa \mid  \om)  \;. 
\enq

\item $\rho(\nu)$ are the scaling dimensions introduced in \eqref{definition scaling dimension};
\item $\mc{F}_{\kappa}\big( \mc{O}_r \big)$ is the properly normalised in the volume thermodynamic amplitude defined in 
\eqref{ecriture definition facteur de forme proprement normalise macroscopique} and taken between excited states satisfying $\ell_{\e{out}}-\ell_{\e{in}}=\kappa$. 
\item $\om_s^{\pm}$ is a phase factor that reads
\beq
\om_s^{\pm} \; = \;  \ex{ \pm 2\i \pi   \f{ x_s }{ L } \a^{\pm} } \;. 
\enq

\end{itemize}

 It follows readily from the discussions carried so far that given two sets 
\beq
\mc{I}^{(s)}_m \; \equiv \; \mc{J}_{ m_{p;+}; m_{h;+} } \cup \mc{J}_{ m_{p;-}; m_{h;-} }
\quad \e{and} \quad 
\mc{I}^{(s+o_s)}_n \; \equiv \; \mc{J}_{ n_{p;+}; n_{h;+} } \cup \mc{J}_{ n_{p;-}; n_{h;-} }
\label{ecriture correspondance etats hilbert initial et hilbert effectif}
\enq
 parametrising critical excited states in the physical Hilbert space $\mf{h}_{\e{phys}}$, one 
has an equality, up to $\e{O}\big( \ln L / L \big)$ corrections between matrix elements:
\bem
\mc{F}_{ \mc{O}_r }\Big(  \mc{I}^{(s)}_m  ;  \mc{I}^{(s+o_r)}_n \mid x_r \Big)  \; =  \; (-1)^{\varkappa}
\Big\{ \bra{ s; \mc{J}_{ m_{p;-}; m_{h;-} }  } \otimes \bra{ s;  \mc{J}_{ m_{p;+}; m_{h;+} } }  \Big\}
\msc{O}_r(\om_r) \\ 
\Big\{ \ket{  \mc{J}_{ n_{p;-}; n_{h;-} } ;  s }\otimes \ket{  \mc{J}_{ n_{p;+}; n_{h;+} } ;  s+o_r } \Big\}  \cdot  \bigg( 1 + \e{O}\Big( \f{ \ln L }{ L }\Big) \bigg) 
\label{ecriture identification asymptotique des FF}
\end{multline}
where
\beq
\varkappa \; = \; m_{p;+}-m_{h;+}-n_{p;+}+n_{h;+} \;. 
\enq
Note that, in order to establish \eqref{ecriture identification asymptotique des FF}, one should use that 
\beq
\bra{ s; \mc{J}_{ m_{p}; m_{h} } } \;  = \; \bra{ \mc{J}_{ m_{p}; m_{h} } } \ex{-s P }  \qquad \e{and} \qquad 
 \ket{ \mc{J}_{ m_{p}; m_{h} } ; s }  \; = \; \ex{ s P } \ket{  \mc{J}_{ m_{p}; m_{h} } } 
\enq
as readily inferred from Lemma \ref{Lemme reecriture de la forme de operateur de translation entre espaces hell}. 

Note also that \eqref{ecriture correspondance etats hilbert initial et hilbert effectif}-\eqref{ecriture identification asymptotique des FF} gives rise to a correspondence
between states belonging to $\ell$-classes in the original Hilbert space and general states in the effective Hilbert space:
\beq
 \ket{\mc{I}_m^{(s)} } \; = \;  \ket{  \mc{J}_{ n_{p;-}; n_{h;-} } ; s } \otimes \ket{  \mc{J}_{ n_{p;+}; n_{h;+} } ; s+o_r  } 
\enq

 \subsection{The effective truncation of $\mf{h}_{\e{phys}}$ and the  correspondence}

 Consider the $r$-point ground-to-ground state expectation value $\big< \mc{O}_1(x_1)\cdots \mc{O}_{r}(x_r)  \big>_{ \mf{h}_{\e{phys}} }$. 
One way to express this quantity is to write its form factor expansion, \textit{viz}. insert the closure relation between 
each operator. According to our hypothesis on the space of states, this recast the correlator as
\beq
\big< \mc{O}_1(x_1)\cdots \mc{O}_{r}(x_r)  \big>_{ \mf{h}_{\e{phys}} }\; = \; \sul{ \big\{ \mc{I}^{(\vsg_s)}_{n^{(\vsg_s)}} \big\} }{}
\pl{s=1}{r} \Big\{ \bra{ \mc{I}^{ (\vsg_{s-1}) }_{ n^{(\vsg_{s-1})} }  } \mc{O}_s(0) \ket{  \mc{I}^{(\vsg_s)}_{ n^{(\vsg_s)} } }     \Big\} 
\cdot \pl{s=1}{r-1}\bigg\{ \exp\Big[ \i (x_{s+1}-x_s) \De\mc{P}\big( \mc{I}^{(\vsg_s)}_{ n^{(\vsg_s)} }  \big)  \Big] \bigg\}
\cdot  
\enq

 Above we have introduced the sequence of spin $\vsg_s\,=\,\sum_{p=1}^{s}o_p$ and the summation runs over all choices of $n^{(\vsg_s)} \in \mathbb{N}$
 and sets $\mc{I}^{(\vsg_s)}_{n^{(\vsg_s)}}$, with $s=1,\dots, r-1$. Note that we have adopted the convention
\beq
 \vsg_0\, = \, \vsg_r \, = \, 0 \qquad \e{and} \qquad \mc{I}^{(\vsg_0)}_{n^{(\vsg_0)}}\, = \,  \mc{I}^{(\vsg_r)}_{n^{(\vsg_r)}} 
      \, =  \, \Big\{  \big\{ \emptyset \big\} \; ; \; \big\{ \emptyset \big\}  \Big\} \; . 
\enq

The above sum is a highly oscillatory sum in the large-distance regime $|x_{s+1}-x_s|p_F>>1$. Therefore, it should localise, in this regime,
either around the saddle-points of the relative excitation momentum  $\De\mc{P}\big( \mc{I}^{(s)}_{ n }  \big)$ defined in \eqref{definition relative ex momentum}
or around the boundaries of the summation region. In the present setting, due to the hypothesis \eqref{ecriture proprietes generales de p et veps},
there are no saddle points, so that the leading in $|x_{s+1}-x_s|p_F>>1$ contribution to the sums will issue from a 
vicinity of the boundaries of summation. This corresponds precisely to the case of the states belonging to critical classes. 
Thus, this reasoning leads to 
\beq
\big< \mc{O}_1(x_1)\cdots \mc{O}_{r}(x_r)  \big>_{ \mf{h}_{\e{phys}} } \; \simeq \; 
\sul{ \Big\{ \mc{J}_{ m_{p;\pm }^{(\vsg_{s})}; m_{h;\pm}^{(\vsg_{s})} }^{ (\vsg_{s}) }   \Big\} }{}
\pl{s=1}{r} \mc{F}_{ \mc{O}_s }\bigg(  \mc{J}_{ m_{p;+}^{(\vsg_{s-1})}; m_{h;+}^{(\vsg_{s-1})} }^{ (\vsg_{s-1}) } \cup 
	\mc{J}_{ m_{p;-}^{(\vsg_{s-1})}; m_{h;-}^{(\vsg_{s-1})} }^{ (\vsg_{s-1}) }     ;  
 \mc{J}_{ m_{p;+}^{(\vsg_{s})}; m_{h;+}^{(\vsg_{s})} }^{ (\vsg_{s}) } \cup 
	\mc{J}_{ m_{p;-}^{(\vsg_{s})}; m_{h;-}^{(\vsg_{s})} }^{ (\vsg_{s}) }   \mid x_s \bigg)    
\cdot  
\enq
Note that the $\simeq$ symbol refers to an equality up to sub-leading corrections to each oscillating factor $\ex{2 \i \ell_s (x_{s+1}-x_s) p_F}$ with 
$s=1,\dots,r-1$ and $\ell_s \in \mathbb{Z}$. Hence invoking the identity \eqref{ecriture identification asymptotique des FF}, using that 
\beq
\pl{a=1}{r} (-1)^{\varkappa_s} \; = \; 1 \qquad \e{with}\quad  \varkappa_s \; = \; m_{p;+}^{(\vsg_{s})} \, - \,  m_{h;+}^{(\vsg_{s})} \, - \, n_{p;+}^{(\vsg_{s})} \, + \, n_{h;+}^{(\vsg_{s})} 
\enq
and ignoring the effect 
of the $ \e{O}\big( \tf{ \ln L }{ L } \big)$ corrections that should vanish in the thermodynamic limit, we re-absorb the sums over the intermediate states 
of $\mf{h}_{\e{eff}}$, hence obtaining
\beq
\big< \mc{O}_1(x_1)\cdots \mc{O}_{r}(x_r)  \big>_{ \mf{h}_{\e{phys}} } \; \simeq  \; 
\big< \msc{O}_1(\om_1)\cdots \msc{O}_{r}(\om_r) \big>_{ \mf{h}_{\e{eff}} } \;. 
\enq
Thus the claimed correspondence holds. Above, we have added the subscripts $\mf{h}_{\e{phys}}$ and $\mf{h}_{\e{eff}}$ so as to insist in which Hilbert space the
expectation values are computed.



\subsection{Some examples}

\subsubsection{The XXZ spin-$1/2$ chain}
\label{SousSection Exemple XXZ}

We shall now illustrate the general framework described above on the example of the XXZ spin-1/2 chain embedded in an external magnetic field $h>0$. 
This model corresponds to the Hamiltonian
 \beq
\mc{H}_{XXZ} \; = \; \sum_{k=1}^{L}\left( \sg^x_{k}\sg^x_{k+1}+\sg^y_{k}\sg^y_{k+1} \, + \, \De(\sg^z_{k}\sg^z_{k+1}-1)\right)-\f{h}{2}S^z \qquad \e{with} \quad S^z\; = \; \sum_{k=1}^{L}\sg^z_{k} \; .
 \enq
Above $\sigma^{x,y,z}_{k}$ are the spin operators (Pauli matrices) acting on the $k$-th site
of the chain,  $h$ is an external magnetic field and the model is subject to periodic boundary conditions. 
It is well known that $ \mc{H}_{XXZ}$ posses different phases depending on the value of  the anisotropy parameter  $\Delta$. 
When $-1<\De<1$, $\mc{H}_{XXZ}$ has a massless spectrum in the $L\tend +\infty$ limit. Below, we shall only consider this regime and adopt the parametrisation $\Delta=\cos \zeta$.  
It is known that, in its massless phase at $h>0$, the excitations in the XXZ-chain can be either
of bound state nature (so-called string solutions) or be of particle-hole type. 
The particle-hole spectrum enjoys of the structure described in the earlier part of this section. The bound states, however,  are described differently. 
None-the-less, one can argue that, in the long-distance regime, they only produce corrections that are exponentially small in the distance;
they can thus be disregarded when studying the correspondence with the free boson model in that it solely involve the algebraically decaying terms.

Owing to $\big[ \mc{H}_{XXZ}, S^z\big]=0$, the role of  bare-particles is played by the number of down spins building up a given eigenstate. 
The local operators of the model consist in products of elementary matrices 
\beq
\mc{O}_k(m_k) \; = \; E^{\eps_{k;1}\eps_{k;1}^{\prime}} _{m_k} \cdots E^{\eps_{k;d_k}\eps_{k;d_k}^{\prime}}_{m_k+d_k}
\label{definition operateur O local pour XXZ}
\enq
The length $d_k$ of the string of elementary matrices  is assumed fixed, \textit{viz}. $m_k$ and $L$ independent. There $E^{\eps \eps^{\prime}}_p$ is the elementary matrix $E^{\eps \eps^{\prime}}$
acting non-trivially on the $p^{\e{th}}$ site of the chain. The operator defined by \eqref{definition operateur O local pour XXZ} carries the spin 
\beq
o_k\; = \; \# \Big\{ a \; : \; \big( \eps_{k;a},\eps_{k;a}^{\prime} \big) \; = \; (2,1) \Big\} \; - \; 
\# \Big\{ a \; : \; \big( \eps_{k;a},\eps_{k;a}^{\prime} \big) \; = \; (1,2) \Big\} \;. 
\enq

One can show using Bethe Ansatz methods that the shift functions -- in the sense of \eqref{ecriture definition fonction comptage} -- take the form 
\beq
F_{\mc{R}_n^{(s)}}(\la) \; = \;  s \Big(   \phi(\la,q)  \; - \;  \f{ Z(\la) }{ 2 } \Big)
\; + \;  \sul{a=1}{n} \Big[ \phi\big(\la,\mu_{p_a} \big) \, - \,  \phi\big(\la,\mu_{h_a} \big)   \Big]\, ,
\enq
in which $\phi$ is the dressed phase,   $Z$ the dressed charge and 
the rapidities $\{\mu_a\}_{a\in \mathbb{Z} }$ are the unique solutions to 
\beq
\xi( \mu_a) \; = \; \f{ a }{ L }  \qquad \e{with} \qquad \xi(\om) \; = \; \f{ p(\om) }{2\pi} \; + \; \f{D}{2} \;. 
\enq
The function $p$ arising in the expression of the asymptotic counting function is the so-called dressed momentum and $D= \lim_{ N \tend + \infty}( N/L )$, $N$
being the number of down spins in the model's ground state. The function $p$ corresponds to the unique odd solution to the linear integro-differential  equation:
\beq
p(\la) \; + \; \ \Int{-q}{q} \th(\la-\mu) \cdot p^{\prime}(\mu) \cdot \f{ \dd \mu }{2\pi} \; = \; \i \ln \bigg( \f{ \sinh( \i \zeta/2 + \la) }{ \sinh( \i \zeta/2 - \la)  } \bigg)   \;. 
\enq
The functions $Z$ and $\phi$ solve the Lieb integral equations
\beq
Z(\la) \; + \;  \Int{-q}{q} \th^{\prime}(\la-\mu) Z(\mu) \cdot \f{ \dd \mu }{ 2\pi } \; = \; 1
\qquad 
\phi(\la,\nu) \; +\;  \Int{-q}{q} \th^{\prime}(\la-\mu) \phi(\mu,\nu) \cdot \f{ \dd \mu }{ 2\pi } \; = \; 
\f{ \i }{ 2\pi }   \ln \bigg( \f{ \sinh(i\zeta + \la-\nu) }{ \sinh(i\zeta - \la+\nu)  } \bigg)  \;. 
\enq
The above integral equation depend on the endpoint $q$ of the model's Fermi zone, which is  fixed once that $D$ is given.

Finally, the relative shift function $\nu_{s}$ associated to critical excited states belonging to the 
$\ell_s$ and $\ell_{s-1}$ classes and differing in $o_s$ in their bare-particle numbers, takes the form 
\beq
\nu_s(\la) \; = \; o_s \Big( \f{ Z(\la) }{ 2 } \; - \;  \phi(\la,q)  \Big)
\; + \; (\ell_{s-1} -\ell_s ) \Big( Z(\la) - 1  \Big) \;. 
\enq
One can then show using certain identities satisfied by $Z(q)$ and $\phi(q,q)$ that 
\beq
\nu_s(q)+\kappa-o_s \; = \; \kappa Z(q) \, - \, \f{o_s}{ 2 Z(q) } \qquad \e{and} \qquad \nu_s(-q)+\kappa \; = \; \kappa Z(q) \, - \, \f{o_s}{ 2 Z(q) } 
\enq
We do stress that numerous lattice operators $O_k(m_k)$ will carry the same spin $o_k$ and thus the same scaling dimensions since these are parametrised 
by  $\nu_s(q)+\kappa-o_s $ and $\nu_s(-q)+\kappa$.

Finally, the behaviour of the form factors of the model follows exactly the form provided in Section \ref{Section presentation gle du modele}. We do not
reproduce the corresponding formulae here but refer the interested reader to the aforementioned literature.

\subsubsection{The Luttinger structure recovered}
\label{SousSoussectionLLStructure}

One can reproduce the structure of the Luttinger liquid-like critical exponents by making additional assumptions on certain ingredients of our general framework. 
We first assume that the shift function  $F_{ \mc{R}_n^{(s)} } (\la)$ satisfies  the reflection property:
\beq
F_{ \mc{R}_n^{(s)} } (-\la) \; = \; s - F_{ \check{\mc{R}}_n^{(s)} } (\la) \qquad \e{with} \qquad  \check{\mc{R}}_n^{(s)} \; = \; \Big\{ \{-\mu_{p_a^{(s)}} \}_1^{n} \; ; \;  \{-\mu_{h_a^{(s)}} \}_1^{n} \Big\} \;. 
\enq
This property can be argued, on the heuristic level, as follows. The physics should be invariant in respect to reflections of the Fermi surface. 
When taking the reflected of an excited state belonging to the $N+s$ bare-particle excitation sector above the ground state, one should
reflect all the rapidities of the particles and holes but also take into account that the $s$ additional bare-particles
that are used to construct this excited state which are initially located on the right boundary are now located on the left one. 
This induces the shift by $-s$. These properties can be directly verified on the example of the XXZ chain introduced earlier on. 

Now, we shall make the additional assumption of a linear response, namely assume that if the macroscopic variables $\mc{R}_{n+n^{\prime}}^{(s+s^{\prime})}$
can be partitioned as
\beq
\mc{R}_{n+n^{\prime}}^{(s+s^{\prime})} \; = \; \mc{R}_n^{(s)} \cup \mc{R}_{n^{\prime}}^{(s^{\prime})}
\enq
then one also has
\beq
F_{ \mc{R}_{n+n^{\prime} }^{ (s+s^{\prime}) } } (\la)  \; = \; F_{ \mc{R}_n^{(s)} } (\la) \; + \;  F_{ \mc{R}_{ n^{\prime} }^{ (s^{\prime}) } }(\la) \;. 
\enq
It is then enough to observe that the set of macroscopic rapidities $\wt{\mc{R}}_{\ell}^{(s)}$ of an $\ell$ class, $\ell \geq 0$ can be decomposed as 
\beq
\wt{\mc{R}}_{\ell}^{(s)} \; = \;   \Big\{ \{ q \}_1^{\ell}\,  ; \, \{-q \}_1^{\ell}   \mid s \Big\}  \; = \; 
\bigcup\limits_{1}^{\ell}  \underbrace{ \Big\{ \{ q \}_1^{1}\,  ; \, \{-q \}_1^{1}   \mid 0 \Big\} }_{ = \mc{R}_{\e{umkp}} } 
\bigcup \limits_{1}^{s}  \underbrace{ \Big\{ \{ \emptyset \} \,  ; \, \{ \emptyset \}_1^{\ell}   \mid 1 \Big\} }_{ = \mc{R}_{\e{spn}} }
\enq
so as to get 
\beq
F_{ \wt{\mc{R}}_{\ell}^{(s)}  } (\la) \; = \; \ell F_{ \mc{R}_{\e{umkp}} }(\la) \; + \; s F_{ \mc{R}_{\e{spn}} }(\la)
\enq
Then the shift function arising in our estimates is recast as
\beq
\nu_{r}(\la) \; = \; F_{\wt{R}^{(s)}_{\ell_{\e{out}}} }(\la) \, - \,  F_{ \wt{R}^{(s+o_r)}_{\ell_{\e{in}}} }(\la) \; = \; 
 \kappa F_{ \mc{R}_{\e{umkp}} }(\la) \; - \; o_r F_{ \mc{R}_{\e{spn}} }(\la)
\enq
where we have set $\kappa=\ell_{ \e{out} }  \, - \, \ell_{ \e{in} }$. 
We do stress that all the above properties are verified in quantum integrable models, the XXZ chain in particular. 

Our relations lead to 
\beq
\nu_{r}(-q) \; = \; v \cdot \Big\{  \kappa K - \f{ o_r }{ K }  \Big\} \quad \e{with} \quad v K \; = \; F_{ \mc{R}_{\e{umkp}} }(-q) \quad \e{and} \quad
\f{v}{K} \; = \;  F_{ \mc{R}_{\e{spn}} }(-q) \;. 
\enq
Then, the reflection property implies that $\nu_{r}(q)=o_r-\nu_{r}(-q)$. All-in-all, this reproduces the scaling dimensions of the
Luttinger liquid model.


\section*{Conclusion}

In the present paper we have provided a first principle-based derivation of the emergence, in the large-separation regime between the operators, 
of an effective description of the model's operators in terms of vertex operators associated with the free boson model. 
Our construction naturally allows one to treat as well the case of time dependent correlations and allows one to recover all the 
features of an effective field theoretic description which are at the very base of the non-linear Luttinger liquid model \cite{GlazmanImambekovSchmidtReviewOnNLLuttingerTheory}. 
In this respect, our method, allows one to justify, starting from the first principles,
the use of the non-linear Luttinger liquid model.




\section*{Acknowledgements}

K.K.K., J.M.M.  are supported by CNRS. 
This work has been partly done within the financing of the grant  ANR grant "DIADEMS".
K.K.K. acknowledges support from the 
the Burgundy region PARI 2013 and 2014 FABER grant "Structures et asymptotiques d'int\'{e}grales multiples". 
K.K.K. would like to thank the laboratory of physics in ENS-Lyon for its warm hospitality.




\appendix 

\section{Special functions}
\label{Appendix fct speciales d interet}

In this section we recall the definition and main properties of the special functions that are of use to our analysis. 
The Euler $\Ga$-function is defined as 
\beq
\Ga(z) \; = \; \Int{0}{+\infty } t^{z-1} \ex{-t} \cdot \dd t \qquad \e{and} \; \e{satisfies} \qquad  
\Ga(z+1) \; = \; z \, \Ga(z) \;. 
\enq
The Barnes $G$-function is a generalisation of the $\Ga$ function, in the sense that it satisfies the functional relation $G(z+1)=\Ga(z) G(z)$. 
It admits the integral representation 
\beq
G(1+z) \; = \; \Big( \sqrt{2\pi} \cdot \Ga(z) \Big)^{z} \cdot  \exp\Bigg\{ \f{z(1-z)}{2} \; - \; \Int{0}{z} \ln \Ga(s) \cdot \dd s   \Bigg\}
\enq
from which one can deduce the reflection formula
\beq
\f{ G(1-z) }{ G(1+z) } \; = \; (2\pi)^{-z} \cdot \exp\Bigg\{ \Int{0}{z} \pi s \cot(\pi s) \cdot \dd s \Bigg\} \;. 
\enq
The latter readily implies the relation 
\beq
G\pab{1+ z , 1-z-\ell }{ 1-z, 1+z+\ell } \; = \; \Big( \f{\sin [\pi z] }{ \pi } \Big)^{\ell} \cdot (-1)^{ \frac{\ell(\ell+1)}{2} } \;. 
\label{Formule de la double redcution pour la fonction de Barnes}
\enq

Above we have introduced the so-called hypergeometric like notation which will be used for products of ratios of Euler $\Ga$ or Barnes $G$-function, \textit{e}.\textit{g}. 
\beq
\Ga \Bigg(  \ba{c} \{v_a\}_1^n \\ \{w_a\}_1^m \ea \bigg) \; = \; 
\Ga \Bigg(  \ba{c} v_1, \dots , v_n  \\ w_1, \dots , w_m \ea \bigg) \; = \;  \f{\pl{a=1}{n} \Ga(v_a) }{ \pl{a=1}{m} \Ga(w_a) }
\qquad \e{and} \qquad 
G \Bigg(  \ba{c} \{v_a\}_1^n \\ \{w_a\}_1^m \ea \bigg)  \; = \;  \f{\pl{a=1}{n} G(v_a) }{ \pl{a=1}{m} G(w_a) }\;. 
\label{introduction notation produit compact gamma functions}
\enq
%
%
%




\section{Some integrals of interest}
\label{Appendix Integrales pertinentes}

In this appendix, we compute some integrals that are of direct application to our study. 
Namely, given $p,h,t \in \mathbb{N}^*$, define 
\beq
\mc{I}_{ht}^{(1)}( \nu \mid \om ) \; = \; 
\Oint{ |\zeta| >|\om| > | \tau | }{} \hspace{-4mm} \f{\dd \zeta  \dd \tau }{ (2\i \pi)^2}   
\f{ \tau^{-t} \zeta^{h-1} }{\zeta -  \tau} \f{ \big( 1- \tf{\om}{\zeta} \big)^{\nu}  }{  \big( 1- \tf{\tau}{\om} \big)^{\nu} }
\enq
and
\beq
\mc{I}_{hp}^{(2)}( \nu \mid \om ) \; = \; 
\Oint{ |\om| > |\zeta| > | \tau | }{} \hspace{-4mm} \f{\dd \zeta  \dd \tau }{ (2\i \pi)^2}   
\f{ \tau^{-p} \zeta^{-h} }{\zeta -  \tau} \f{  \big( 1- \tf{\tau}{\om} \big)^{\nu} }{ \big( 1- \tf{\zeta}{\om} \big)^{\nu}  } \;. 
\enq
Two other integrals appear in the course of the analysis, namely 
\beq
\wt{\mc{I}}_{pk}^{(1)}( \nu \mid \om ) \; = \; 
\Oint{ |\zeta| >|\om| > | \tau | }{} \hspace{-4mm} \f{\dd \zeta  \dd \tau }{ (2\i \pi)^2}   
\f{ \tau^{-k} \zeta^{p-1} }{\zeta -  \tau} \f{  \big( 1- \tf{\tau}{\om} \big)^{\nu} }{ \big( 1- \tf{\om}{\zeta} \big)^{\nu}  }
\quad \e{and} \quad 
\wt{\mc{I}}_{ph}^{(2)}( \nu \mid \om ) \; = \; 
\Oint{ | \tau |  > |\zeta| >|\om| }{} \hspace{-4mm} \f{\dd \zeta  \dd \tau }{ (2\i \pi)^2}   
\f{ \tau^{p-1} \zeta^{h-1} }{ \tau- \zeta} \f{  \big( 1- \tf{\om}{\zeta} \big)^{\nu} }{ \big( 1- \tf{\om}{\tau} \big)^{\nu}  } \;. 
\enq
These two new integrals are however related to the first two introduced above. 
Indeed, one has
\beq
 \wt{\mc{I}}_{ht}^{(1)}( \nu \mid \om ) \; = \; \mc{I}_{ht}^{(1)}( -\nu \mid \om ) \qquad \e{and} \qquad
\wt{\mc{I}}_{ph}^{(2)}( \nu \mid \om ) \; = \;  \om^{2(p+h-1)} \mc{I}_{hp}^{(2)}( -\nu \mid \om ) \;. 
\enq
The first equality is straightforward whereas the second follows from the change of variables $(\tau,\zeta)\mapsto (\om^2/x, \om^2/y)$.

\begin{lemme}
The double integrals  $\mc{I}_{ht}^{(1)}( \nu \mid \om )$ and  $\mc{I}_{hp}^{(2)}( \nu \mid \om )$ can be explicitly computed as
\beq
 \mc{I}_{ht}^{(1)}( \nu \mid \om )  \; =  \;  \f{  \sin [\pi \nu] \cdot \om^{h-t} }{ \pi ( t-h+\nu ) } \cdot 
	    \Ga\bigg(\ba{ccc} h-\nu \, , &  t+\nu \\
			      h \;\;\;\; \; ,  & t \ea \bigg)  
\label{formule pour I1ht}
\enq
and
\beq
\mc{I}_{hp}^{(2)}( \nu \mid \om ) \;  =  \;  \f{  \sin [\pi \nu] \cdot \om^{1-h-p} }{ \pi ( h+p-1 ) } \cdot 
	    \Ga\bigg(\ba{ccc} h+\nu \, , &  p-\nu \\
			      h  \; \;\;\; \; ,  & p \ea \bigg)  \;. 
\label{formule pour I2hp}
\enq

\end{lemme}

\Proof 

The calculation can be done by means of the method proposed in Appendix A of \cite{AlexandrovZaborodinTechniquesOfFreeFermions}
which allows for an effective separation of the integrals. First of all, one observes that 
\beq
\big(\zeta \Dp{ \zeta } + \tau \Dp{\tau}  \big) \bigg\{ \f{\zeta}{\zeta-\tau} \cdot \zeta^{\nu} \cdot \f{ (1-\zeta^{-1})^{\nu} }{ (1-\tau)^{\nu} } \bigg\} \; = \; 
 \nu \zeta^{\nu} \cdot \f{ (1-\zeta^{-1})^{\nu-1} }{ (1-\tau)^{\nu+1} } \;. 
\enq
Then, upon a rescaling of the integration variables, one gets 
\beq
\mc{I}_{ht}^{(1)}( \nu \mid \om ) \; = \; \f{ \om^{h-t} }{ h-t-\nu}
\Oint{ |\zeta| >1 > | \tau | }{} \hspace{-4mm} \f{\dd \zeta }{ 2\i \pi \zeta}\f{\dd \tau }{ 2\i \pi \tau} 
\bigg\{ \f{\zeta}{\zeta-\tau} \cdot \zeta^{\nu} \cdot \f{ (1-\zeta^{-1})^{\nu} }{ (1-\tau)^{\nu} } \bigg\}  
\, \times \, \big(\zeta \Dp{ \zeta } + \tau \Dp{\tau}  \big) \cdot \big\{ \tau^{1-t} \zeta^{h-1-\nu} \big\} \;. 
\enq
An integration by parts then leads to 
\bem
\mc{I}_{ht}^{(1)}( \nu \mid \om ) \; = \; \f{ - \om^{h-t} }{ h-t-\nu}
\Oint{ |\zeta| >|\om| > | \tau | }{} \hspace{-4mm} \f{\dd \zeta  \dd \tau }{ (2\i \pi)^2}   
  \tau^{-t} \zeta^{h-2}   \f{ (1-\zeta^{-1})^{\nu-1} }{ (1-\tau)^{\nu+1} } \\
\;= \; \f{ - \om^{h-t} }{ h-t-\nu} \Oint{ |\zeta|<1 }{}  \zeta^{-h} (1-\zeta)^{\nu-1} \cdot \f{\dd \zeta }{ 2\i \pi }  
\times \Oint{ |\tau|<1 }{}  \tau^{-t} (1-\tau)^{-\nu-1}  \cdot \f{\dd \tau }{ 2\i \pi } \;. 
\end{multline}
The two integrals can then be evaluated by means of the series expansion
\beq
(1-z)^{\a} \; = \; \f{\sin [\pi \a] }{ \pi } \cdot \sul{ n \geq 0}{} \; z^n  	\cdot     \Ga\bigg(\ba{ccc} \a+1 \, , &  n-\a \\
			      1 \;\;\;\; \; ,  & n+1 \ea \bigg) 
\enq
hence leading to \eqref{formule pour I1ht}.

The integral $\mc{I}_{ph}^{(2)}( \nu \mid \om )$ is computed along the same lines on the basis of the identity 
\beq
\big(\zeta \Dp{ \zeta } + \tau \Dp{\tau}  \big) \bigg\{ \f{\tau}{\zeta-\tau} \cdot \f{ (1-\tau)^{\nu} }{ (1-\zeta)^{\nu} } \bigg\} \; = \; 
 \nu \tau \cdot \f{ (1-\tau)^{\nu-1} }{ (1-\zeta)^{\nu+1} } \;. 
\enq
The details are left to the reader. \qed

\section{Proof of Proposition \ref{Proposition expectation value J-J+ Ops}}
\label{Appendix Preuve proposition valeur expectation ops vertex non shifte}

The exponents of current operators $\ex{ \msc{J}_{\pm}(\nu,\om) }$ preserve the charge of a state. Therefore, the expectation value
$\bra{ \mc{J}_{ n_{p}; n_{h} } }  \ex{ \msc{J}_-(\nu,\om) } \ex{  \msc{J}_+(\nu,\om)  } \ket{  \mc{J}_{ n_{k}; n_{t} } } $ is non-zero only if 
\beq
n_p-n_h=n_k-n_t\;. 
\enq
This explains the occurrence of the Kronecker symbol in \eqref{ecriture representation VO pour la valeur moyenne a shift r nul}. 

Observe that the role played by $\psi_k$ and $\psi_k^{*}$ in the whole construction of the Hilbert space $\mf{h}$ is symmetric. In particular, 
this ensures that the result will be invariant under the transformation $(\psi_k,\psi_k^{*} ) \mapsto (\psi_{-1-k}^{*},\psi_{-1-k} ) $. 
We stress that this transformation leaves the vacuum $\ket{0}$ unchanged. Since this transform maps 
the local current operators as $J_k \mapsto -J_k$, we get the symmetry 
\beq
\bra{ \mc{J}_{ n_{p}; n_{h} } }  \ex{ \msc{J}_-(\nu,\om) } \ex{  \msc{J}_+(\nu,\om)  } \ket{  \mc{J}_{ n_{k}; n_{t} } } 
\; = \; (-1)^{n_t+n_h}\bra{ \mc{J}_{ n_{h}; n_{p} } }  \ex{ - \msc{J}_-(\nu,\om) } \ex{  - \msc{J}_+(\nu,\om)  } \ket{  \mc{J}_{ n_{t}; n_{k} } } \;. 
\enq
where the sign prefactor issues from the change in the order of products of the $\psi$'s and $\psi^{*}$'s. 
It is easy to check that indeed, as soon as the condition  $n_p-n_h=n_k-n_t$ is imposed, the \textit{rhs} of
\eqref{definition microscopic form factor} is indeed invariant under the simultaneous transformations
\beq
 \Big( \mc{J}_{ n_{p}; n_{h} } ,  \mc{J}_{ n_{k}; n_{t} }, \nu \Big) \quad \mapsto  \quad    \Big( \mc{J}_{ n_{h}; n_{p} } ,  \mc{J}_{ n_{t}; n_{k} }, -\nu \Big) 
\enq
followed by a multiplication by $(-1)^{n_t+n_h}$. 
The above thus ensures that it is enough to establish the validity of the representation \eqref{definition microscopic form factor}
solely in the case when $n_p-n_h=n_k-n_t\geq 0$. Hence, from now on, we focus on this case only and, so as to lighten the notation, 
we introduce the shorthand notations
\beq
n_h=n \quad n_p=n+\ell \qquad n_t=d \quad \e{and} \quad n_{k}=d+\ell \;. 
\enq

In order to compute the expectation value, we first reduce the problem to a simpler one
where the application of Wick's theorem is straightforward. This boils down the problem to the computation of determinants 
whose entries are given by bi- and quadri-linear expectation values in the fermions. 
The expectation values can be expressed in therms of the integrals computed in Appendix \ref{Appendix Integrales pertinentes}. 
Building on these results, we compute the various determinants by using their relation to Cauchy determinants. 

\subsubsection*{$\bullet$ Reduction to a simpler problem}

Observe that, for any $r \geq 0$, one has  
\beq
\bra{0} \, \psi_{-r} \, \psi^{*}_{-r} \; = \; \bra{0} \;.
\label{ecriture propriete action sur bra0 de op nb parts}
\enq
Let $r_1,\dots, r_{\ell}$ be distinct integers satisfying to the constraints
\beq
r_b \; \geq  \; \max\{h_a\; : \; a=1,\dots n \}  \qquad \e{for} \; \e{any} \quad  b=1,\dots, \ell\;. 
\enq
Then, in virtue of \eqref{ecriture propriete action sur bra0 de op nb parts}, we get
\bem
\hspace{-5mm} \bra{ \mc{J}_{ n_{p}; n_{h} } } \ex{ \msc{J}_-(\nu,\om)  } \ex{ \msc{J}_+(\nu,\om)  } \ket{  \mc{J}_{ n_{k}; n_{t} } } \; = \; 
\bra{0} \pl{a=1}{\ell}  \Big\{ \psi_{-r_{a} } \psi^{*}_{-r_{a} } \Big\} \cdot 
\psi_{p_1-1}^* \cdots \psi_{p_{n+\ell}-1}^*\cdot \psi_{-h_n} \cdots \psi_{-h_1} \ex{ \msc{J}_-(\nu,\om)  } \ex{ \msc{J}_+(\nu,\om)  } \ket{  \mc{J}_{ n_{k}; n_{t} } } 
 \\
\; = \; (-1)^{\ell n} \bra{0} \psi_{p_1-1}^* \cdots \psi_{p_{n+\ell}-1}^*\cdot \psi_{-r_{\ell} } \cdots \psi_{-r_1} \cdot 
\psi_{-h_n} \cdots \psi_{-h_1} \psi^{*}_{-r_{1} } \cdots \psi^{*}_{-r_{\ell} } \ex{ \msc{J}_-(\nu,\om)  } \ex{ \msc{J}_+(\nu,\om)  } \ket{  \mc{J}_{ n_{k}; n_{t} } } \;. 
\end{multline}
We now move the operators $\psi_{-r_a}^*$ through the exponents of current operators. For this purpose, we first 
represent $ \psi^{*}_{-r }$ in terms of the conjugate field $\Psi^*(\tau)$ by means of a contour integral around $0$ and then we apply
the exchange relation \eqref{ecriture relations echange vars Miwa}:
\bem
 \psi^{*}_{-r_{1} } \cdots \psi^{*}_{-r_{\ell} } \cdot \ex{ \msc{J}_-(\nu,\om)  } \ex{ \msc{J}_+(\nu,\om)  } \; = \; 
\Oint{}{} \f{ \dd^{\ell} \tau }{ (2\i \pi)^{\ell} } \pl{a=1}{\ell} \Big\{ \tau_a^{-r_a-1} \Big\} \cdot 
\Psi^*(\tau_1) \cdots \Psi^{*}(\tau_{\ell}) \cdot \ex{ \msc{J}_-(\nu,\om)  } \ex{ \msc{J}_+(\nu,\om)  }  \\
\; = \;  \ex{ \msc{J}_-(\nu,\om)  } \ex{ \msc{J}_+(\nu,\om)  }  \cdot \Phi_{-r_{1};\nu}^{*} \cdots  \Phi_{-r_{\ell};\nu}^{*} 
\qquad \e{with} \qquad
\Phi_{-r;\nu}^{*} \; = \; \Oint{}{} \tau^{-r-1} \f{ (1-\tf{\tau}{\om})^{\nu}  }{ (1-\tf{\om}{\tau})^{\nu} } \Psi^*(\tau) \cdot  \f{ \dd \tau }{ 2\i \pi } \;. 
\end{multline}
Thus, upon extending the sequence $h_a$ as $h_{n+b}=r_b$ for $b=1,\dots,  \ell$ and after setting 
\beq
\mf{w}_a^*\; = \; \psi_{-t_a}^* \quad a=1,\dots, d  \quad  \mf{w}_{d+a}^*\; = \; \Phi_{-r_{a};\nu}^{*} \quad a=1,\dots, \ell \qquad \e{and} \qquad
\mf{v}_a\; = \; \psi_{k_a-1} \quad a=1,\dots, d + \ell \; ,
\enq
the expectation value we started with reduces to 
\bem
\bra{ \mc{J}_{ n_{p}; n_{h} } } \ex{ \msc{J}_-(\nu,\om)  } \ex{ \msc{J}_+(\nu,\om)  } \ket{  \mc{J}_{ n_{k}; n_{t} } } \; = \; (-1)^{\ell(n+d)}
\bra{ \mc{J}_{ n+\ell; n + \ell } } \ex{ \msc{J}_-(\nu,\om)  } \ex{ \msc{J}_+(\nu,\om)  }
\mf{w}^*_1\cdots \mf{w}^*_{d+\ell} \cdot \mf{v}_{d+\ell}\cdots \mf{v}_1  \ket{0}  \\
\; = \;  (-1)^{\ell(n+d)} \bra{ \mc{J}_{ n+\ell; n + \ell } } \ex{ \msc{J}_-(\nu,\om)  } \ex{ \msc{J}_+(\nu,\om)  } \ket{0} \cdot \det_{d+\ell}\big[ M_{ab} \big] \;. 
\label{ecriture reduction vers det M du facteur de forme J-J+}
\end{multline}
The set $ \mc{J}_{ n+\ell; n + \ell } $ appearing above is defined in terms of the extended sequence $\{h_a\}_1^{n+\ell}$ as
\beq
\mc{J}_{ n+\ell; n + \ell } \; = \; \Big\{ \{p_a\}_1^{n+\ell} \; ; \;   \{h_a\}_1^{n+\ell} \Big\} \;. 
\enq
We also specify that we have obtained the second line in \eqref{ecriture reduction vers det M du facteur de forme J-J+} by applying Wick's theorem. 
Finally, the matrix $M$ arising in  \eqref{ecriture reduction vers det M du facteur de forme J-J+} reads
\beq
M_{ab} \; = \;   \f{ \bra{ \mc{J}_{ n+\ell; n + \ell } } \ex{ \msc{J}_-(\nu,\om)  } \ex{ \msc{J}_+(\nu,\om)  } \psi^*_{-t_a} \psi_{k_b-1} \ket{0} }
{   \bra{ \mc{J}_{ n+\ell; n + \ell } } \ex{ \msc{J}_-(\nu,\om)  } \ex{ \msc{J}_+(\nu,\om)  } \ket{0}   }
\enq
for $a=1,\dots,d$ and 
\beq
M_{a+d\, b} \; = \; \Oint{}{} \tau^{-r_a-1} \f{ (1-\tf{\tau}{\om})^{\nu}  }{ (1-\tf{\om}{\tau})^{\nu} }  
\cdot 
\f{ \bra{ \mc{J}_{ n+\ell; n + \ell } } \ex{ \msc{J}_-(\nu,\om)  } \ex{ \msc{J}_+(\nu,\om)  } \Psi^*(\tau) \psi_{k_b-1} \ket{0}  }
{  \bra{ \mc{J}_{ n+\ell; n + \ell } } \ex{ \msc{J}_-(\nu,\om)  } \ex{ \msc{J}_+(\nu,\om)  } \ket{0}  }
\cdot  \f{ \dd \tau }{ 2\i \pi } 
\enq
for $a=1,\dots, \ell$.

\subsubsection*{ $\bullet$ Evaluation of $ \bra{ \mc{J}_{ n+\ell; n + \ell } } \ex{ \msc{J}_-(\nu,\om)  } \ex{ \msc{J}_+(\nu,\om)  } \ket{0} $  }

The expectation value $ \bra{ \mc{J}_{ n+\ell; n + \ell } } \ex{ \msc{J}_-(\nu,\om)  } \ex{ \msc{J}_+(\nu,\om)  } \ket{0} $ can be computed 
by applying, again, Wick's theorem:
\beq
 \bra{ \mc{J}_{ n+\ell; n + \ell } } \ex{ \msc{J}_-(\nu,\om)  } \ex{ \msc{J}_+(\nu,\om)  } \ket{0}   \; = \; 
\det_{n+\ell}\Big[ \bra{0 } \psi_{p_a-1}^* \psi_{-h_b}\ex{ \msc{J}_-(\nu,\om)  }  \ket{0} \Big]
\enq
where we have used that $ \ex{ \msc{J}_+(\nu,\om)  } \ket{0} \, = \, \ket{0}$  and $\bra{0}\ex{ \msc{J}_-(\nu,\om)  } \, = \, \bra{0}$. 
The entries of the matrix arising in the determinant coincide with one of the integrals studied in Appendix \ref{Appendix Integrales pertinentes}. 
Indeed, one has 
\bem
\bra{0 } \psi_{p-1}^* \psi_{-h} \cdot \ex{ \msc{J}_-(\nu,\om)  } \ket{0}  \; = \; 
\Oint{|\tau|>|\zeta|}{}  \tau^{p-2} \zeta^{h-1}  \bra{0} \Psi^{*}(\tau) \Psi(\zeta) \cdot \ex{ \msc{J}_-(\nu,\om)  } \ket{0} \cdot \f{\dd \tau \dd \zeta}{ (2\i\pi)^2} \\
\; = \; \Oint{|\tau|>|\zeta|> |\om| }{} \hspace{-3mm}  \f{ \tau^{p-1} \zeta^{h-1} }{ \tau- \zeta } \cdot
\f{ \big(1-\tf{\om}{\zeta}\big)^{\nu} }{ \big(1-\tf{\om}{\tau}\big)^{\nu} } \cdot \f{\dd \tau \dd \zeta}{ (2\i\pi)^2} 
\; = \; \wt{\mc{I}}^{(2)}_{ph}(\nu \mid \om) \; . 
\label{calcul facteur de form a deux points}
\end{multline}
In the intermediate calculations, we have used the exchange relation \eqref{ecriture relations echange vars Miwa} along with 
\beq
\bra{0} \Psi^{*}(\tau) \Psi(\zeta)  \ket{0} \; = \;  \f{ \tau }{ \tau - \zeta} \qquad \e{for} \qquad |\tau| \; > \;  |\zeta| \;. 
\enq
Hence, we arrive to the determinant representation
\beq
 \bra{ \mc{J}_{ n+\ell; n + \ell } } \ex{ \msc{J}_-(\nu,\om)  } \ex{ \msc{J}_+(\nu,\om)  } \ket{0}   \; = \; 
\det_{n+\ell}\Big[ \wt{\mc{I}}^{(2)}_{p_a\, h_b}(\nu \mid \om)  \Big] \;. 
 \label{ecriture reprentation det FF J ell ell J-J+ gs}
\enq

\subsubsection*{ $\bullet$ Evaluation of $ \bra{ \mc{J}_{ n+\ell; n + \ell } } \ex{ \msc{J}_-(\nu,\om)  } \ex{ \msc{J}_+(\nu,\om)  } \psi^*_{-t} \psi_{k-1} \ket{0} $  }

Since the operator $\ex{ \msc{J}_-(\nu,\om)  } \ex{ \msc{J}_+(\nu,\om)  } \psi^*_{-t} \psi_{k-1}$ is a group-like element, Wick's theorem ensures that 
\bem
\bra{ \mc{J}_{ n+\ell; n + \ell } } \ex{ \msc{J}_-(\nu,\om)  } \ex{ \msc{J}_+(\nu,\om)  } \psi^*_{-t} \, \psi_{k-1} \ket{0}  \\
 \; = \; \bra{ 0 } \ex{ \msc{J}_+(\nu,\om)  } \psi^*_{-t}  \, \psi_{k-1} \ket{0} \cdot 
\det_{n+\ell}\Bigg[  \f{ \bra{0 } \psi_{p_a-1}^*  \, \psi_{-h_b}\ex{ \msc{J}_-(\nu,\om)  }   \ex{ \msc{J}_+(\nu,\om)  } \psi^*_{-t}  \, \psi_{k-1}\ket{0} }
{  \bra{ 0 } \ex{ \msc{J}_+(\nu,\om)  } \psi^*_{-t}  \, \psi_{k-1} \ket{0}   }\Bigg] \;. 
\end{multline}
The two-fermion expectation value can be computed analogously to \eqref{calcul facteur de form a deux points}
\beq
\bra{ 0 } \ex{ \msc{J}_+(\nu,\om)  } \psi^*_{-t}  \, \psi_{k-1} \ket{0}  \; = \; 
\Oint{ |\zeta| > |\tau| > |\om| }{}  \hspace{-3mm} \f{ \tau^{-k} \, \zeta^{-t} }{ \zeta - \tau } \cdot \f{ \big(1-\tf{\tau}{\om}\big)^{\nu} }{ \big(1-\tf{\zeta}{\om}\big)^{\nu} } 
\cdot \f{\dd \tau \dd \zeta}{ (2\i\pi)^2} \; = \; \mc{I}^{(2)}_{tk}(\nu \mid \om)  \;. 
\enq
The four point function is recast as
\bem
 \bra{0 } \psi_{p-1}^*  \, \psi_{-h}\ex{ \msc{J}_-(\nu,\om)  }   \ex{ \msc{J}_+(\nu,\om)  } \psi^*_{-t}  \, \psi_{k-1}\ket{0}  \\ 
 \; = \;   \hspace{-4mm}
 \Oint{  |z|>|\zeta|>|\om|>|\tau|>|\xi| }{} \hspace{-6mm}
\f{ z^{p-2} \zeta^{h-1} }{\tau^{t+1}\xi^{k} }\cdot   
\f{ \big(1-\tf{\om}{\zeta}\big)^{\nu}  \big(1-\tf{\xi}{\om}\big)^{\nu} }{  \big(1-\tf{\om}{z}\big)^{\nu} \big(1-\tf{\tau}{\om}\big)^{\nu} }
\cdot \bra{0} \Psi^{*}(z) \Psi(\zeta)  \Psi^{*}(\tau)  \Psi(\xi)  \ket{0} \cdot
\f{\dd z \dd \zeta \dd \tau \dd \xi}{ (2\i\pi)^4 } \;. 
\label{ecriture rep. Int. pour fct a quatre pts}
\end{multline}
A straightforward computation shows that 
\beq
\bra{0} \Psi^{*}(z) \Psi(\zeta)  \Psi^{*}(\tau) \Psi(\xi)  \ket{0} \; = \; 
\f{ \tau \cdot  z }{ (\zeta - \tau) \cdot (z-\xi) } \; + \; \f{ \tau \cdot  z }{ ( \tau- \xi) \cdot (z-\zeta) } 
\enq
provided that the variables satisfy to the ordering $|z|\, > |\zeta| \, > \,  |\tau| \, > \,  |\xi|$. 
After inserting the above expectation value in \eqref{ecriture rep. Int. pour fct a quatre pts}, we are led to 
\beq
 \bra{0 } \psi_{p-1}^*  \, \psi_{-h}\ex{ \msc{J}_-(\nu,\om)  }   \ex{ \msc{J}_+(\nu,\om)  } \psi^*_{-t}  \, \psi_{k-1}\ket{0}  \\ 
 \; = \; \mc{I}^{(1)}_{ht}(\nu\mid \om) \cdot \wt{\mc{I}}^{(1)}_{pk}(\nu\mid \om)  \; + \; 
  \mc{I}^{(2)}_{tk}(\nu\mid \om) \cdot \wt{\mc{I}}^{(2)}_{ph}(\nu\mid \om) \;. 
\enq
All this allows us to represent the expectation value of interest as 
\beq
\bra{ \mc{J}_{ n+\ell; n + \ell } } \ex{ \msc{J}_-(\nu,\om)  } \ex{ \msc{J}_+(\nu,\om)  } \psi^*_{-t} \, \psi_{k-1} \ket{0} \; = \;
 \mc{I}^{(2)}_{tk}(\nu\mid \om) \cdot 
\det_{n+\ell} \Bigg[ \wt{\mc{I}}^{(2)}_{p_ah_b}(\nu\mid \om) \; + \; 
\f{ \mc{I}^{(1)}_{h_bt}(\nu\mid \om) \cdot \wt{\mc{I}}^{(1)}_{p_a k}(\nu\mid \om) }{  \mc{I}^{(2)}_{tk}(\nu\mid \om)  }   \Bigg] \;. 
\enq
This representation involves the determinant of the matrix $\mc{N}_{ab}\,=\,\wt{\mc{I}}^{(2)}_{p_ah_b}(\nu\mid \om)$
perturbed by a rank $1$ matrix. Since the matrix $\mc{N}$ is of Cauchy type, its inverse is explicitly computable and takes the form
\beq
\Big( \mc{N}^{-1} \Big)_{ab} \; = \; - \f{ \pi \om^{1-p_b-h_a} }{ \sin[\pi \nu] } \cdot 
	  \Ga\bigg( \ba{ccc} p_b \;\;\;\;\;  , & h_a \\  
			    p_b +\nu  \; , & h_a - \nu   \ea \bigg) \cdot \f{1}{h_a+p_b-1} \cdot 
  \f{\pl{c=1}{n+\ell} \Big\{ (h_c+p_b-1) \cdot (h_a+p_c-1) \Big\}   }
		  {  \prod_{ \substack{c=1 \\ \not=b} }^{n+\ell} (p_c-p_b) \cdot  \prod_{ \substack{c=1 \\ \not=a} }^{n+\ell} (h_c-h_a)} \; . 
\enq
Thus, recalling the representation \eqref{ecriture reprentation det FF J ell ell J-J+ gs} we are led to 
\beq
\bra{ \mc{J}_{ n+\ell; n + \ell } } \ex{ \msc{J}_-(\nu,\om)  } \ex{ \msc{J}_+(\nu,\om)  } \psi^*_{-t} \, \psi_{k-1} \ket{0} \; = \;
 \mc{I}^{(2)}_{tk}(\nu\mid \om)\cdot \bra{ \mc{J}_{ n+\ell; n + \ell } } \ex{ \msc{J}_-(\nu,\om)  } \ex{ \msc{J}_+(\nu,\om)  }  \ket{0} 
\cdot \mc{S}_{t,k} \Big( \mc{J}_{ n+\ell; n + \ell } \Big) \;. 
\enq
The set function $ \mc{S}_{t,k} \Big( \mc{J}_{ n+\ell; n + \ell } \Big) $ takes into account the contribution issuing from the rank $1$ perturbation of 
$\mc{N}$ and takes the form:
\bem
\mc{S}_{t,k} \Big( \mc{J}_{ n+\ell; n + \ell } \Big) \; = \; 1\, + \, 
\sul{a,b=1}{n+\ell} \f{ \mc{I}^{(1)}_{h_bt}(\nu\mid \om) \cdot \big(\mc{N}^{-1} \big)_{ba} \cdot \wt{\mc{I}}^{(1)}_{p_a k}(\nu\mid \om) }{  \mc{I}^{(2)}_{tk}(\nu\mid \om)  }  \\
\; = \; 1 \; + \;  \sul{a,b=1}{n+\ell}   \f{t+k-1}{(h_a+p_b-1)(t-h_b+\nu)(k-p_a-\nu)} \cdot 
  \f{\pl{c=1}{n+\ell} \Big\{ (h_c+p_a-1) \cdot (h_b+p_c-1) \Big\}   }
		  {  \prod_{ \substack{c=1 \\ \not=a} }^{n+\ell} (p_c-p_a) \cdot  \prod_{ \substack{c=1 \\ \not=b} }^{n+\ell} (h_c-h_b)} \\
\; = \; 1 \; + \;  \Oint{ \Ga\big(\{p_a\}\big) }{} \hspace{-3mm} \f{ \dd s }{2\i \pi} \hspace{-2mm} \Oint{ \Ga\big(\{h_a\}\big) }{}\hspace{-3mm} \f{ \dd z }{2\i \pi}  
\f{t+k-1}{(s+z-1)(t-z+\nu)(k-s-\nu)} \cdot \pl{c=1}{n+\ell}   \bigg\{ \f{  (h_c+s-1) \cdot (z+p_c-1)    }{  (p_c-s) \cdot (h_c-z) } \bigg\} \;. 
\end{multline}
Above, $\Ga\big(\{p_a\}\big)$, resp. $\Ga\big(\{h_a\}\big)$, stands for a small counterclockwise loop around $p_1,\dots, p_{n+\ell}$, resp. $h_1,\dots, h_{n+\ell}$,  
that avoids all other singularities of the integrand. Note that the contour integral should be understood as an encased integral. 
As a consequence, the $z$-integrand does \textit{not} contain poles at $z=1-s$. The integral can be evaluated by taking the residues located outside 
of the contours of integration:
\bem
\mc{S}_{t,k} \Big( \mc{J}_{ n+\ell; n + \ell } \Big) \; = \; 1 \; + \hspace{-2mm}  \Oint{ \Ga\big(\{p_a\}\big) }{} \hspace{-3mm} \f{ \dd s }{2\i \pi} 
\f{t+k-1}{(s+t+\nu-1)(k-s-\nu)} \cdot \pl{c=1}{n+\ell} \bigg\{  \f{  (h_c+s-1) \cdot (t+ \nu + p_c-1)    }{  (p_c-s) \cdot (h_c-t-\nu) } \bigg\} \\
\; =\; \pl{c=1}{n+\ell}  \bigg\{ \f{  (h_c + k- \nu - 1 ) \cdot (t + \nu + p_c-1)    }{  ( p_c - k + \nu ) \cdot ( h_c - t - \nu ) } \bigg\} \;. 
\end{multline}
All-in-all, we are thus led to the representation
\beq
\f{ \bra{ \mc{J}_{ n+\ell; n + \ell } } \ex{ \msc{J}_-(\nu,\om)  } \ex{ \msc{J}_+(\nu,\om)  } \psi^*_{-t} \, \psi_{k-1} \ket{0} }
{  \bra{ \mc{J}_{ n+\ell; n + \ell } } \ex{ \msc{J}_-(\nu,\om)  } \ex{ \msc{J}_+(\nu,\om)  }  \ket{0}  }
\; = \;
 \mc{I}^{(2)}_{tk}(\nu\mid \om)\cdot \pl{c=1}{n+\ell}  \bigg\{ \f{  (h_c + k- \nu - 1 ) \cdot (t + \nu + p_c-1)    }{  ( p_c - k + \nu ) \cdot ( h_c - t - \nu ) } \bigg\}  \;. 
\enq

\subsubsection*{ $\bullet$ Evaluation of $ M_{a+d\, b} $ for $a=1,\dots, \ell$  }

It follows from Wick's theorem that one has the representation 
\bem
\bra{ \mc{J}_{ n+\ell; n + \ell } } \ex{ \msc{J}_-(\nu,\om)  } \ex{ \msc{J}_+(\nu,\om)  }  \ket{0} \cdot M_{a+d\, b}  \\ 
\; = \; \Oint{}{} \f{ (1-\tf{\tau}{\om})^{\nu}  }{ (1-\tf{\om}{\tau})^{\nu} }  \f{ g_{k_b}^{(2)}(\tau) }{ \tau^{r_a+1}  } 
\cdot \det_{n+\ell}\Bigg[ \f{ \bra{ 0 } \, \psi^{*}_{p_q-1}\, \psi_{-h_m} \ex{ \msc{J}_-(\nu,\om)  } \ex{ \msc{J}_+(\nu,\om)  } \Psi^*(\tau) \, \psi_{k_b-1} \ket{0}  }
{   g_{k_b}^{(2)}(\tau)   } \Bigg] \cdot  \f{ \dd \tau }{ 2\i \pi } 
\label{equation rep int pour M a+d b}
\end{multline}
where $q,m$ label the entries of the matrix whose determinant is to be evaluated while the function $g_k^{(2)}(\tau)$ is defined by means of a contour integral:
\beq
g_k^{(2)}(\tau) \; =\; \bra{0} \ex{ \msc{J}_+(\nu,\om)  } \Psi^*(\tau) \, \psi_{k_b-1} \ket{0}  \; = \; 
\Oint{ |\om| > |\tau|> |\zeta|  }{} \hspace{-3mm} \f{\tau \cdot \zeta^{-k} }{ \tau-\zeta } \cdot  \f{ (1-\tf{\zeta}{\om})^{\nu} }{ (1-\tf{\tau}{\om})^{\nu} }
\cdot \f{\dd \zeta }{ 2\i \pi } \;. 
\enq
The expectation value in the determinant is computed along the lines already discussed at length and reads:
\beq
\bra{ 0 }\psi^{*}_{p-1}\, \psi_{-h} \ex{ \msc{J}_-(\nu,\om)  } \ex{ \msc{J}_+(\nu,\om)  } \Psi^*(\tau) \, \psi_{k-1} \ket{0}  \; = \; 
\wt{\mc{I}}^{(1)}_{pk}(\nu\mid\om) \cdot g_h^{(1)}(\tau)  \; + \; \wt{\mc{I}}^{(2)}_{ph}(\nu\mid\om) \cdot g_k^{(2)}(\tau) 
\enq
where the function $g_h^{(1)}(\tau)$ is defined as 
\beq
g_h^{(1)}(\tau) \; =\; 
\Oint{ |\zeta| > |\om|> |\tau| }{} \hspace{-2mm} \f{\tau \cdot \zeta^{h-1} }{ \zeta-\tau } \cdot 
\f{ (1-\tf{\om}{\zeta})^{\nu} }{ (1-\tf{\tau}{\om})^{\nu} } \cdot \f{\dd \zeta }{ 2\i \pi }  \;. 
\enq
Upon factorising the matrix $(\mc{N})_{ab}\, = \, \wt{\mc{I}}^{(2)}_{p_ah_b}(\nu\mid\om) $ from the determinant in \eqref{equation rep int pour M a+d b}
we are led to the representation
\bem
 M_{a+d\, b} \; = \; \Oint{ |\tau|> |\om|>|\zeta| }{} \hspace{-4mm} \f{\tau^{-r_a} \cdot \zeta^{-k_b} }{ \tau-\zeta } \cdot  \f{ (1-\tf{\zeta}{\om})^{\nu} }{ (1-\tf{\om}{\tau})^{\nu} }
\cdot \f{\dd \zeta \dd \tau }{ (2\i \pi)^2 }  \\
\; + \; \sul{m,q=1}{n+\ell} \big( \mc{N}^{-1} \big)_{mq} \cdot \wt{\mc{I}}^{(1)}_{p_q k_b}(\nu\mid\om)
\Oint{ |\zeta|> |\tau| > |\om|}{} \hspace{-4mm} \f{\tau^{-r_a} \cdot \zeta^{h_m} }{ \tau-\zeta } \cdot  \f{ (1-\tf{\om}{\zeta})^{\nu} }{ (1-\tf{\om}{\tau})^{\nu} }
\cdot \f{\dd \zeta \dd \tau }{ (2\i \pi)^2 } \;. 
\end{multline}
The first integral vanishes as can be seen by deforming the $\tau$-contour to $\infty$. Likewise, by deforming the $\tau$
contour to $\infty$ in the integral arising under the double summation symbol, one solely gets the contribution of the 
pole at $\tau=\zeta$. The latter yields the Kronecker symbol $\de_{m,n+a}$ which is there so as to enforce the constraint
that $m$ should be such that $h_m=r_a$. After some algebra, one
recasts $ M_{a+d\, b} $ in the form 
\beq
 M_{a+d\, b} \; = \; \om^{1-k_b-h_{n+a}} \cdot  \Ga\bigg( \ba{ccc} h_{n+a} & , &  k_b-\nu \\ 
								h_{n+a}-\nu  & , &  k_b \ea \bigg) \cdot 
 \f{\pl{c=1}{n+\ell}  (h_{n+a}+p_c-1)    }{  \prod_{ \substack{c=1 \\ \not=n+a} }^{n+\ell} (h_c-h_{n+a})} \cdot \msc{S}_{n+a,b}
\enq
where 
\bem
\msc{S}_{n+a,b} \; = \; \sul{q=1}{n+\ell}  \f{1}{(k_b-p_q-\nu)\cdot (h_{n+a}+p_q-1)} \cdot 
\f{\pl{c=1}{n+\ell} (h_c+p_q-1)   } {  \prod_{ \substack{c=1 \\ \not=q} }^{n+\ell} (p_c-p_q) }  \\
\; = \;  \Oint{  \Ga(\{p_c\}) }{}  \f{- 1}{(k_b-s-\nu)\cdot (h_{n+a}+s-1)} \cdot \pl{c=1}{n+\ell} 
\f{(h_c+s-1)   } {  (p_c-s) } \cdot \f{ \dd s }{ 2\i \pi } \; = \; 
\f{-1}{h_{n+a}+k_b-\nu-1} \cdot \pl{c=1}{n+\ell} \f{ (h_c+k_b-\nu-1)   } {  (p_c-k_b+\nu) } 
\end{multline}

\subsubsection*{ $\bullet$ Calculation of $ \det_{ d +\ell}\big[  M_{a\, b} \big] $ }

Summarising the results obtained so far, we get 
\beq
M_{ab} \; = \;  \f{  \sin[\pi \nu] \cdot \om^{1-k_b-t_a} }{  \pi  (t_a + k_b - 1)  }  \cdot  
	  \Ga\bigg( \ba{ccc} k_b - \nu  \, , & t_a +\nu \\  
			      k_b  \;\;\;\;  , &  t_a    \ea \bigg) \cdot \pl{c=1}{n+\ell} \bigg\{ 
  \f{ (p_c + t_a + \nu - 1) \cdot (h_c+k_b-\nu-1)   }{  (h_c-t_a-\nu)  (p_c-k_b+\nu)  } \bigg\} 
\enq
for $a=1,\dots,d$ and $b=1,\dots,d+\ell$ while 
\beq
 M_{a+d\, b} \; = \; \f{-\om^{1-k_b-h_{n+a}} }{ h_{n+a}+k_b-\nu-1}  \cdot  \Ga\bigg( \ba{ccc} h_{n+a} \; \; \; \;  , &  k_b-\nu \\ 
								h_{n+a}-\nu  \, , &  k_b \ea \bigg) \cdot 
 \f{ \prod_{c=1}^{n+\ell} \big\{ (h_c+k_b-\nu-1)   (h_{n+a}+p_c-1)  \big\}  } 
{ \prod_{c=1}^{n+\ell}  (p_c-k_b+\nu) \cdot   \prod_{ \substack{c=1 \\ \not=n+a} }^{n+\ell} (h_c-h_{n+a})} 
\enq
We now recast the matrix $M_{ab}$ into a uniform representation. Having this in mind, we introduce two sequences $\a_a$, and $\be_a$
with $a=1,\dots, d+\ell$. The sequences $\a_a$ reads 
\beq
\a_a \; = \; \f{  \sin[\pi \nu] }{  \pi  \cdot \om^{t_a}  }  \cdot  
	  \Ga\bigg( \ba{ccc} t_a +\nu \\  
			       t_a    \ea \bigg) \cdot \pl{c=1}{n+\ell} \bigg\{ \f{ (p_c + t_a + \nu - 1)  }{  (h_c-t_a-\nu)   } \bigg\} 
\enq
and
\beq
\a_{d+s} \; = \; -\om^{-h_{n+s}}  \cdot  \Ga\bigg( \ba{ccc} h_{n+s} \\ 
								h_{n+s}-\nu   \ea \bigg) \cdot 
 \f{ \prod_{c=1}^{n+\ell}    (h_{n+s}+p_c-1)  } 
{   \prod_{ \substack{c=1 \\ \not=n+a} }^{n+\ell} (h_c-h_{n+s})} 
\enq
where the index $a$ runs over $\{1,\dots,d\}$ while the index $s$ runs over $\{1,\dots,\ell\}$. Furthermore, let 
\beq
\be_b \; = \;  \om^{1-k_b}   \cdot  
	  \Ga\bigg( \ba{ccc} k_b - \nu   \\  
			      k_b     \ea \bigg) \cdot \pl{c=1}{n+\ell} \bigg\{ \f{ (h_c+k_b-\nu-1)   }{ (p_c-k_b+\nu)  } \bigg\} 
\qquad \e{for} \quad b \; = \; 1,\dots, d+\ell \;.  			      
\enq
Then,  upon extending the sequence $t_a$, for $a=d+1,\dots,d+\ell$, as
\beq
t_{a+d} \; = \; h_{n+a}-\nu
\enq
we see that $M$ is closely related to a Cauchy matrix
\beq
M_{ab} \; = \; \f{ \a_a \cdot \be_{b}  }{ k_b+t_a-1 }  \qquad \e{for} \qquad a,b \in \{ 1, \dots , d+\ell \} \;. 
\enq

\subsubsection*{ $\bullet$ Synthesis of the calculations}

By gathering all together the results obtained in the course of the proof, one arrives at the representation
\bem
\bra{ \mc{J}_{ n_{p}; n_{h} } } \ex{ \msc{J}_-(\nu,\om) } \ex{ \msc{J}_+(\nu,\om)  } \ket{  \mc{J}_{ n_{k}; n_{t} } } 
\; = \; (-1)^{\ell (n+d)} \det \bigg[ \f{  - \sin[\pi \nu] \cdot \om^{h_b+p_a-1} }{  \pi  (p_a + h_b - 1)  }  \cdot  
	  \Ga\bigg( \ba{ccc} h_b - \nu  \,  , & p_a +\nu \\  
			      h_b  \;\;\;\;\; , &  p_a    \ea \bigg)   \bigg]  \\
\times \pl{a=1}{d+\ell} \Big\{ \a_a \cdot \be_a \Big\}
\cdot \det_{n+\ell}\Big[ \f{1}{p_a+h_b-1} \Big] \cdot \det_{d+\ell}\Big[ \f{1}{t_a+k_b-1} \Big] \;. 
\end{multline}
It is then a matter of straightforward calculations to obtain the representation \eqref{definition microscopic form factor}. 
In particular, all the dependence on the auxiliary integers $\{r_a\}_1^{\ell}$ completely disappears. \qed

 \section{The multi-point restricted sums}
\label{Appendix connection avec sommes restreintes} 
 
The multi-point restricted sums were first introduced in \cite{KozKitMailTerMultiRestrictedSums}. 
This name refers to a multi-dimensional summation formula that is necessary for the computation of the localised
form factor series expansion. More precisely, given $|z_1|>\dots > |z_r|$ and generic complex numbers $\nu_1,\dots,\nu_r$
it holds 
\bem
\msc{S}\big( \{\nu_a\}_1^{r}; \{z_a\}_1^r \big) \; = \; 
\pl{s=1}{r-1} \bigg\{ \sul{ \substack{ n_p^{(s)}, n_h^{(s)} =0  \\ n_p^{(s)} - n_h^{(s)} =  \ell_s  } }{+\infty}   \sul{ \mc{J}_{ n_p^{(s)}; n_h^{(s)} } }{ } \bigg\}
\; \pl{s=1}{r-1} \mc{R} \Big( \mc{J}_{ n_p^{(s)}; n_h^{(s)} } \big| \nu_s, \nu_{s+1} ; \f{z_{s+1}}{z_s} \Big)
 \pl{s=2}{r-1}  \varpi\Big( \mc{J}_{ n_p^{(s-1)}; n_h^{(s-1)} }; \mc{J}_{ n_p^{(s)}; n_h^{(s)} } \mid  \pm \nu_s\Big)  \\
 \; = \; 
\pl{s=1}{r-1} \Bigg\{  \Big(\f{z_{s+1}}{z_s} \Big)^{ \f{ \ell_s(\ell_s+1) }{2} }  
 G\bigg( \ba{cc}   1+\ell_s-\nu_s , 1 + \ell_s + \nu_{s+1}  \\  
						1-\nu_s , 1+\nu_{s+1} \ea \bigg)   \Bigg\} \\
\times 						
\pl{s=2}{r-1} G\bigg( \ba{cc}   1 + \nu_s , 1 + \ell_{s-1}-\ell_s + \nu_s  \\  
						1-\ell_s + \nu_s , 1 +\ell_{s-1} + \nu_{s}  \ea \bigg)
\; \cdot \; \pl{b>a}{r} \Big( 1-\f{z_b}{z_a}  \Big)^{(\nu_a + \kappa_a)(\nu_b + \kappa_b)		} \;. 
\label{definition sommes restreintes type plus et moins}
\end{multline}
The function $\mc{R}$ appearing above takes the form 
\bem
\mc{R}  \Big( \mc{J}_{n_p;n_h} \mid \nu, \eta ; z  \Big)  =
\bigg(- \f{\sin[\pi \nu] }{ \pi } \cdot \f{\sin[\pi \eta] }{ \pi } \bigg)^{n_h} 
				 \cdot \f{ \pl{a<b}{n_p} (p_a-p_b)^2 \cdot \pl{a<b}{n_h} (h_a-h_b)^2 }
{ \pl{a=1}{n_p} \pl{b=1}{n_h} (p_a+h_b-1)^2 }   \\
\times  \pl{a=1}{n_p} \bigg\{ z^{p_a} \Ga\pab{p_a-\nu, p_a + \eta }{p_a, p_a} \bigg\}
\cdot \pl{a=1}{n_h} \bigg\{ z^{h_a-1} \Ga\pab{h_a+\nu, h_a - \eta}{h_a, h_a} \bigg\}  \;. 
\nonumber
\end{multline}

This summation formula has been argued in \cite{KozKitMailTerMultiRestrictedSums} on the basis of comparing the large-$N$ asymptotic expansion 
of a Toeplitz determinant generated by a symbol having Fisher-Hartwig singularities obtained by two different means. In this section of the appendix, we derive the 
above formula by using the free fermion formalism that we have developed. For this purpose we compute the expectation value
\beq
\bra{0}\msc{V}\big(\nu_1,\kappa_1 \mid z_1\big) \cdots \msc{V}\big(\nu_r,\kappa_r \mid z_r\big) \ket{0} 
\enq
in two ways. First of all, by inserting the decomposition of the identity
\beq
\sul{n_p, n_h}{} \sul{  \mc{J}_{n_p;n_h}  }{} \ket{ \mc{J}_{n_p;n_h} }\bra{ \mc{J}_{n_p;n_h} } \; = \; \e{id}
\enq
between each vertex operator and this, for the specific choice of the sequence $\kappa_a$:
\beq
\kappa_s \; = \; \ell_{s-1}\, - \, \ell_s \qquad \e{where}  \qquad \ell_0 \; = \; \ell_{r} \; = \; 0 \;. 
\label{ecriture parametrisation des kappa s}
\enq
Second, by applying repeatedly the exchange relation
\beq
\ex{ \msc{J}_+(\nu,\om) } \cdot \ex{ \msc{J}_-(\mu,z) } \;= \; \bigg( 1 \, - \,  \f{z}{\om}  \bigg)^{\mu \nu} \cdot \ex{ \msc{J}_-(\mu,z) } \cdot \ex{ \msc{J}_+(\nu,\om) } \;. 
\label{ecriture relation echange J+ et J -}
\enq
 satisfied by the current operators. The second method yields
\beq
\bra{0}\msc{V}\big(\nu_1,\kappa_1 \mid z_1\big) \cdots \msc{V}\big(\nu_r,\kappa_r \mid z_r\big) \ket{0}   \; = \; \pl{b>a}{r} \Big( 1-\f{z_b}{z_a}  \Big)^{(\nu_a + \kappa_a)(\nu_b + \kappa_b)}
\label{ecriture calcul valeur moyenne produit op de vertex}
\enq
while the first one leads to 
\beq
\bra{0}\msc{V}\big(\nu_1,\kappa_1 \mid z_1\big) \cdots \msc{V}\big(\nu_r,\kappa_r \mid z_r\big) \ket{0}  \; \; = \hspace{-3mm}
\sul{ \substack{ n_p^{(s)}-n_h^{(s)}  = \\  \kappa_{s+1} + n_p^{(s+1)}-n_h^{(s+1)}  } }{}
\sul{  \mc{J}_{n_p^{(s)} ;n_h^{(s)} }  }{} 
\pl{s=1}{r}\bigg\{  \bra{ \mc{J}_{n_p^{(s-1)};n_h^{(s-1)}} }  \msc{V}\big(\nu_s,\kappa_s \mid z_s\big)  \ket{ \mc{J}_{n_p^{(s)};n_h^{(s)}} } \bigg\}
\enq
Above, we have adopted the convention that 
\beq
n_p^{(0)}=n_h^{(0)}=n_p^{(r)}=n_h^{(r)}=0 \qquad viz. \qquad \ket{ \mc{J}_{n_p^{(0)};n_h^{(0)}} } \; = \; \ket{ \mc{J}_{n_p^{(r)};n_h^{(r)}} } \; = \; \ket{0} \;. 
\enq
The constraints on  the differences in the numbers of particles and holes in the excitations of the intermediary states
issue from charge conservation requirements and can be recast, due to the parametrisation \eqref{ecriture parametrisation des kappa s},
\beq
n_p^{(s)}-n_h^{(s)}  \; = \;   \ell_s \quad s=1,\dots, r-1 \;. 
\enq
Then a straightforward calculation recasts the sum as
\beq
\bra{0}\msc{V}\big(\nu_1,\kappa_1 \mid z_1\big) \cdots \msc{V}\big(\nu_r,\kappa_r \mid z_r\big) \ket{0}  \cdot C\big( \{\nu_a\}_1^{r}; \{z_a\}_1^r \big)  \; = \;
 \msc{S}\big( \{\nu_a\}_1^{r}; \{z_a\}_1^r \big)
\enq
where the constant $C\big( \{\nu_a\}_1^{r}; \{z_a\}_1^r \big)$ takes the form
\beq
C\big( \{\nu_a\}_1^{r}; \{z_a\}_1^r \big) \; = \; \pl{s=1}{r}\bigg\{ G\pab{ 1  -  \nu_s - \kappa_s  }{ 1 - \nu_s} \cdot 
\Big( -\f{\sin[\pi \nu_s]}{\pi}  \Big)^{ - \ell_{s-1} } 
\cdot \f{ z_s^{  \frac{1}{2}\kappa_{s}(\kappa_{s}-1) + \kappa_s \ell_s }   }
{ (-1)^{ \frac{\ell_{s-1}(\ell_{s-1}+1)}{2} + \frac{\kappa_{s}(\kappa_{s}+1)}{2} } }  \bigg\} \cdot 
\pl{s=1}{r-1} \bigg\{  \Big( \f{ z_{s+1} }{ z_{s} } \Big)^{\ell_s}  \bigg\} \;. 
\enq
By using the reduction property of the Barnes function \eqref{Formule de la double redcution pour la fonction de Barnes}
and the parametrisation of $\kappa_s$ in terms of $\ell_s$, one gets that 
\bem
\pl{s=1}{r}\bigg\{ G\pab{ 1  -  \nu_s - \kappa_s  }{ 1 - \nu_s} \cdot 
\Big( -\f{\sin[\pi \nu_s]}{\pi}  \Big)^{ - \ell_{s-1} } 
\cdot (-1)^{ - \frac{\ell_{s-1}(\ell_{s-1}+1)}{2} - \frac{\kappa_{s}(\kappa_{s}+1)}{2} }   \bigg\}  \\ 
\; =\; \pl{s=1}{r-1} \Bigg\{ 
 G\bigg( \ba{cc}   1+\ell_s-\nu_s , 1 + \ell_s + \nu_{s+1}  \\  
						1-\nu_s , 1+\nu_{s+1} \ea \bigg)   \Bigg\} 
\pl{s=2}{r-1} G\bigg( \ba{cc}   1 + \nu_s , 1 + \ell_{s-1}-\ell_s + \nu_s  \\  
						1-\ell_s + \nu_s , 1 +\ell_{s-1} + \nu_{s}  \ea \bigg) \; , 	
\end{multline}
and
\beq
 \pl{s=1}{r} \Big\{  z_s^{  \frac{1}{2}\kappa_{s}(\kappa_{s}-1) + \kappa_s \ell_s }  \Big\} \cdot 
\pl{s=1}{r-1} \bigg\{  \Big( \f{ z_{s+1} }{ z_{s} } \Big)^{\ell_s}  \bigg\} \; = \; 
\pl{s=1}{r-1} \Bigg\{  \Big(\f{z_{s+1}}{z_s} \Big)^{ \f{ \ell_s(\ell_s+1) }{2} }  \Bigg\}  \;. 
\enq
Putting all of this together leads to \eqref{definition sommes restreintes type plus et moins}. \qed


\begin{thebibliography}{10}

\bibitem{AlexandrovZaborodinTechniquesOfFreeFermions}
A.~Alexandrov and A.~Zabrodin, \emph{{"Free fermions and tau-functions."}},
  J.Geom.Phys. \textbf{67} (2013), 37--80.

\bibitem{BarnesDoubleGaFctn1}
E.W. Barnes, \emph{{"Genesis of the double gamma function."}}, Proc. London
  Math. Soc. \textbf{\bf{31}} (1900), 358--381.

\bibitem{BeliavinPolyakovZalmolodchikovCFTin2DQFT}
A.~A. Beliavin, A.~M. Polyakov, and Z.~B. Zalmolodchikov, \emph{{"Infinite
  conformal symmetry in two-dimensional quantum field theory"}}, Nucl. Phys.
  \textbf{B \bf 241} (1984), 333--380.

\bibitem{BenfattoFalcoMastropietroUniversalityOfPerturbedFreefermionsModelsCritExpAndRespondseFcts}
G.~Benfatto, P.~Falco, and V.~Mastropietro, \emph{{"Universality of
  one-dimensional Fermi systems, I. Response functions and critical
  exponents."}}, Comm. Math. Phys. \textbf{\bf 330} (2014), 153--215.

\bibitem{BenfattoFalcoMastropietroUniversalityOfPerturbedFreefermionsModelsLuttLiqStructure}
\bysame, \emph{{"Universality of one-dimensional Fermi systems, II. The
  Luttinger liquid structure."}}, Comm. Math. Phys. \textbf{\bf 330} (2014),
  217--282.

\bibitem{BetheSolutionToXXX}
H.~Bethe, \emph{{"Zur Theorie der Metalle: Eigenwerte und Eigenfunktionen der
  linearen Atomkette."}}, Zeitschrift f$\ddot{u}$r Physik \textbf{\bf 71}
  (1931), 205--226.

\bibitem{BloteCardyNightingalePredictionL-1correctionsEnergyAscentralcharge}
H.W.J. Bl\"{o}te, J.L. Cardy, and M.P. Nightingale, \emph{{"Conformal
  invariance, the central charge, and universal finite-size amplitudes at
  criticality."}}, Phys. Rev. Lett. \textbf{\bf 56} (1986), 742--745.

\bibitem{BogoliubiovIzerginKorepinBookCorrFctAndABA}
N.~M. Bogoliubov, A.~G. Izergin, and V.~E. Korepin, \emph{{"Quantum inverse
  scattering method, correlation functions and algebraic Bethe Ansatz."}},
  Cambridge mono. on math. phys., 1993.

\bibitem{CardyConformalExponents}
J.L. Cardy, \emph{{"Conformal invariance and universality in finite-size
  scaling."}}, J. Phys. A: Math. Gen. \textbf{\bf 17} (1984), L385--387.

\bibitem{DestriDeVegaAsymptoticAnalysisCountingFunctionAndFiniteSizeCorrectionsinTBAFiniteMagField}
C.~Destri and H.~J. DeVega, \emph{{"Unified approach to Thermodynamic Bethe
  Ansatz and finite size corrections for lattice models and field theories."}},
  Nucl. Phys. B \textbf{\bf{438}} (1995), 413--454.

\bibitem{DeVegaWoynarowichFiniteSizeCorrections6VertexNLIEmethod}
H.J. DeVega and F.~Woynarowich, \emph{{"Method for calculating finite size
  corrections in Bethe Ansatz systems- Heisenberg chains and 6-vertex
  model."}}, Nucl. Phys. B \textbf{\bf{251}} (1985), 439--456.

\bibitem{DiFrancescoMathieuSenechalCFTKniga}
P.~DiFrancesco, P.~Mathieu, and D.~S\'{e}n\'{e}chal, \emph{{"Conformal field
  theory."}}, Graduate texts in contemporary physics, Springer, 1997.

\bibitem{GaudinFonctionOndeBethe}
M.~Gaudin, \emph{La fonction d'onde de {B}ethe}, Collection du commisariat
  \`{a} l'\'{e}nergie atomique, Masson, 1990.

\bibitem{GiulianiGreenblattMastropietroScalingbehavior2nFctsofEnergyCorrsNonIntIsing}
A.~Giuliani, R.~L. Greenblatt, and V.~Mastropietro, \emph{{"The scaling limit
  of the energy correlations in non-integrable Ising models."}}, J. Math. Phys.
  \textbf{\bf 53} (2012), 095214.

\bibitem{GiulianiMastropietroProofOfCntralCharge1/2forPertOf2DIsing}
A.~Giuliani and V.~Mastropietro, \emph{{"Universal finite-size corrections and
  the central charge in non-solvable Ising models."}}, Comm. Math. Phys.
  \textbf{\bf 324} (2013), 179--214.

\bibitem{GlazmanImambekovSchmidtReviewOnNLLuttingerTheory}
L.I. Glazman, A.~Imambekov, and T.L. Schmidt, \emph{{"One-dimensional quantum
  liquids: beyond the Luttinger liquid paradigm."}}, Rev. Mod. Phys.
  \textbf{84} (2012), 1253--1306.

\bibitem{GriffithsUniversalityAndExponentsParameterDepending}
R.B. Griffiths, \emph{{"Dependence of critical indicies on a parameter."}},
  Phys. Rev. Lett. \textbf{\bf 24} (1970), 1479--1482.

\bibitem{HaldaneCritExponentsAndSpectralPropXXZ}
F.D.M. Haldane, \emph{{"General relation of correlation exponents and spectral
  properties of one-dimensional Fermi systems: Application to the anisotropic s
  = 1/2 Heisenberg chain."}}, Phys. Rev. Lett. \textbf{\bf 45} (1980),
  1358--1362.

\bibitem{HaldaneLuttingerLiquidCaracterofBASolvableModels}
\bysame, \emph{{"Demonstration of the ``Luttinger liquid`` character of
  Bethe-Ansatz soluble models of 1-D quantum fluids"}}, Phys. Lett. A
  \textbf{\bf 81} (1981), 153--155.

\bibitem{HaldaneStudyofLuttingerLiquid}
\bysame, \emph{{"Luttinger liquid theory of one-dimensional quantum fluids: I.
  Properties of the Luttinger model and their extension to the general 1D
  interacting spinless Fermi gas"}}, J. Phys. C: Solid State Phys. \textbf{\bf
  14} (1981), 2585--2609.

\bibitem{JimMiwaSatoQuantumFieldsI}
M.~Jimbo, T.~Miwa, and M.~Sato, \emph{{"Holonomic quantum fields I."}}, Publ.
  RIMS \textbf{\bf{14}} (1977), 223--267.

\bibitem{KadanoffGotzeHAmblenHechtLexisPAiCiauskasRaylSwiftComparaisonUniversalitesurResultatsExperiencePlusDiscussion}
L.~P. Kadanoff, W.~Gotze, R.~Hecht D.~Hamblen, E.~A.~S. Lewis, V.~V.
  Palciauskas, M.~Rayl, J.~Swift, D.~Aspens, and J.~Kane, \emph{{"Static
  phenomena near critical points: theory and experiment"}}, Rev. Mod. Phys.
  \textbf{\bf 39} (1967), 395-431.

\bibitem{KozKitMailSlaTerXXZsgZsgZAsymptotics}
N.~Kitanine, K.K. Kozlowski, J.-M. Maillet, N.A. Slavnov, and V.~Terras,
  \emph{{"Algebraic Bethe Ansatz approach to the asymptotics behavior of
  correlation functions."}}, J. Stat. Mech: Th. and Exp. \textbf{04} (2009),
  P04003.

\bibitem{KozKitMailSlaTerEffectiveFormFactorsForXXZ}
\bysame, \emph{{"On the thermodynamic limit of form factors in the massless XXZ
  Heisenberg chain."}}, J. Math. Phys. \textbf{50} (2009), 095209.

\bibitem{KozKitMailSlaTerRestrictedSums}
\bysame, \emph{{"A form factor approach to the asymptotic behavior of
  correlation functions in critical models."}}, J. Stat. Mech. : Th. and Exp.
  (2011), P12010.

\bibitem{KozKitMailSlaTerThermoLimPartHoleFormFactorsForXXZ}
\bysame, \emph{{"Thermodynamic limit of particle-hole form factors in the
  massless XXZ Heisenberg chain."}}, J. Stat. Mech. : Th. and Exp. (2011),
  P05028.

\bibitem{KozKitMailSlaTerRestrictedSumsEdgeAndLongTime}
\bysame, \emph{{"Form factor approach to dynamical correlation functions in
  critical models."}}, J. Stat. Mech. (2012), P09001.

\bibitem{KozKitMailTerMultiRestrictedSums}
N.~Kitanine, K.K. Kozlowski, J.-M. Maillet, and V.~Terras,
  \emph{{"Long-distance asymptotic behavior of multi-point correlation
  functions in massless quantum integrable models."}}, J. Stat. Mech. (2014),
  P05011.

\bibitem{KMTFormfactorsperiodicXXZ}
N.~Kitanine, J.-M. Maillet, and V.~Terras, \emph{{"Form factors of the XXZ
  Heisenberg spin-1/2 finite chain."}}, Nucl. Phys. B \textbf{554} (1999),
  647--678.

\bibitem{KlumperBatchelorNLIEApproachFiniteSizeCorSpin1XXZIntroMethod}
A.~Kl\"{u}mper and M.~T. Batchelor, \emph{{"An analytic treatment of
  finite-size corrections of the spin-1 antiferromagnetic XXZ chain."}}, J.
  Phys. A: Math. Gen. \textbf{\bf 23} (1990), L189.

\bibitem{KlumperBatchelorPearceCentralChargesfor6And19VertexModelsNLIE}
A.~Kl\"{u}mper, M.~T. Batchelor, and P.~A. Pearce, \emph{{"Central charges for
  the 6- and 19-vertex models with twisted boundary conditions."}}, J. Phys. A:
  Math. Gen. \textbf{\bf 24} (1991), 3111--3133.

\bibitem{KlumperWehnerZittartzConformalSpectrumofXXZCritExp6Vertex}
A.~Kl\"{u}mper, T.~Wehner, and J.~Zittartz, \emph{{"Conformal spectrum of the
  6-vertex model."}}, J. Phys. A: Math. Gen. \textbf{\bf 26} (1993),
  2815--2827.

\bibitem{KorepinSlavnovApplicationDualFieldsFredDets}
V.~E. Korepin and N.~A. Slavnov, \emph{{"Correlation functions of fields in
  one-dimensional Bose gas."}}, Comm. Math. Phys. \textbf{\bf 136} (1991),
  633--644.

\bibitem{KozFFConjFieldNLSELatticeSpacingGoes0}
K.K. Kozlowski, \emph{{"On form factors of the conjugated field in the
  non-linear Schr\"{o}dinger model."}}, J. Math. Phys. \textbf{52} (2011),
  083302.

\bibitem{KozReducedDensityMatrixAsymptNLSE}
\bysame, \emph{{"Large-distance and long-time asymptotic behavior of the
  reduced denisty matrix in the non-linear Schr\"{o}dinger model."}}, Ann.
  Henri-Poincar\'{e} \textbf{16} (2015), 437--534.

\bibitem{LiebExcitationStructureBoseGas}
E.H. Lieb, \emph{{"Exact analysis of an interacting Bose gas II: the excitation
  spectrum."}}, Phys. Rev. \textbf{\bf{130}} (1967), 1616--1624.

\bibitem{LutherPeschelCriticalExponentsXXZZeroFieldLuttLiquid}
A.~Luther and I.~Peschel, \emph{{"Calculation of critical exponents in two
  dimensions from quantum field theory in one dimension."}}, Phys. Rev. B
  \textbf{\bf{12}(9)} (1975), 3908--3917.

\bibitem{MastropietroScalingofEnergyEnergyCorrFctsNonIntIsing}
V.~Mastropietro, \emph{{"Ising models with four spin interactions at
  criticality"}}, Comm. Math. Phys. \textbf{\bf 244} (2004), 595--642.

\bibitem{NozieresTheoryIntFermiSyst}
P.~Nozi\`{e}res, \emph{{"Theory of interacting Fermi systems."}}, Advanced book
  classics, Westview press, 1997.

\bibitem{OkunkovSomeAlgebraicManipulationWithFreeFermion}
A.~Okunkov, \emph{{"Infinite wedge and random partitions."}}, Selecta
  Mathematica. \textbf{\bf 7} (2001), 57-81. 

\bibitem{OotaInverseProblemForFieldTheoriesIntegrability}
T.~Oota, \emph{{"Quantum projectors and local operators in lattice integrable
  models."}}, J. Phys. A: Math. Gen. \textbf{\bf 37} (2004), 441--452.

\bibitem{PatashinskiiPokrovskiiSecondOrderPhaseTransitionBoseFluid}
A.~Z. Patashinskii and V.~L. Pokrovskii, \emph{{"Second order phase transitions
  in a Bose fluid."}}, Sov. Phys. JETP \textbf{\bf 19} (1964), 677--691.

\bibitem{PinsonSpencerFirstApproachToUniversalityInPertOf2DIsing}
H.~Pinson and T.~Spencer, \emph{{"Universality in 2D Critical Ising Model."}},
  preprint.

\bibitem{PolyakovArgumentAbouMicoOriginOfUniversality}
A.~M. Polyakov, \emph{{"Microscopic description of critical phenomena."}},
  Soviet Phys. JETP \textbf{\bf 28} (1969), 533--539.

\bibitem{PolyakovArgumentAboutCFTInvarianceCriticalCorrelators}
\bysame, \emph{{"Conformal symmetry of critical fluctuations."}}, Pis. ZhETP
  \textbf{\bf 12} (1970), 538--541.

\bibitem{PolyakovArgumentAboutpropertiesofLongDistAsympt}
\bysame, \emph{{"Properties of long and short range correlations in the
  critical region."}}, Soviet Phys. JETP \textbf{\bf 30} (1970), 151--157.

\bibitem{CauxGlazmanImambekovShasiNonUniversalPrefactorsFromFormFactors}
A.~Shashi, L.~I. Glazman, J.-S. Caux, and A.~Imambekov, \emph{{"Non-universal
  prefactors in correlation functions of 1D quantum liquids."}}, Phys. Rev. B.
  \textbf{84} (2011), 045408.

\bibitem{SlavnovFormFactorsNLSE}
N.~A. Slavnov, \emph{{"Non-equal time current correlation function in a
  one-dimensional Bose gas."}}, Theor. Math. Phys. \textbf{\bf 82} (1990),
  273--282.

\bibitem{SmirnovProofOfConformalInvarianceInIsing}
S.~Smirnov, \emph{{"Conformal invariance in random cluster models. I.
  Holomorphic fermions in the Ising model."}}, Ann. of Math. \textbf{\bf 172} (2010), 1435-1467.

\bibitem{SuzukiCorrespondence(D+1)StatPhysDQuantumHamiltonians}
M.~Suzuki, \emph{{"Relationship between $d$-dimensional quantal spin systems
  and $(d+1)$-dimensional Ising systems."}}, Prog. Theor. Phys. \textbf{\bf 56}
  (1976), 1454--1469.

\end{thebibliography}
\end{document}